\documentclass[11pt]{article}
\usepackage{amsfonts}
\usepackage{amssymb}
\usepackage{graphicx}
\usepackage{amsmath}
\usepackage{amsthm}
\usepackage{color}
\usepackage{multirow}
\usepackage{url}
\usepackage{float}
\usepackage{booktabs} 
\usepackage{makecell}
\usepackage{listings}
\usepackage{subcaption}
\usepackage{algorithm}
\usepackage{enumerate}
\usepackage{tikz}
\newcommand{\impactline}[1]{
	\begin{tikzpicture}
		\draw (0,0) -- (5,0); 
		\draw (0,-0.1) -- (0,0.1); 
		\node at (0,-0.3) {\tiny 850}; 
		\draw (1.25,-0.05) -- (1.25,0.05); 
		\draw (2.5,-0.1) -- (2.5,0.1); 
		\draw (3.75,-0.05) -- (3.75,0.05); 
		\draw (5,-0.1) -- (5,0.1); 
		\node at (5,-0.3) {\tiny 1050}; 
		\foreach \x/\num in {#1} {
			\fill[red] (\x*0.05,0) circle (0.05cm); 
		}
	\end{tikzpicture}
}

\newcommand{\impactlines}[1]{
	\begin{tikzpicture}
		\draw (0,0) -- (5,0); 
		\draw (0,-0.1) -- (0,0.1); 
		\node at (0,-0.3) {\tiny 275}; 
		\draw (1.25,-0.05) -- (1.25,0.05); 
		\draw (2.5,-0.1) -- (2.5,0.1); 
		\draw (3.75,-0.05) -- (3.75,0.05); 
		\draw (5,-0.1) -- (5,0.1); 
		\node at (5,-0.3) {\tiny 560}; 
		\foreach \x/\num in {#1} {
			\fill[blue] (\x*0.008757,0) circle (0.05cm); 
		}
	\end{tikzpicture}
}

\usepackage{algpseudocode}
\lstnewenvironment{Rcode}{\lstset{language=R}}{}
\newtheorem{theorem}{Theorem}[section]

\newtheorem{remark}[theorem]{Remark}

\definecolor{battleshipgrey}{rgb}{0.52, 0.52, 0.51}
\definecolor{navyblue}{rgb}{0.0, 0.0, 0.5}
\definecolor{arsenic}{rgb}{0.23, 0.27, 0.29}
\definecolor{oldmauve}{rgb}{0.4, 0.19, 0.28}
\usepackage[colorlinks=TRUE]{hyperref}
\hypersetup{
	colorlinks=true,
	linkcolor=navyblue,
	filecolor=blue,      
	urlcolor=oldmauve,
	citecolor=navyblue,
	pdftitle={Overleaf Example},
	pdfpagemode=FullScreen,
}
\usepackage{natbib}
\setcitestyle{authoryear}
\begin{document}
	\title{ \bf fsemipar: an R package for SoF  semiparametric regression}
	\author{Silvia Novo{*} \hspace{2pt} Germ\'{a}n Aneiros{**}   \\	
		{\normalsize*Statistics Department, Universidad Carlos III de Madrid}\\
		{\normalsize **Mathematics Department, CITIC,  Universidade da Coruña}}
	
	\date{}
	\maketitle
	\begin{abstract} 
		Functional data analysis  
		has become a tool of interest in applied areas such as  
		economics, medicine, and chemistry.  
		Among the techniques developed in recent literature, functional semiparametric regression stands out for its balance between flexible modelling and output interpretation.
		Despite the large variety of research papers dealing with scalar-on-function (SoF) semiparametric models, 
		there is a notable gap in software tools for their implementation.
		This article introduces the R package \texttt{fsemipar}, tailored for these models. \texttt{fsemipar} not only estimates functional single-index models using kernel smoothing techniques
		but also estimates and selects relevant scalar variables in semi-functional models with multivariate linear components. 
		A standout feature is its ability to identify impact points of a curve on the response, even in models with multiple functional covariates, and to integrate both continuous and pointwise effects of functional predictors within a single model.
		In addition, it allows the use of location-adaptive estimators based on the $k$-nearest-neighbours approach for all the semiparametric models included.
		Its flexible interface empowers users to customise a wide range of input parameters
		and includes the standard S3 methods for prediction, statistical analysis, and estimate visualization (\texttt{predict}, \texttt{summary}, \texttt{print}, and \texttt{plot}), enhancing clear result interpretation. Throughout the article, we illustrate the functionalities and the practicality of \texttt{fsemipar} using two chemometric datasets.
	\end{abstract}
	\vspace{1cm}
	\noindent \textit{Keywords: } R package; functional data analysis; semiparametric regression;  functional single-index model; semi-functional models; variable selection; impact point selection; $k$-nearest neighbours
	
	\section{Introduction}
	The technological advances of recent decades have enabled the recording of vast amounts of data, leading to a revolution in data analysis. Applied disciplines now require tools to handle complex datasets, posing ongoing challenges for statisticians in both methodological and computational aspects.  For example,  real-world applications often involve an extensive number of random variables. Data may even be infinite-dimensional; that is, the observations of these variables are measured either continuously or in a finely-spaced grid over a time interval, over a surface or even more intricate structures. For instance, in economics, intraday stock price can be considered as a curve (see, e.g., \citealt{Riceetal2020}); in chemometrics, the measurement of a spectrometer results in a curve (see, e.g., \citealt{DaiGenton2018}); in medicine, magnetic resonance imaging outputs are images (see, e.g. 
	 \citealt{lietal2019}). In such cases, data ``atoms'' are random functions, leading to the categorization of this data type as \emph{functional data}.
	
	The term \emph{Functional Data Analysis} (FDA) encompasses those statistical methods designed to handle functional data. Currently, functional regression has garnered significant attention within the statistical community. In particular, scalar-on-function (SoF) regression, i.e. the case of scalar response and functional covariates, has emerged as a prolific area of research (see \citealt{Greven2017} and \citealt{Reissetal2017} for recent reviews). Several SoF regression models have been formulated, including parametric, nonparametric, and semiparametric approaches  (if we follow the terminology of conventional finite-dimensional regression). Parametric models, like the functional linear model (see, e. g., chapter 12 of \citealt{RamsaySilverman2005} for a comprehensive description and \citealt{Febreroetal2017} for a discussion on estimation techniques), provide a straightforward practical interpretation but are limited in terms of flexibility. In contrast, nonparametric models (see, e. g., \citealt{FerratyVieu2006} for a detailed development and \citealt{LingVieu2018} for a review) provide a high level of flexibility, but the outputs are challenging to interpret. As a middle point, semiparametric regression (see \citealt{LingVieu2020} for a recent survey) provides a balance between parametric and nonparametric approaches allowing for flexible models while yielding interpretable results.

The practical application of functional linear and functional nonparametric regression has been facilitated by computational tools designed for their implementation, typically developed within the R programming environment (\citealt{RCoreTeam}). Among these, the R packages \texttt{fda} (see \citealt{Ramsayetal2022}) and \texttt{fda.usc} (see \citealt{FebreroOviedo2012}) stand out as key references in the FDA field. In the context of SoF regression, 	the former allows for the estimation of the functional linear model using the  techniques described in \citealt{RamsaySilverman2005}.  The latter includes several parametric and nonparametric functional regression
models. It can estimate the functional linear model using several approaches (including those described in \citealt{Cardotetal1999}, \citealt{RamsaySilverman2005} and \citealt{PredaSaporta2005}),  the functional nonparametric model (described in \citealt{FerratyVieu2006}) and various generalised additive models for functional data (described in \citealt{Febrero2013}).
There are also two other major packages designed specifically for regression problems involving functional variables:  \texttt{FDBoost} (see \citealt{FDBoost2020}) and \texttt{refund} (see \citealt{refund2023}). In the SoF context, the package
\texttt{FDBoost} fits additive models using a component-wise gradient boosting algorithm. It allows for functional and scalar covariates, but the effects of functional predictors currently implemented are functional linear and smooth interactions. Regarding the \texttt{refund} package, it  implements penalised functional regression, including 
generalised additive models for functional data using spline-based methods (developed in \citealt{McLeanetal2014}). 

While there are other packages tailored to specific models and applications (see \citealt{CranTask2022} for an overview of R packages for functional data analysis), to the best of our knowledge there was no R package dealing with the estimation of functional single-index models prior to the introduction of \texttt{fsemipar}. It is worth noting that despite the practical advantages of functional semiparametric regression over parametric and nonparametric approaches, implementation is usually complex and requires advanced knowledge of mathematical programming. 
To address this gap, we created the R package \texttt{fsemipar} (see \citealt{AneirosNovo2023}) aiming to simplify the use of SoF semiparametric models. It can estimate functional single-index models (see \citealt{Amatoetal2006}) and can also simultaneously estimate and select relevant scalar variables in models composed of a multivariate linear component and either a functional single-index part (see \citealt{NovoAneirosVieu2020}) or a functional nonparametric part (see \citealt{AneirosFerratyVieu2015}). 
For that, it allows both kernel-based and $k$-nearest-neighbours-based estimators, allowing the user to choose between a global smooth or a location-adaptive one. A remarkable feature of \texttt{fsemipar} it is the selection of impact points of a curve on the scalar response variable (see \citealt{AneirosVieu2014}). Essentially, this means the package can incorporate discretised observations from a curve into the regression model. From these observations, it then identifies and selects the most predictive points of the discretisation interval, termed ``impact points'' or ``points of impact''. As a consequence, it allows for the inclusion of both pointwise and continuous effects of functional variables in the response (see \citealt{AneirosVieu2015} and \citealt{NovoVieuAneiros2021}), leading to interpretability, model simplification and good predictive performance.
An important feature of its functions is their ability to accommodate a wide range of customisable input parameters. This flexibility enables expert users to achieve a more precise fit. At the same time, the functions require only the data as a mandatory argument, thereby simplifying usage for non-experts. 
 
In this tutorial, we present the R package \texttt{fsemipar}, available from the Comprehensive R Archive Network (CRAN) at \url{https://CRAN.R-project.org/package=fsemipar}. The remainder of the paper is structured as follows.
Section 2 addresses SoF semiparametric regression with a single functional covariate. Section 3 deals with SoF semiparametric models incorporating multiple scalar predictors and a functional covariate. Section 4 discusses SoF semiparametric regression with multiple scalar predictors derived from the discretization of a curve and a functional covariate. The three sections share the same structure (I=2,3,4): Section I.1 introduces the modeling approaches implemented in \texttt{fsemipar}, Section I.2 summarizes the techniques for estimating such models, and Section I.3 presents the software infrastructure, introducing the main functions and classes of the package \texttt{fsemipar} tailored for each model. To highlight the capabilities of the package, Section I.3 also contains practical applications using two real datasets from the chemometric field. Finally, Section 5 includes a discussion about the functions included and some directions for future development. Due to space constraints, we have not delved into every facet of the package. For a detailed description of each function, readers are directed to the package manual.

	\section{Single functional covariate}
	In Section 2, we focus on SoF semiparametric models with single functional predictor. Section 2.1 introduces the functional single-index model as a competitor of the linear model and the nonparametric model. In Section 2.2, we discuss the estimation techniques implemented in \texttt{fsemipar}. Finally, Section 2.3 presents the associated functions of the package and illustrates their use with a real chemometric dataset. 
		
	\subsection{The functional single-index model}
Let $Y$ be a scalar random variable, $\mathcal{X}$ (also denoted as $\mathcal{X}(t)$ with $t\in\mathcal{I}$ and $\mathcal{I}\subset\mathbb{R}$) a functional covariate valued in some infinite-dimensional space $\mathcal{H}$, and $\varepsilon$, a random error. We assume that we observe data pairs $\{(Y_i,\mathcal{X}_i), i=1,\dots,n\}$ independent and identically distributed (i.i.d.) to $(Y,\mathcal{X})$. We are interested in predict $Y$ using $\mathcal{X}$.

 Literature in FDA focused on three different regression strategies to address this situation:
\begin{itemize}
\item \emph{The linear (parametric) approach}. The functional linear model (FLM) is formulated under the assumption of a linear relationship between the functional covariate and the response, represented as:
$Y_i=\int_\mathcal{I}\theta_0(t)\mathcal{X}_i(t)dt+\varepsilon_i$ $i=1,\dots,n$.
Here, $\mathcal{H}=L^2(\mathcal{I})$ and   $\theta_0$ is an unknown parameter-function defined on $\mathcal{I}$.  One of the strengths of this model is its ability to provide an interpretable output, $\theta_0$, and its relative ease of estimation compared to alternative models. The main disadvantage is that it may not be reliable in some real applications due to its dependence on the linearity assumption.
\item \emph{The nonparametric approach}. To address the inflexibility of the FLM, a natural step is to move away from the strict linear hypothesis and instead adopt a basic smooth condition. This lead us to the functional nonparametric model (FNM) given by the expression: $Y_i=m(\mathcal{X}_i)+\varepsilon_i$ $i=1,\dots,n$. In this context, $\mathcal{H}$ is a semimetric space and $m(\cdot)$ an unknown, smooth, non-linear functional operator. The FNM is more reliable than the FLM in practice, but $m(\cdot)$ is harder to represent and interpret, and also more difficult to estimate.
\item \emph{The semiparametric approach}. To combine the strengths of both parametric and nonparametric approaches, an appealing technique is to assume that the functional predictor acts on the response only through its projection onto some one-dimensional subspace. This leads to the functional single-index model (FSIM), represented as:
	\begin{equation}
		Y_i=r\left(\left<\theta_0,\mathcal{X}_i\right>\right)+\varepsilon_i, \quad i=1,\dots,n\label{FSIM}.
	\end{equation}
In this case, $\mathcal{H}$ is a separable Hilbert space and $\left\langle \cdot, \cdot \right\rangle$ denotes its the inner product; $\theta_0 \in \mathcal{H}$ is the functional index: a function-parameter that allows to summarize the information carried in $\mathcal{X}_i$ to predict $Y_i$,  and $r$ is the unknown, non-linear, smooth link function. 
The FSIM can be viewed as an extension of the FLM, offering enhanced flexibility due to the smooth link function \( r \). Conversely, it can be considered a specific instance of the FNM, but with the advantage of dimensionality reduction, given that \( r \) operates on a scalar rather than a function like \( m(\cdot) \). As a consequence, the FSIM offers interpretable and representable outputs  (\(\theta_0\) and \(r\)).

\end{itemize}
We can implement the FLM using several R functions: \texttt{fRegress} (estimation based on  basis representation)  of the \texttt{fda} R package, as well as the functions \texttt{fregre.basis} (estimation based on basis representation), \texttt{fregre.pc} (functional principal component estimation) and \texttt{fregre.pls} (partial least-squares estimation) of the \texttt{fda.usc} package. For the implementation of the FNM, we can use the function \texttt{fregre.np} (nonparametric kernel estimation) of the \texttt{fda.usc} package. We can also refer to the codes created by F. Ferraty, available on his website: \url{https://www.math.univ-toulouse.fr/~ferraty/SOFTWARES/NPFDA/index.html}. However, to the best of our knowledge, there is no function available for the direct implementation of the FSIM.
In the subsequent section, we will introduce the techniques for estimating the FSIM that are included in the \texttt{fsemipar} package.

\subsection{Link function and functional index estimation}\label{sec:FNM}

Since the FSIM can be seen as a particular case of the FNM, we can estimate the former by adapting the estimation techniques used for the latter. In this section, we will briefly discuss how this is achieved. It is important to note that the package can also estimate models with functional nonparametric components. This capability is precisely why we have explored functional nonparametric estimation in depth throughout this section. 

\noindent\citet{FerratyVieu2006} proposed the functional extension of the classical Nadaraya-Watson kernel estimator (see \citealp{Nadaraya1964} and \citealp{Watson1964}) for  the FNM. 
The following expression provides the kernel estimator for $m(\cdot)$:
\begin{equation}
	\widehat{m}_h(\chi)=\frac{\sum_{i=1}^nY_iK\left(h^{-1}d(\mathcal{X}_i,\chi)\right)}{\sum_{i=1}^nK\left(h^{-1}d(\mathcal{X}_i,\chi)\right)}, \ \forall \chi\in\mathcal{H}.
	\label{kernel}
\end{equation}
Here, $h\in\mathbb{R}^+$ represents the bandwidth, $d$ is a general semi-metric, which allows the computation of the proximity between functional data, and $K$ is a real-valued asymmetrical kernel.

\noindent Alternatively, \cite{Burbaetal2009} studied the functional version of the $k$NN estimator  (see \citealp{Collomb1979} and \citealp{Devroyeetal1994}, among others) for $m(\cdot)$. For each element $\chi$ within $\mathcal{H}$, it calculates the regression solely based on the $k$ sample observations that exhibit the closest proximity to said element. Specifically, the $k$NN estimator for $m(\cdot)$ is defined as follows
\begin{equation}
	\widehat{m}_k^\ast(\chi)=\frac{\sum_{i=1}^nY_iK\left(H_{k,\chi}^{-1}d(\mathcal{X}_i,\chi)\right)}{\sum_{i=1}^nK\left(H_{k,\chi}^{-1}d(\mathcal{X}_i,\chi)\right)}, \ \forall \chi\in\mathcal{H}.
	\label{kNN}
\end{equation}
\noindent Here, $k\in \mathbb{Z}^+$ is the tuning parameter and $H_{k,\chi}=\min\left\{h\in \mathbb{R}^+ \mbox{\text{ such that }} \sum_{i=1}^n1_{B(\chi,h)}(\mathcal{X}_i)=k\right\}$, 
where 	$B(\chi,h)=\left\{z\in \mathcal{H}:d\left(\chi,z\right) \leq h\right\}$, the derived local bandwidth which depends on $k$.
Unlike the kernel case, in the $k$NN estimator (\ref{kNN}) the smoothing parameter $H_{k,\chi}$ depends on  $\chi$ (and $k$), which provides the desired location-adaptive property.  Furthermore, this local smoothing estimator exclusively relies on a discrete parameter $k$ that takes values from a finite set $\{1,\dots,n\}$. This attribute presents an additional advantage over the kernel estimator, where selecting the tuning parameter $h$ needs consideration of a continuous interval.

 Regarding the tuning parameters, in FDA the typical approach for addressing the selection of $h$ and $k$ is cross-validation (CV), specifically, leave-one-out CV (LOOCV), where the objective functions are defined as
\begin{equation*}CV(h)=n^{-1}\sum_{j=1}^n\left(Y_j-\widehat{m}_{h}^{(-j)}(\mathcal{X}_j)\right)^2 \mbox{\text{ and }} CV^\ast(k)=n^{-1}\sum_{j=1}^n\left(Y_j-\widehat{m}_{k}^{\ast(-j)}(\mathcal{X}_j)\right)^2,
\end{equation*}
respectively. Here, $\widehat{m}_{h}^{(-j)}(\cdot)$ and $\widehat{m}_{k}^{\ast(-j)}(\cdot)$ are the leave-one-out versions of $\widehat{m}_{h}(\cdot)$ and $\widehat{m}^\ast_{k}(\cdot)$, respectively. Then, we select
$\widehat{h}=\arg \min_{h\in[a,b]}CV\left(h\right)$ and  $\widehat{k}=\arg \min_{k\in \{k_1,\dots,k_2\} }CV^\ast\left(k\right)$. The theoretical validation of both the data-driven selectors was showed in \cite{karaetal2017a,karaetal2017b}. In practise, it is usual to minimize over an interval $[a, b]\subset\mathbb{R}^+$ (respectively, over a set $\{k_1,\dots,k_2\}\subset\{1,\dots,n\}$) that encompasses a range of reasonable values for $h$ ($k$, respectively). The issue of automatically selecting the interval $[a, b]$ remains unresolved in one-dimensional nonparametric statistics. However, it is often considered of lesser importance due to the tendency of the CV function to exhibit relative flatness around its minimum, see \cite{NovoAneirosVieu2019}. In Section \ref{sec:mainfsim} we delve into the options included in the \texttt{fsemipar} package.

\noindent In the functional context, we must pay attention to the selection of the semi-metric $d$: since $\mathcal{H}$ represents an infinite-dimensional space, the equivalence between norms fails (in contrast to what happens in the finite-dimensional Euclidean space).  \cite{FerratyVieu2006} provided some alternatives of semi-metric, like the based on the functional principal component expansion, on the Fourier expansion, and on the B-spline expansion of the derivatives. We can compute these semi-metrics through the functions  \texttt{semimetric.pca}, \texttt{semimetric.fourier},  and \texttt{semimetric.deriv}, respectively, included in the \texttt{fda.usc} package.

 \noindent An interesting special case arises when the semi-metric is contingent upon a parameter. This circumstance is observed in the  FSIM (\ref{FSIM}),  where  we consider the operator $r_{\theta_0}(\cdot):\mathcal{H}\longrightarrow\mathbb{R}$ defined as
 $r_{\theta_0}(\chi)=r\left(\left<\theta_0,\chi\right>\right), \ \forall \chi \in \mathcal{H}.$ In this special case, if we want to quantify the proximity between elements of the functional space to obtain a nonparametric estimator for $r_{\theta_0}(\cdot)$, the semi-metric adopts the following form: 
 \begin{equation} d_{\theta_0}\left(\chi_1,\chi_2\right)=\left|\left<\theta_0, \chi_1\right>-\left<\theta_0,\chi_2\right>\right|=\left|\left<\theta_0,\chi_1-\chi_2\right>\right|, \ \ \ {\mbox{for}} \ \chi_1,\chi_2\in\mathcal{H}, \label{semimetric}
 \end{equation}
 thus relying on the function-parameter  $\theta_0$. When $\theta_0$ is unknown, the
 expressions (\ref{kernel}) and (\ref{kNN}) become statistics (and not estimators) since they also depend on a parameter and can't be directly employed in practice. That is,  for each $\theta\in\mathcal{H}$, they turn into  
 \begin{equation}
 	\widehat{r}_{h,\theta}(\chi)=\frac{\sum_{i=1}^nY_iK\left(h^{-1}d_{\theta}(\mathcal{X}_i,\chi)\right)}{\sum_{i=1}^nK\left(h^{-1}d_{\theta}(\mathcal{X}_i,\chi)\right)}, \ \ \ 	\widehat{r}_{k,\theta}^\ast(\chi)=\frac{\sum_{i=1}^nY_iK\left(H_{k,\chi,\theta}^{-1}d_{\theta}(\mathcal{X}_i,\chi)\right)}{\sum_{i=1}^nK\left(H_{k,\chi,\theta}^{-1}d_{\theta}(\mathcal{X}_i,\chi)\right)}, \ \
 	\forall \chi\in\mathcal{H}.
 	\label{expr:StatFSIM}
 \end{equation}
Here, $
H_{k,\chi,\theta}=\min\left\{h\in \mathbb{R}^+ \mbox{\text{ such that }} \sum_{i=1}^n1_{B_\theta(\chi,h)}(\mathcal{X}_i)=k\right\}
$
with $B_{\theta}(\chi,h)=\left\{z\in \mathcal{H}:d_{\theta}\left(\chi,z\right) \leq h\right\}$
 In this scenario, the semi-metric selection is direct; however, as a trade-off, we are required to estimate $\theta_0$. To address the estimation of the regression in the FSIM (selection of tuning parameter and estimation of $\theta_0$),  we can use LOOCV.  
 That is, considering the following objective functions
 \begin{equation} CV(h,\theta)=n^{-1}\sum_{j=1}^n\left(Y_j-\widehat{r}_{h,\theta}^{(-j)}(\mathcal{X}_j)\right)^2 \mbox{\text{ and }} CV^\ast(k,\theta)=n^{-1}\sum_{j=1}^n\left(Y_j-\widehat{r}_{k,\theta}^{\ast(-j)}(\mathcal{X}_j)\right)^2,\label{CV-func}\end{equation}
respectively, and minimising them respect to both parameters. Consequently, 
 the kernel-based and $k$NN-based estimators of $\theta_0$  are 
 \begin{equation}
 	\widehat{\theta}_h=\arg \min_{\theta \in \Theta}CV(h,\theta),\ \ \ \widehat{\theta}^\ast_k=\arg \min_{\theta \in \Theta}CV^\ast(k,\theta),\label{min_theta}
 \end{equation}
 where $\Theta\subset\mathcal{H}$, while the selectors of ${h}$ and ${k}$ are, respectively,
 $	\widehat{h}=\arg \min_{h\in[a,b]}CV\left(h,\widehat{\theta}_h\right), 
 	\mbox{\text{ and }}  \widehat{k}=\arg \min_{k_1\leq k\leq k_2}CV^\ast\left(k,\widehat{\theta}_k\right).$ 
 By substituting these estimators into (\ref{expr:StatFSIM}), we obtain $\widehat{r}_{\widehat{h},\widehat{\theta}_{\widehat{h}}}(\chi)$ and $\widehat{r}^\ast_{\widehat{k},\widehat{\theta}^\ast_{\widehat{k}}}(\chi)$, respectively. The theoretical validation of both the data-driven selectors for the tuning parameters ($h$ or $k$) and the estimators for $\theta_0$ was established by \cite{NovoAneirosVieu2019}. 
 
\noindent The focal aspect in practical application is the implementation of the minimization problems described in equation (\ref{min_theta}) to estimate $\theta_0$. Note that the minimisation is performed on a subset of the functional space $\Theta\subset\mathcal{H}$, so the question of how to build this subset seems challenging. \citet{AitSaidietal2008} or \citet{Ferratyetal2013}  proposed to solve this issue using B-Spline approximation  of the functional directions and reduce the infinite-dimensional optimization problem to a multivariate one. Specifically, each direction $\theta \in \Theta$ is obtained from a $d$-dimensional space generated by B-spline basis functions, denoted as $\{e_1(\cdot),\ldots,e_{d}(\cdot)\}$. Consequently, our attention is focused on directions expressed as
 \begin{equation}
 	\theta(\cdot)=\sum_{j=1}^{d}\alpha_j e_j(\cdot) \ \mbox{where} \ (\alpha_1,\ldots,\alpha_{d}) \in \mathcal{V}. \label{theta-exp}
 \end{equation}
and the problem in expression (\ref{min_theta}) turns into: 
\begin{equation}
	\widehat{\theta}_h=\arg \min_{(\alpha_1,\ldots,\alpha_{d}) \in \mathcal{V}}CV(h,\theta), \quad \ \ \widehat{\theta}^\ast_k=\arg \min_{(\alpha_1,\ldots,\alpha_{d}) \in \mathcal{V}}CV^\ast(k,\theta).
\end{equation}
 To construct the set of coefficients $\mathcal{V}$ in (\ref{theta-exp}), it is necessary to consider 
the model identifiability conditions (see \citealp{Ferratyetal2003}). Such restrictions hold true when $\theta \in \Theta$ satisfies the constraints $\left<\theta,\theta\right>=1$ and $\theta(t_0)>0$ for some arbitrary $t_0$ in the domain of $\theta_0$ (see \citealp{AitSaidietal2008}).

\noindent Taking into account the outlined considerations, the \texttt{fsemipar} package includes two procedures to carry out the minimisation of the CV functions defined in (\ref{CV-func}): (i) a direct method based on \cite{AitSaidietal2008}; (ii) an iterative algorithm based on \cite{Ferratyetal2013}.

\subsubsection{Procedure in \cite{AitSaidietal2008}}\label{sec:ait}

\cite{AitSaidietal2008} proposed the joint minimisation in the tuning parameter and the functional direction in (\ref{CV-func}).
As a consequence, their method requires intensive computation. Thus, it is necessary to strike a balance between the size of $\Theta$ and the performance of the estimators.
The following lines describe how to construct the set of coefficients $\mathcal{V}$ in (\ref{theta-exp}):
 \begin{enumerate}	
 	\item Choose the dimension of the B-spline basis $d$. We can decompose $d$ as $d=\ell+n_r$, where $\ell$ is the order of the B-spline functions (degree $\ell-1$) and $n_r$ is the number of regularly spaced interior knots (see \citealt{deBoor2001}).
 	\item For each $(\beta_1,\ldots,\beta_{d}) \in \mathcal{C}^{d}$, where $\mathcal{C}=\{c_1,\ldots,c_J\} \subset \mathbb{R}^J$ denotes a set of $J$ ``seed-coefficients", build the initial functional direction
 	\begin{equation} 
 	\theta_{init}(\cdot)=\sum_{j=1}^{d}\beta_j e_j(\cdot). \label{theta_ini}
 	\end{equation}
 	\item For each $\theta_{init}$ in Step 1 that verifies the condition $\theta_{init}(t_0)>0$, where $t_0$ denotes a fixed value in the domain of $\theta_{init}(\cdot)$, compute $\left<\theta_{init},\theta_{init}\right>$ and construct \begin{equation*}(\alpha_1,\ldots,\alpha_{d})=\frac{(\beta_1,\ldots,\beta_{d})}{\left<\theta_{init},\theta_{init}\right>^{1/2}}.
 	\end{equation*}
 	\item Construct $\mathcal{V}$ as the set of vectors $(\alpha_1,\ldots,\alpha_{d})$ obtained in Step 2.
 \end{enumerate}

 Therefore, the final set of eligible functional directions is given by
 $$\Theta=\left\{\theta(\cdot)=\sum_{j=1}^{d}\alpha_j e_j(\cdot); \ (\alpha_1,\ldots,\alpha_{d}) \in \mathcal{V}\right\}.$$ Subsequently, the goal is to compute the value of the CV functions in (\ref{CV-func}) for each pair $(\theta, h)$ (respectively $(\theta,k)$) where $\theta\in\Theta$, and select pair that minimises (\ref{CV-func}).
 \begin{remark} 
To achieve a balance between computation time and accurate estimation, \citealt{AitSaidietal2008} recommended using the seed-coefficients $\mathcal{C}=\{1,0,-1\}$. This set has been shown to offer good practical performance (see \citealp{NovoAneirosVieu2019}). Regarding the dimension of the B-spline basis, $d$, the usual procedure is to set the order of the B-spline functions ($\ell$) and try different number of internal knots, $n_r$.  Then, we can choose the value of $n_r$ that minimises the CV function: $\widehat{n}_r=\arg\min_{n_r\in\mathbb{Z}^+} CV(\widehat{\theta}_{\widehat{h}},\widehat{h})$ (respectively, $\widehat{n}_r^*=\arg\min_{n_r\in\mathbb{Z}^+} CV^*(\widehat{\theta}_{\widehat{k}},\widehat{k}$)). 
 \end{remark}
\subsubsection{Procedure in \cite{Ferratyetal2013}} \label{sec:fer}
Instead of the direct joint minimisation, \cite{Ferratyetal2013} and \cite{Chan2023} proposed an iterative algorithm. For a given $d$:  
\begin{enumerate}
	\item Choose an initial set of coefficients  $\pmb{\gamma}=(\gamma_1,\ldots,\gamma_{d})\in\mathbb{R}^d$ to give rise to a functional direction $\theta^{(0,m)}$ as in (\ref{theta_ini}). Calibrate them to verify the identifiability conditions of the model. 
	As a result you obtain another set of coefficients $\pmb{\alpha}=(\alpha_1,\ldots,\alpha_{d})$ corresponding to a functional direction $\theta^{(m)}$.
	\item Select the optimal tuning parameter setting $\theta^{(m)}$ in (\ref{CV-func}): $\widehat{h}^{(m)}=\arg\min_{h\in[a_m,b_m]} CV(h,\theta^{(m)})$ (respectively, $\widehat{k}^{(m)}=\arg\min_{k_{m_1}<k<k_{m_2}} CV^*(k,\theta^{(m)})$).
	\item Estimate the direction setting $\widehat{h}^{(m)}$ (respectively, $\widehat{k}^{(m)}$) in (\ref{CV-func}): $\widehat{\theta}^{(0, m+1)}=\arg \min_{\pmb{\gamma} \in \mathbb{R}^d}CV(\widehat{h}^{(m)},\theta)$, ($\widehat{\theta}^{\ast (0,m+1)}=\arg \min_{\pmb{\gamma} \in \mathbb{R}^d}CV^\ast(\widehat{k}^{(m)},\theta)$).
	\item Update $\theta^{(0,m)}=\widehat{\theta}^{(0,m+1)}$ ($\theta^{(0,m)}=\widehat{\theta}^{\ast (0,m+1)}$).
    Continue the process until convergence is achieved: the algorithm halts when the change in CV from one iteration to the next (scaled by the variance of the response) is positive and falls below a predetermined threshold.
\end{enumerate}
\begin{remark}
	In step 3 of the method, the minimisation can be carried out using general-purpose optimisation algorithms for multidimensional spaces, such as Nelder and Mead (1965), included in the R function \texttt{optim}.
\end{remark}
 \subsection{Fitting the FSIM with the \texttt{fsemipar} package}
 The \texttt{fsemipar} package contains several functions for estimating the FSIM. It provides the flexibility to perform both kernel and $k$NN fitting, with implementation options based on either \cite{AitSaidietal2008} or \cite{Ferratyetal2013}.
 The primary related functions are summarised in the Table \ref{tab:FSIM}. Each primary function produces an object that belongs to an S3 class. These S3 classes are designed to include essential S3 methods, such as \texttt{print}(), \texttt{summary}(), \texttt{plot}() and \texttt{predict}() to facilitate the posterior analysis. Each function offers a wide range of customisable parameters for the user, while also providing  a straightforward option to use default settings for the input parameters.
 
 \begin{table}[H]
 	\centering
 	\begin{tabular}{lll}
 		\cmidrule[1.5pt](lrr){1-3} 
 		\textbf{Function} & \textbf{S3 class} &\textbf{Description}\\
 		\cmidrule[1.5pt](lrr){1-3}
 		\texttt{fsim.kernel.fit} & `fsim.kernel' & kernel estimation based on \cite{AitSaidietal2008} \\
 		\texttt{fsim.kNN.fit} & `fsim.kNN' & $k$NN estimation based on \cite{AitSaidietal2008} \\
 			\texttt{fsim.kernel.fit.optim}  & `fsim.kernel' & kernel estimation based on \cite{Ferratyetal2013} \\
 		\texttt{fsim.kNN.fit.optim} & `fsim.kNN' & $k$NN estimation based on \cite{Ferratyetal2013} \\
 		
 		\cmidrule[1.5pt](lrr){1-3}  
 	\end{tabular}
 	\caption{Summary of main functions for the FSIM and associated S3 classes.}
 	\label{tab:FSIM}
 \end{table}
\subsubsection{Main fitting functions} \label{sec:mainfsim}
The main fitting functions for the FSIM (Table \ref{tab:FSIM}) implement either the procedure detailed in \cite{AitSaidietal2008} (\texttt{fsim.kernel.fit()} and \texttt{fsim.kNN.fit()}) or the procedure outlined in \cite{Ferratyetal2013} (\texttt{fsim.kernel.fit.optim()} and \texttt{fsim.kNN.fit.optim()}). In each case, these procedures are combined with either kernel or $k$NN estimation.  

Regarding the functional predictor, all the functions of the package \texttt{fsemipar} require the values of the curves evaluated on a grid of equally-spaced discretisation points. The functions internally represent the discretised curves in a B-spline basis of dimension that can be customised by the user.

\noindent Functions in Table \ref{tab:FSIM} share two required arguments: the functional predictor \texttt{x}, a matrix of dimension $(n,p)$ containing $n$ curves discretised in $p$ points, and the scalar response variable \texttt{y}, a vector of dimension $n$. 

\begin{verbatim}
fsim.kernel.fit(x,y,...)
fsim.kNN.fit(x,y,...)
fsim.kernel.fit.optim(x,y,...)
fsim.kNN.fit.optim(x,y,...)
\end{verbatim}
\noindent However, these functions also take several optional arguments for finer control:
\begin{itemize} 
	\item \emph{Optional arguments for the nonparametric fit}. These arguments are summarised in Table \ref{tab:nonpar} and refer to the type of kernel used and the choice of the tuning parameters: $h$ (kernel), or $k$ ($k$NN). In the kernel-based functions \texttt{fsim.kernel.fit()} and \texttt{fsim.kernel.fit.optim()}, there is a default option for constructing the sequence of bandwidths from which $\widehat{h}$ will be selected. This default option takes into account the proximity between sample curves using the semimetric (in the FSIM, the projection semimetric). Specifically, the endpoints of the sequence are determined by calculating quantiles of a certain order, controlled by the user through the arguments \texttt{min.q.h} (default 0.05) and \texttt{max.q.h} (default 0.5), of the distances between curves. Additionally, the number of bandwidths in the sequence can also be customised using \texttt{num.h} (default 10). Alternatively, we can directly specify the sequence of bandwidths with the argument \texttt{h.seq}. When \texttt{h.seq} is provided, the arguments \texttt{min.q.h}, \texttt{max.q.h} and \texttt{num.h} are ignored.
Regarding $k$NN-based functions  \texttt{fsim.kNN.fit()} and \texttt{fsim.kNN.fit.optim()}, the user has the flexibility to choose the endpoints of the sequence for $k$ (number of neighbours) using  \texttt{min.knn} (default $2$) and \texttt{max.knn} (default integer division $n/5$). In addition, the user can specify the step between two consecutive elements of the sequence using the argument \texttt{step} (default smallest integer greater than or equal to $n/100$). Alternatively, we can directly provide the sequence of number of neighbours using \texttt{knearest}. It is important to note that if we specify the \texttt{knearest} argument, \texttt{min.knn}, \texttt{max.knn} and \texttt{step} are ignored.

The four primary functions share the argument \texttt{kind.of.kernel} that allows user to choose the type of kernel function. Currently, only Epanechnikov kernel (``quad'') is available.

\begin{table}[h]
	\centering
	\begin{tabular}{p{3.8cm}lp{7cm}}
		\cmidrule[1.5pt](lr){1-3}  
		\textbf{Function} & \textbf{Argument} & \textbf{Description} \\
		\cmidrule[1.5pt](lr){1-3}
		\multirow{4}{*}{\makecell[l]{\small \texttt{fsim.kernel.fit}\\ \small \texttt{fsim.kernel.fit.optim}}} & \texttt{min.q.h} & Minimum quantile order for the distances. \\
		& \texttt{max.q.h} & Maximum quantile order for the distances. \\
		& \texttt{num.h} & Number of $h$ values in the sequence. \\
		& \texttt{h.seq} & Sequence of $h$ values. \\
		\cmidrule[0.5pt](lr){1-3}
		\multirow{5}{*}{\makecell[l]{\small\texttt{fsim.kNN.fit}\\ \small \texttt{fsim.kNN.fit.optim}}} & \texttt{min.knn} & Minimum $k$ for $k$NN fitting. \\
		& \texttt{max.knn} & Maximum $k$ for $k$NN fitting. \\
		& \texttt{step} & Step between elements of the sequence. \\
		& \texttt{knearest} & Sequence of $k$ values. \\
	
			\cmidrule[0.5pt](lr){1-3}
		All & \texttt{kind.of.kernel} & Type of kernel used. \\
		\cmidrule[1.5pt](lr){1-3}  
	\end{tabular}
	\caption{Optional input arguments related to nonparametric estimation (kernel or $k$NN).}
	\label{tab:nonpar}
\end{table}

	\item \emph{Optional arguments for B-spline expansions}. These arguments are summarised in Table \ref{tab:bspline} and refer to the B-spline expansion for the functional directions and predictor.   The four functions allow to customize the order of the B-spline functions, $\ell$,  through the argument \texttt{order.Bspline} (default 3), the number of interior knots, $n_r$, for directions in $\Theta$ using \texttt{nknot.theta} (default 3), and the number of interior knots for the sample curves with \texttt{nknot} (default floor $(p-\ell - 1)/2$), and their definition interval with \texttt{range.grid}.
	
	In addition, functions \texttt{fsim.kernel.fit()} and \texttt{fsim.kNN.fit()} allow to choose the set of seed coefficients $\mathcal{C}$  through the argument \texttt{seed.coeff} (see Section \ref{sec:ait}).   Furthermore, the functions \texttt{fsim.kernel.fit()} and \texttt{fsim.kNN.fit.optim()} enable the choice of the set of initial coefficients $\gamma\in\mathbb{R}^d$ through the argument \texttt{gamma}.

	\item \emph{Optional arguments for the parallel computation}. For functions \texttt{fsim.kernel.fit()} and \texttt{fsim.kN\newline N.fit()}  the implementation is computationally expensive. For that, they both allow the parallel computation  (internally, they use the functions \texttt{makeCluster()}, \texttt{registerDoParallel()}, and \texttt{foreach()} of the R packages \texttt{parallel}, \texttt{doParallel} and \texttt{foreach}, respectively). Users can explicitly specify the number of CPU cores to use for parallel execution with the argument \texttt{n.core}. By default, the both functions take \texttt{n.core} as the number of available cores minus 1 if more than 1 is available, otherwise 1 (this is done internally using the function \texttt{availableCores()} of the R package \texttt{parallely}). For sequential execution, it is sufficient to set \texttt{n.core=1}.
	
	\item \emph{Optional arguments for controlling the iterative algorithm}. The functions \texttt{fsim.kernel.fit.opt\newline im()} and \texttt{fsim.kNN.fit.optim()} are built upon an iterative procedure (see Section \ref{sec:fer}). Users have the flexibility to customize the convergence threshold using the argument \texttt{threshold} (default: 5e-3).

\end{itemize}

	\begin{table}[h]
		\centering
		\begin{tabular}{llp{7cm}}
			\cmidrule[1.5pt](lr){1-3}  
			\textbf{Function} & \textbf{Arguments} & \textbf{Description} \\
			\cmidrule[1.5pt](lr){1-3}
			\multirow{5}{*}{\makecell[l]{\small \texttt{fsim.kernel.fit} \\ \small \texttt{fsim.kNN.fit}\\ \texttt{fsim.kernel.fit.optim}\\ \texttt{fsim.kNN.fit.optim}}}  & \texttt{order.Bspline} & Order of the B-spline basis functions. \\
			& \texttt{nknot.theta} & Number of interior knots for the functional directions in $\Theta$. \\
			& \texttt{nknot} & Number of interior knots for the curves. \\
			& \texttt{range.grid} & Endpoints of the definition interval for the curves. \\
				\cmidrule[0.5pt](lr){1-3}
		\makecell[l]{\small \texttt{fsim.kernel.fit} \\ \small \texttt{fsim.kNN.fit}}	& \texttt{seed.coeff} & Set of seed coefficients $\mathcal{C}$.\\
			  \cmidrule[0.5pt](lr){1-3}
				\makecell[l]{\small \texttt{fsim.kernel.fit.optim} \\ \small \texttt{fsim.kNN.fit.optim}} & \texttt{gamma}& Set of initial coefficients $\gamma$.\\
				\cmidrule[1.5pt](lr){1-3}
		\end{tabular}
		\caption{Optional input arguments related to B-spline expansions.}
		\label{tab:bspline}
	\end{table}
	\subsubsection{Case study I, part I: Tecator dataset}\label{sec:tecator}
The Tecator dataset is a well-known benchmark database in FDA, as evidenced by the numerous research works in the field that have used it to illustrate their techniques (see, e.g., \citealt{Amatoetal2006}, \citealt{FerratyVieu2006}, \citealt{Galeanoetal2015}, among others).  This dataset contains spectral measurements of $215$ pork samples collected using a Tecator Infratec Food and Feed Analyser. The spectrum includes absorbances at $100$ wavelengths in the range 850 to 1050 nm, which are commonly considered to be functional data. In addition, the dataset records the fat, protein and moisture content of each meat sample. 

\begin{figure}[h] 
		\centering
		\includegraphics[width=0.5\textwidth]{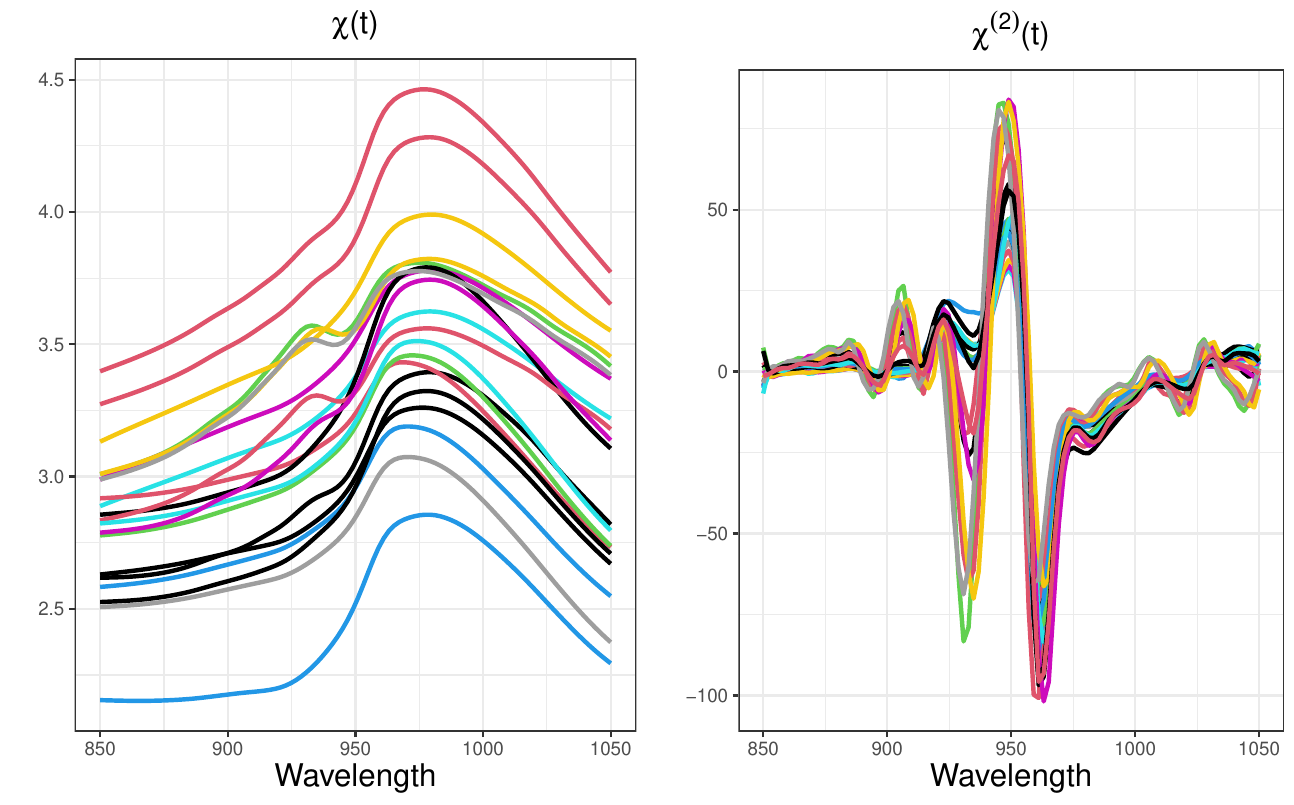}
		\caption{Sample of 20 spectrometric curves (left panel) and their second derivatives (right panel).}
		\label{fig:tecator}
	\end{figure}

\begin{figure}[h] 
	\centering
	\includegraphics[width=0.4\textwidth]{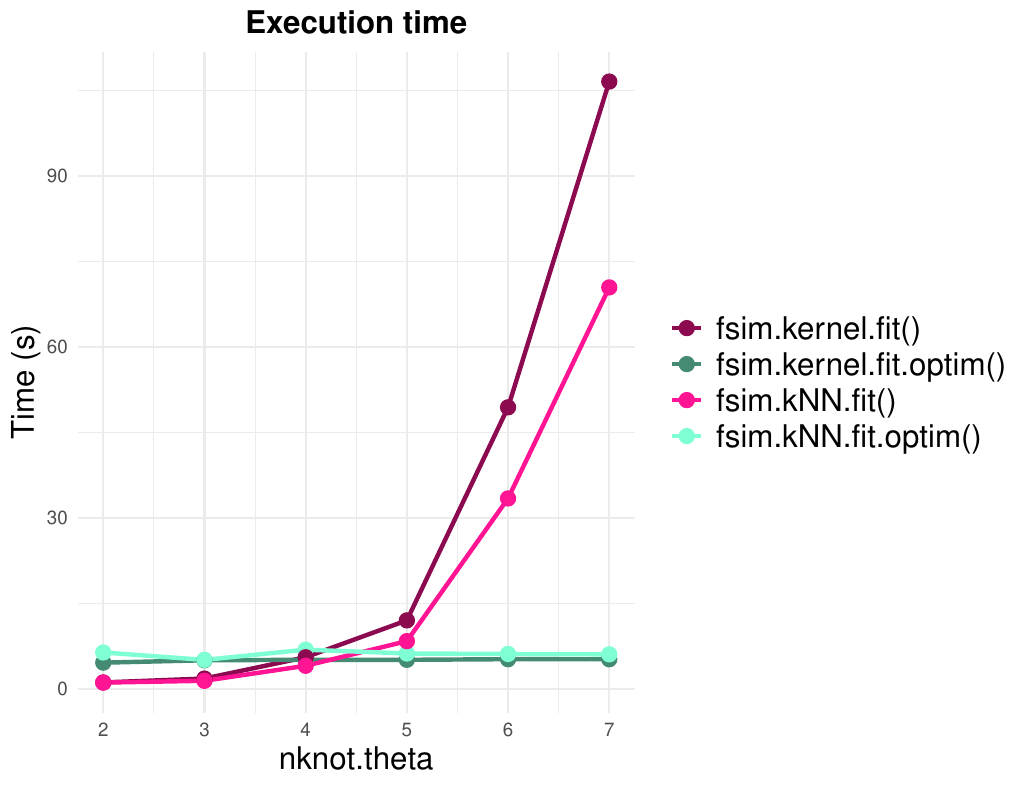}
	\caption{Execution time for functions in Table \ref{tab:FSIM} over $n_r$ obtained in a computer with the following features: Intel Core i9-10850K, 64GB, 1TB SSD and 2TB HDD, RTX 3090 24GB.}
	\label{fig:fsim}
\end{figure}

	The Tecator dataset is included in the \texttt{fsemipar} package. In addition to the discretised original spectrometric curves, the first and second derivatives are included. Figure \ref{fig:tecator} includes a graphical representation of the first $20$ curves (left panel) and its second derivative (right panel). 
	We illustrate the primary FSIM fitting functions of the package by predicting the fat percentage using the available spectrometric data, specifically the second derivative of the curves (usual choice in the literature, see, e. g.,  \citealt{AneirosVieu2015}, \citealt{NovoAneirosVieu2019}). Note that spectrometric data is more readily obtainable, as analysing the chemical composition typically involves cost-intensive and lengthier chemical experiments. To evaluate the performance of the procedures we split the sample into two subsamples: the training sample is composed of the first $160$ entries in the dataset, whereas the test sample includes the remaining $55$ observations.
	 \begin{verbatim}
	 	> data(Tecator)
	 	> str(Tecator)
	 	List of 6
	 	$ fat           : num [1:215] 22.5 40.1 8.4 5.9 25.5 42.7 42.7 10.6 19.9 19.9 ...
	 	$ protein       : num [1:215] 16.7 13.5 20.5 20.7 15.5 13.7 13.7 19.3 17.7 17.7 ...
	 	$ moisture      : num [1:215] 60.5 46 71 72.8 58.3 44 44 69.3 61.4 61.4 ...
	 	$ absor.spectra : num [1:215, 1:100] 2.62 2.83 2.58 2.82 2.79 ...
	 	$ absor.spectra1: num [1:215, 1:100] 0.0352 0.4171 0.1772 0.1776 0.168 ...
	 	$ absor.spectra2: num [1:215, 1:100] 1.118 -4.412 -4.764 -6.778 -0.601 ...
	 \end{verbatim}
As an example, the following code shows the implementation of the FSIM using the \texttt{fsim.kNN.fit()} function. In addition to the data, we have provided some specific input arguments to improve the fit: we have narrowed the sequence in which the number of neighbours is selected (setting \texttt{max.knn=15}), discarding the largest values which in this case lead to oversmoothing and an increment of the computational time. We also set \texttt{step=1} to try all the values in this sequence. Additionally, we have specified the number of interior knots for the B-spline representation of the unknown functional direction (\texttt{nknot.theta=4}) and for the curves (\texttt{nknot=20}), also providing the range in which the curves were observed (\texttt{range.grid=c(850,1050)}). It is important to note that it is reasonable to seek a very precise representation of the curves on the B-spline basis, which leads to the use of a B-spline basis of greater dimension than in the case of the functional directions. In fact, large values of \texttt{nknot.theta} will tend to undersmoothing and capture the noise of the training sample. Furthermore, the computational cost of estimating the functional direction is strongly related to \texttt{nknot.theta} (greater \texttt{nknot.theta} more intensive computation), as we can deduce from Figure \ref{fig:fsim}. For these reasons, it is usually recommended to take \texttt{nknot.theta} between $2$ and $8$.
 \begin{verbatim}
 		> y<-Tecator$fat
 		> x<-Tecator$absor.spectra2
 		> train=1:160
 		> test=161:215
 		> fit<-fsim.kNN.fit(y[train],x=x[train,],max.knn=15,nknot.theta=4,nknot=20,
 		 range.grid=c(850,1050),step=1)
 \end{verbatim} 
The fitted object of the functions  listed in the Table \ref{tab:FSIM} contains useful information about the estimation performed: the fitted values for the response (\texttt{fitted.values}), the residuals (\texttt{residuals}), the coefficients of $\widehat{\theta}$ in the B-spline basis (\texttt{theta.est}), the selected value of the tuning parameter, $\hat{h}$ (\texttt{h.opt}) or $\hat{k}$ (\texttt{k.opt}), the coefficient of determination (\texttt{r.squared}), the residual variance (\texttt{var.res}), the degrees of freedom (\texttt{df}) or the minimum value of the CV function (\texttt{CV.opt}).
The following code shows the information contained in the output of the function \texttt{fsim.kNN.fit()}.
\begin{verbatim}
		> names(fit)
		[1] "fitted.values"  "residuals"      "theta.est"      "k.opt"         
		[5] "r.squared"      "var.res"        "df"             "yhat.cv"       
		[9] "CV.opt"         "CV.values"      "H"              "m.opt"         
		[13] "theta.seq.norm" "k.seq"          "call"           "y"             
		[17] "x"              "n"              "kind.of.kernel" "range.grid"    
		[21] "nknot"          "order.Bspline"  "nknot.theta"  
		> class(fit)
		 [1] "fsim.kNN"  
\end{verbatim}
Each fitted object belongs to an S3 class (see Table \ref{tab:FSIM}), which implements S3 methods. In the particular case of \texttt{fsim.kNN.fit()} (class \texttt{fsim.kNN}), the returned object can be used in the functions \texttt{print.fsim.kNN()} and \texttt{summary.fsim.kNN()} to display summaries of the fitted model. It can also  be employed in \texttt{predict.fsim.kNN()} to obtain predictions for new curves (which can be provided using the \texttt{newdata} argument), and the mean squared error of prediction (MSEP) if the actual responses are provided via the \texttt{y.test} argument. The MSEP is calculated as follows: 
\begin{equation}
	\label{MSEP}
	\textrm{MSEP}=\frac{1}{n_{test}}\sum_{i=n+1}^{n+n_{test}}\left(Y_i - \widehat{Y}_i\right)^2,
\end{equation}
where $n_{test}$ is the size of the test sample and  $\widehat{Y}_i$, the predicted values.  The code below illustrates these functionalities:
	\begin{verbatim}
		> summary(fit)        
		*** FSIM fitted using kNN estimation with Nadaraya-Watson weights ***
		
		-Call: fsim.kNN.fit(x = x[1:160, ], y = y[1:160], nknot.theta = 4, 
		max.knn = 15, step = 1, range.grid = c(850, 1050), nknot = 20)
		
		-Number of neighbours (k): 9
		-Theta coefficients in the B-spline basis: 0.1656316 -0.1656316 0 
		0.1656316 -0.1656316 0 0.1656316
		-CV: 3.921117
		-R squared: 0.9824381
		-Residual variance: 3.323535 on 132.9207 degrees of freedom

		> predict(fit,newdata=x[test,],y.test=y[test])$MSEP
		[1] 2.68585
	\end{verbatim}
Regarding the prediction, Table \ref{tab:fsimMSEP} displays the MSEP corresponding to each primary fitting function when using the respective cross-validation (CV) selectors for $n_r$, $h$, or $k$ and $\theta_0$. In this example, the $k$NN-based functions yield lower MSEP values than the kernel-based ones. Furthermore, the procedure described in \cite{AitSaidietal2008} results in a lower MSEP than the one presented in \cite{Ferratyetal2013}. However, for the functions based on the former procedure, when we increase $n_r$ (\texttt{nknot.theta}), the computation time also increases, while for the functions based on the latter, it remains constant. This fact can be observed in Figure \ref{fig:fsim}, where we show the execution times for the four functions over $n_r$. Note that  \texttt{fsim.kernel.fit()} and \texttt{fsim.kNN.fit()} use parallel computation and were executed with \texttt{n.core=19}. For each function we set all the remaining optional input parameters (in particular, \texttt{order.Bspline=3}) while we vary \texttt{nknot.theta}.

\begin{table}[h]
	\centering
	\begin{tabular}{p{4.5cm}p{1.5cm}p{1.5cm}}
		\cmidrule[1.5pt](lr){1-3}  
		\textbf{\small Function} & \textbf{$\widehat{n}_k$} & \textbf{MSEP} \\
		\cmidrule[1.5pt](lr){1-3}
		\texttt{\small fsim.kernel.fit()} & 4 & 3.49 \\ 
		\texttt{\small fsim.kNN.fit()} & 4 & 2.69 \\ 
		\texttt{\small fsim.kernel.fit.optim()} & 2 & 3.64 \\ 
		\texttt{\small fsim.kNN.fit.optim()} & 4 & 2.81 \\ 
		\cmidrule[1.5pt](lr){1-3}  
	\end{tabular}
	\caption{MSEP obtained for each function using the respective CV selectors for tuning parameters, functional index, and $n_r$.}
	\label{tab:fsimMSEP}
\end{table}

\begin{figure}[h] 
		\centering
		\begin{subfigure}{0.45\textwidth}
			\includegraphics[width=\linewidth]{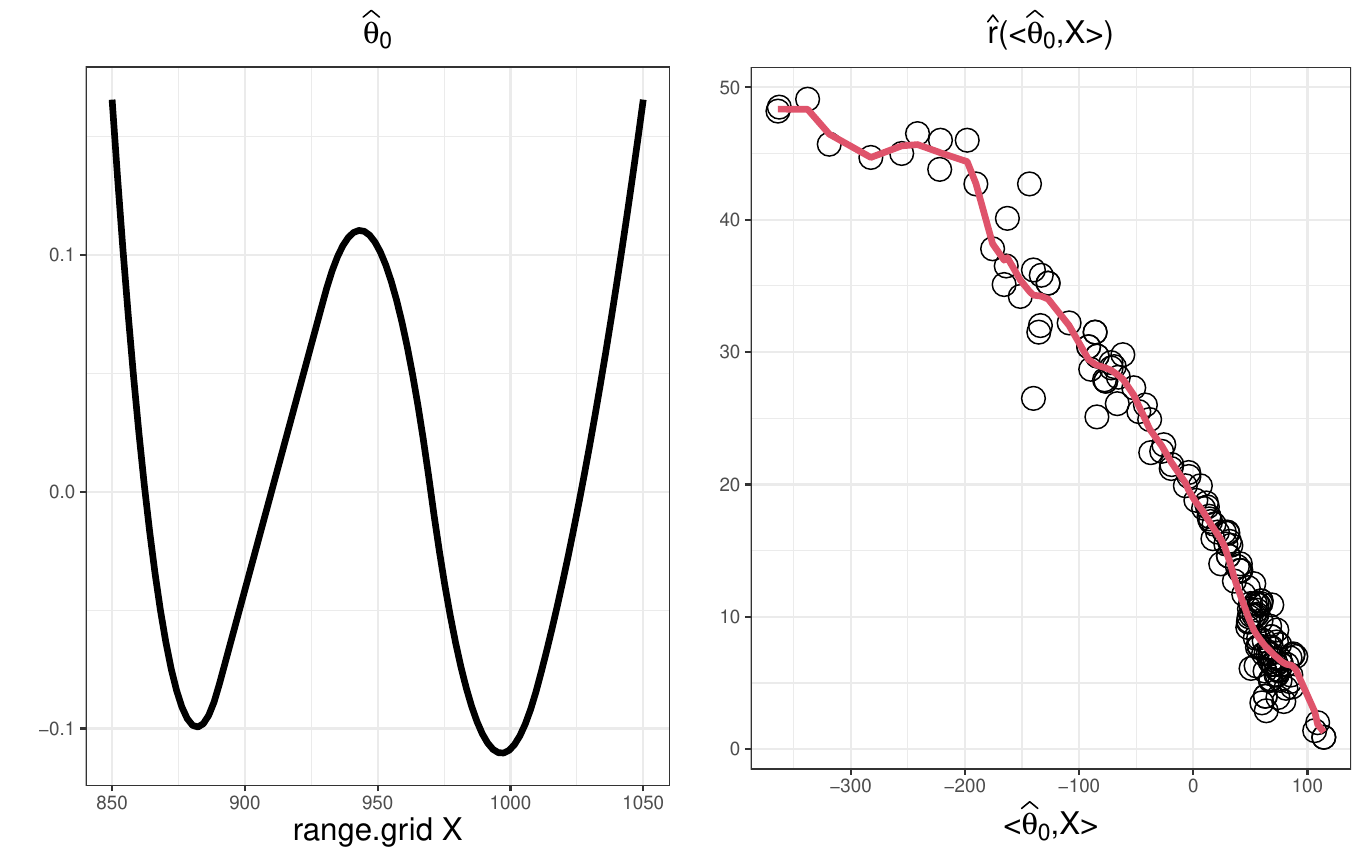}
			\caption{\texttt{fsim.kernel.fit()} ($\widehat{n}_r=4$)}
		\end{subfigure}
		\begin{subfigure}{0.45\textwidth}
			\includegraphics[width=\linewidth]{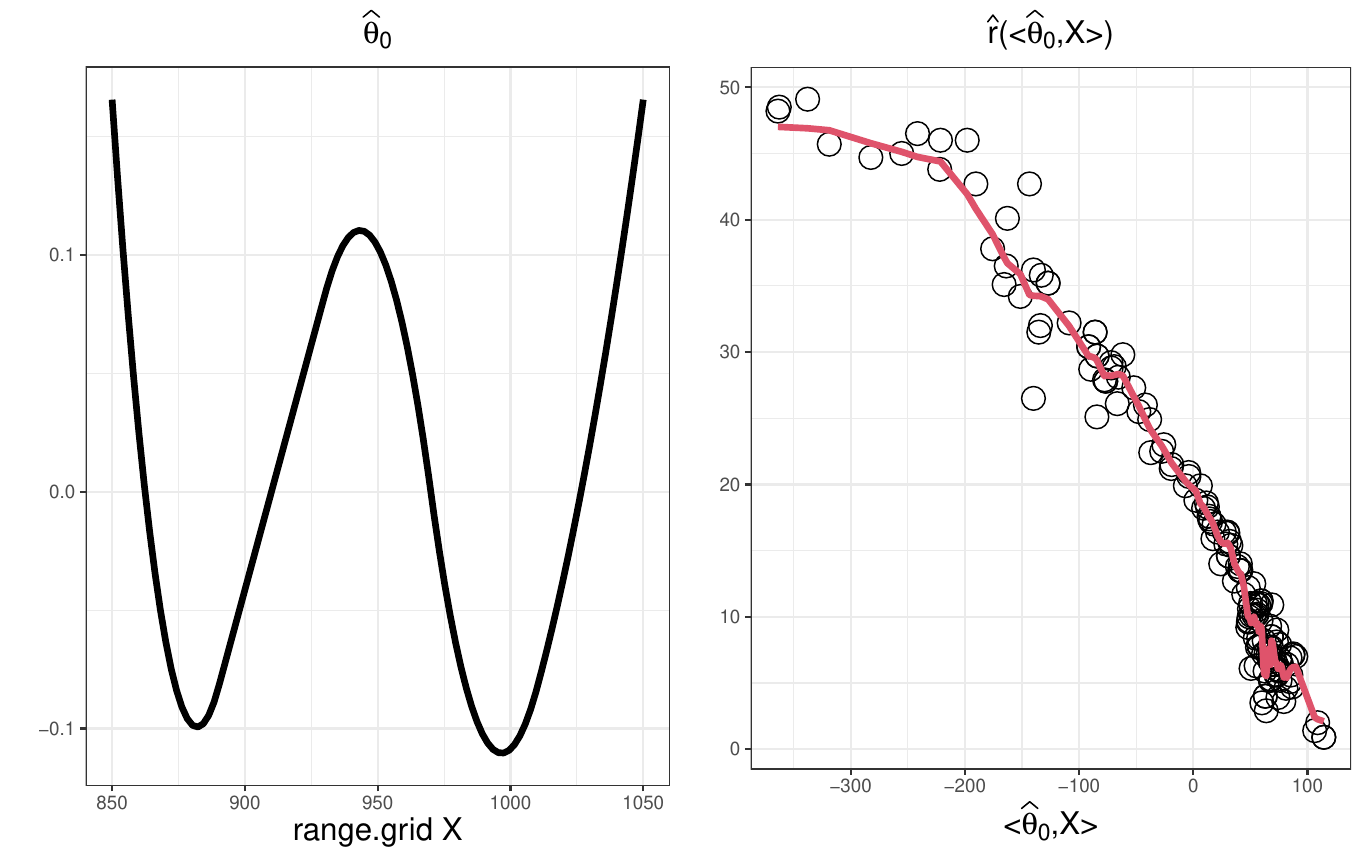}
			\caption{\texttt{fsim.knn.fit()} ($\widehat{n}_r^*=4$)}
		\end{subfigure}
		\begin{subfigure}{0.45\textwidth}
			\includegraphics[width=\linewidth]{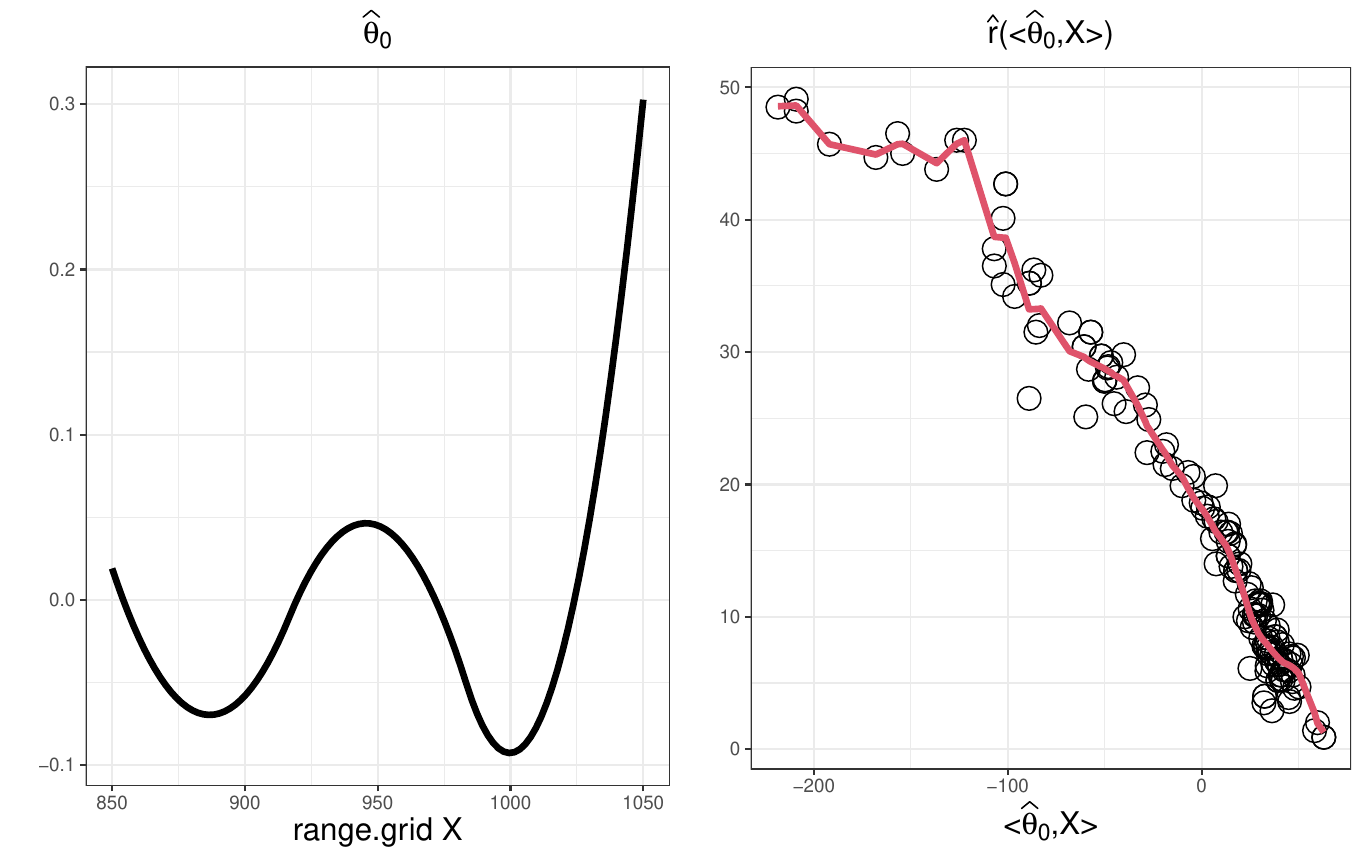}
			\caption{\texttt{fsim.kernel.fit.optim()} ($\widehat{n}_r=2$)}
		\end{subfigure}
		\begin{subfigure}{0.45\textwidth}
			\includegraphics[width=\linewidth]{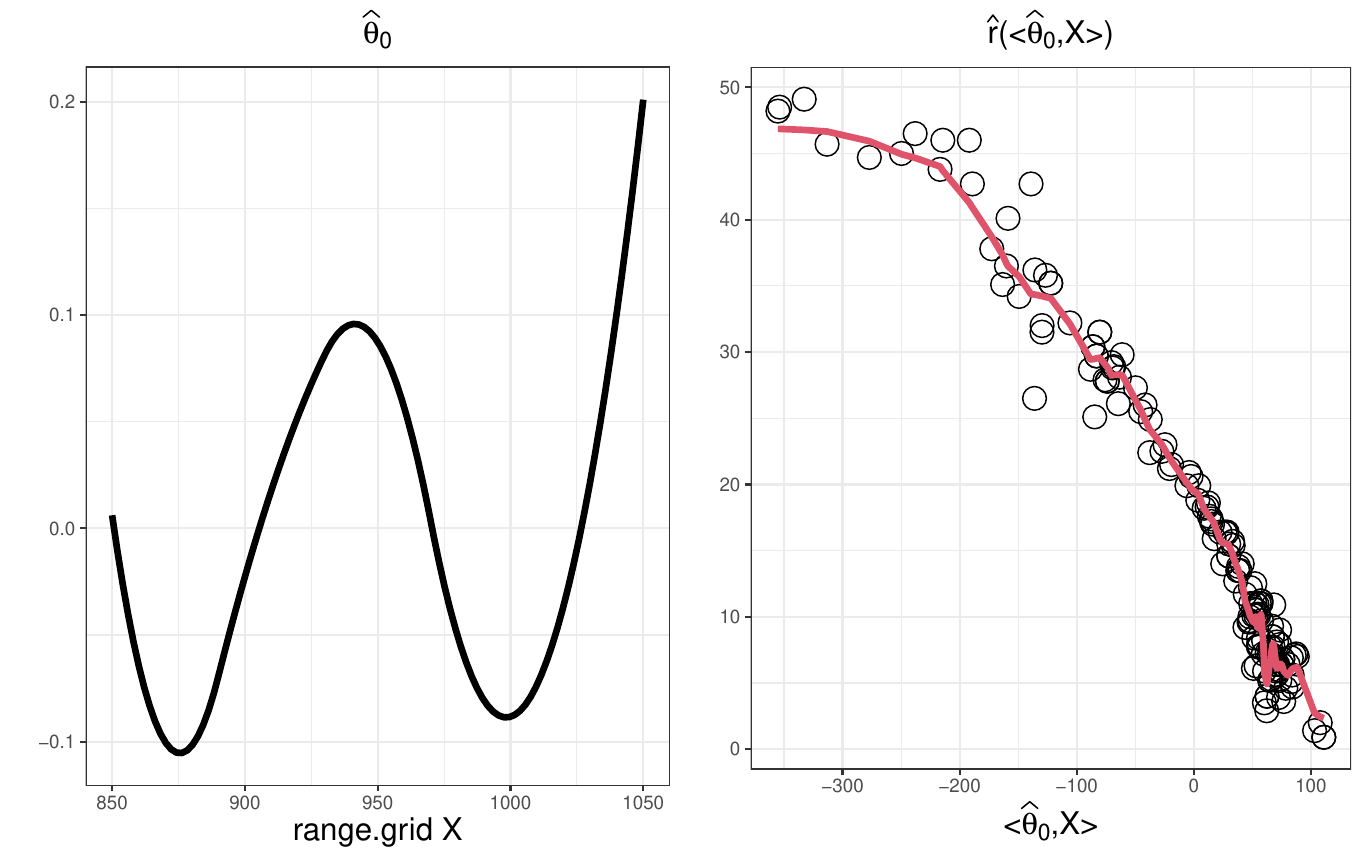}
			\caption{\texttt{fsim.knn.fit.optim()} ($\widehat{n}_r^*=4$)}
		\end{subfigure}
		\caption{Outputs of the S3 method \texttt{plot()} applied to the fitted objects of functions in the Table \ref{tab:FSIM}, using the corresponding CV estimator for $\theta_0$ and CV selectors for the tuning parameters and $n_r$.} 
		\label{fig:fsim2}
	\end{figure}

The S3 classes listed in Table \ref{tab:FSIM} also implement the S3 method \texttt{plot()}. The routines implementing this S3 method use internally the R package \texttt{ggplot2}, trying to produce elegant and high quality charts.  The output of this method represents the regression fit and the estimated functional index. In Figure \ref{fig:fsim2}, we provide the outputs of the 4 main functions when using the CV selectors for tuning parameters, functional index, and $n_r$.
		\section{Multiple scalar predictors and a functional covariate}
	In Section 3, we focus on SoF semiparametric regression including as predictors multiple scalar variables and a functional one. Section 3.1 introduces the modelling alternatives based on the partial linear structure. In Section 3.2, we discuss the techniques implemented in \texttt{fsemipar} for simultaneous estimation and variable selection in the linear component. Finally, Section 3.3 presents the associated functions of the package and illustrates their use with the Tecator dataset (see Section \ref{sec:tecator}).
	\subsection{Semi-functional partial linear regression}
	In practical applications, it is common to encounter more than one covariate. Particularly, situations arise where, in addition to a functional predictor, several scalar variables are linked to the response. Take, for example, the Tecator dataset (see Section \ref{sec:tecator}), which includes measurements of not only fat content but also the percentage of protein and moisture for each meat sample. These additional variables can be valuable in predicting fat content.
	
 Statistically, this implies that, in addition to $Y$ and $\mathcal{X}$, there exists a vector $\pmb{Z}=(Z_1,\dots,Z_{p_n})^{\top}$ of scalar variables potentially related to the response. We observe tuples $\{(Y_i,\pmb{Z}_i,\mathcal{X}_i), i=1,\dots,n\}$ i.i.d. to $(Y,\pmb{Z},\mathcal{X})$ and our goal is to predict $Y$ using the information carried by the remaining variables. In these complex scenarios, combining  the partial linear approach with functional nonparametric or functional single-index modelling  typically yields interpretable results with good predictive performance. Consequently, literature in FDA focused on two regression strategies:
\begin{itemize}	
\item 	The \emph{semi-functional partial linear model} (SFPLM) (see \citealt{AneirosVieu2006}, \citealt{AneirosFerratyVieu2015}, \citealt{Boente2017}, among others), which is given by the expression
	\begin{equation}
		\label{SFPLM}
		Y_i=Z_{i1}\beta_{01}+\dots+Z_{ip_n}\beta_{0p_n}+m\left(\mathcal{X}_i\right)+\varepsilon_i \quad i=1,\dots,n.
	\end{equation}
Here,  
$\pmb{\beta}_0=(\beta_{01},\dots,\beta_{0p_n})^{\top}$ is a vector of unknown real coefficients and  $m(\cdot)$ is an  unknown functional operator with the same features as in the FNM. 
		
\item	The \emph{semi-functional partial linear single-index model} (SFPLSIM) (\citealt{Wang2016}, \citealt{NovoAneirosVieu2020}), which is given by the relationship
	\begin{equation}
		Y_i=Z_{i1}\beta_{01}+\dots+Z_{ip_n}\beta_{0p_n}+r\left(\left<\theta_0,\mathcal{X}_i\right>\right)+\varepsilon_i \quad i=1,\dots,n.\label{SFPLSIM}
	\end{equation}
Here, the vector $\pmb{\beta}_0=\left(\beta_{01},\dots,\beta_{0p_n}\right)^{\top}\in\mathbb{R}^p$ is assumed to be unknown, as is the functional index $\theta_0\in \mathcal{H}$ and the real-valued link function $r(\cdot)$, which both have the same characteristics as in the FSIM.
\end{itemize}	

\noindent An interesting situation arises when only some of the $p_n$ scalar covariates have a real effect on the response. Then estimation techniques must be combined with a variable selection tool. 

The SFPLM, along with the kernel-based estimation procedure introduced in \citealt{AneirosVieu2006}, can be implemented in R using the function \texttt{fregre.plm} from the \texttt{fda.usc} package. However, it is worth noting that this function does not support variable selection in the linear component or $k$NN-based estimation. Furthermore, as far as we know, there is currently no  function available for estimating the SFPLSIM and performing variable selection excluding those belonging to the \texttt{fsemipar} package.
Consequently, in the subsequent section we introduce the techniques implemented in the \texttt{fsemipar} related to the SFPLM and SFPLSIM.
	
	\subsection{Estimation and variable selection in the linear component}\label{sec:EstVS}
\cite{AneirosFerratyVieu2015} introduced a procedure based on penalised least-squares (PeLS) for simultaneous estimation and variable selection in the linear component for the SFPLM (\ref{SFPLM}). This procedure entails transforming the model into a linear model using nonparametric regression. Specifically:
\begin{enumerate} 
\item The first step is to extract the effect of the functional covariate from the response and predictors:
	\begin{equation}
		Y_i-\mathbb{E}\left(Y_i|\mathcal{X}_i\right)=\left(\pmb{Z}_i-\mathbb{E}\left(\pmb{Z}_i|\mathcal{X}_i\right)\right)^{\top}\pmb{\beta}_0+\varepsilon_i, \quad i=1,\dots,n,\label{mod_lineal}
	\end{equation}
where $\pmb{Z}_i=(Z_{i1},\dots,Z_{ip_n})$. Given that the expectations in the expression (\ref{mod_lineal}) are unknown, it becomes necessary to derive estimations for them. To accomplish this, we can employ functional nonparametric regression, utilizing either the kernel (\ref{kernel}) or the $k$NN (\ref{kNN}) estimators; so we need to choose an appropriate semimetric (see Section \ref{sec:FNM}). After that, an approximate linear model is obtained:
	\begin{equation}
		\widetilde{\pmb{Y}}\approx\widetilde{\pmb{Z}}\pmb{\beta}_0+\pmb{\varepsilon}, \label{linear}
	\end{equation}
	where $\widetilde{\pmb{Y}}=(\widetilde{Y_1},\dots, \widetilde{Y_n})^{\top}$ and $\widetilde{\pmb{Z}}=(\widetilde{\pmb{Z}_1},\dots, \widetilde{\pmb{Z}_n})^{\top}$ are the residuals of the regressions of the response variable and the predictors, respectively, over the functional covariate, and $\pmb{\varepsilon}=\left(\varepsilon_1,\dots,\varepsilon_n\right)^{\top}$.  
\item The second step entails the application of the PeLS approach to model (\ref{linear}), minimising the profile least-squares function with respect to $\pmb{\beta}=(\beta_1,\ldots,\beta_{p_n})^{\top}$:
	\begin{equation}
		\mathcal{Q}\left(\pmb{\beta}\right)=\frac{1}{2}\left(\widetilde{\pmb{Y}}-\widetilde{\pmb{Z}}\pmb{\beta}\right)^{\top}\left(\widetilde{\pmb{Y}}-\widetilde{\pmb{Z}}\pmb{\beta}\right)+n 
		\sum_{j=1}^{p_n}\mathcal{P}_{\lambda_{j}}\left(\beta_j\right).
		\label{expr:func_minimizar}
	\end{equation}
	Here $\mathcal{P}_{\lambda_{j}}\left(\cdot\right)$ represents a penalty function and $\lambda_{j} > 0$ is a tuning parameter that controls the amount of penalisation. In fact, under suitable conditions on $\mathcal{P}_{\lambda}$ (see, e.g., \citealt{FanLi2001}), the PeLS estimators produce sparse solutions, that is, many estimated coefficients are zero (see \citealt{AneirosNovoVieu2022} for a review on variable selection in the functional setting using shrinkage methods). Therefore, we perform simultaneous estimation and variable selection in the linear component. 
	\item The third step is the estimation of the functional operator $m(\cdot)$. After obtaining $\widehat{\pmb{\beta}}_0$, a natural way to perform this estimation is to smooth the partial residuals, $Y_i-\pmb{Z}_i^{\top}\widehat{\pmb{\beta}}_0$. For that it is enough to replace $Y_i$ by $Y_i-\pmb{Z}_i^{\top}\widehat{\pmb{\beta}}_0$ in   (\ref{kernel}) for kernel estimation,  or in (\ref{kNN}) for $k$NN techniques.	
\end{enumerate} 
Regarding the choice of the penalty function $\mathcal{P}_{\lambda}\left(\cdot\right)$ in (\ref{expr:func_minimizar}), there is a large list of possibilities in the literature for linear models. The LASSO (\citealt{Tibshirani1996}), defined as $\mathcal{P}_{\lambda}(\beta_j)=\lambda \left\lvert\beta_j\right\lvert$, is the most famous one, despite not satisfying oracle properties (see \citealt{FanLi2001}). A prominent alternative to norm-based penalties like the LASSO is the SCAD (\citealt{Fan97}), 
defined,  for $a>2$, as  
	\begin{equation}
		 \mathcal{P}_{\lambda}(\beta_j)=\left\{\begin{aligned}
			&\lambda\left|\beta_j\right|&  \quad |\beta_j|<\lambda,\\
			&\frac{(a^2-1)\lambda^2-(|\beta_j|-a\lambda)^2}{2(a-1)}&\quad \lambda\leq |\beta_j|<a\lambda, \\
			&\frac{(a+1)\lambda^2}{2}&\quad |\beta_j|\geq a\lambda.
		\end{aligned}
		\right.
		\label{SCAD}
	\end{equation} 
	(\citealt*{FanLi2001} suggested taking $a=3.7$)
The SCAD satisfies the oracle properties and, in general, improves the features of the LASSO (\citealt{FanLi2001}). \cite{AneirosFerratyVieu2015} demonstrated the consistency of variable selection achieved using the PeLS procedure combined with the SCAD penalty. They also derived the convergence rates for the estimators and established the oracle property in the context of the SFPLM.

Both penalty functions, LASSO and SCAD, are included in the \texttt{fsemipar} package. In fact, the package features a generalization of both penalties. Specifically, it facilitates dealing with grouped partial linear models. In this scenario, the scalar regressors are divided into $M$ groups, denoted as $\pmb{Z}_{im}=(Z_{im1},\dots,Z_{imv_m})^\top$ for $m=1,\dots,M$, where $\sum_{m=1}^M v_{m}=p_n$. In such situations, the focus is not on selecting variables individually, but rather on choosing important factors, each corresponding to a group of covariates.
\citet{YuanLin2005} proposed the group LASSO penalty defined as $P_{\lambda}(\pmb{\beta})=\lambda\sum_{m=1}^M\sqrt{\pmb{\beta}_m^{\top}K_m\pmb{\beta}_m}$, where $K_m$ are positive definite matrices for $m\in{1,\dots,M}$. They used $K_m=v_mI_{v_m}$, with $m\in{1,\dots,M}$, where $I_{v_m}$ is the identity matrix of dimension $v_m$.
\citet{Wangetal2007} proposed the group SCAD penalty, defined as
$P_{\lambda}(\pmb{\beta})=\lambda\sum_{m=1}^M\mathcal{P}_{\lambda}\left(\sum_{r=1}^{v_m}\beta_{mr}^{2}\right)$,
where $\mathcal{P}_{\lambda}(\cdot)$ was defined in (\ref{SCAD}). Both group LASSO and group SCAD penalties are included in the \texttt{fsemipar} package. It is important to note that when $M=1$, individual selection is performed.

Another important aspect is the selection of the parameter $\lambda_j$ ($j=1,\dots,p_n$) on which the penalty functions depend (see expression (\ref{expr:func_minimizar})).
To reduce  the number of parameters to select, we consider $\lambda_j = \lambda \widehat{\sigma}_{\beta_{0,j,OLS}}$, where $\beta_{0,j,OLS}$ denotes the ordinary least-squares (OLS) estimate of $\beta_{0,j}$ and $\widehat{\sigma}_{\beta_{0,j,OLS}}$ is the estimated standard deviation. 
\begin{remark}
Note that  in step 1 we must perform nonparametric regression, which involves a tuning parameter, $h$ (kernel) or $k$ ($k$NN). Therefore we compute the nonparametric fit for a grid of values for the tuning parameter, with each value leading to a different linear model in expression (\ref{linear}).
In the package \texttt{fsemipar}, the PeLS objective function (\ref{expr:func_minimizar}) is minimised using the function \texttt{grpreg} 
of the package \texttt{grpreg} (\citealt{grpreg}).  Specifically, for each linear model (\ref{linear}) (setting the value of $h$ or $k$), the function \texttt{grpreg} fits the regularization path  over a grid of values for $\lambda$ (obtains the value of $\beta$ that minimises (\ref{expr:func_minimizar}) for each $\lambda$ in the grid). This optimisation is carried out using the coordinate descent approach in \cite{Breheny2011}. Then, the function \texttt{select} (\texttt{grpreg} package) is used to determine the optimal value of $\lambda$  for each linear model using a predefined objective criterion, such as Generalised CV (GCV), $k$-folds CV or Bayesian Information Criterion (BIC). Finally, the same criterion is used to select the optimal value for $\widehat{h}$  or $\widehat{k}$. Therefore, once selected the linear model, $\widehat{\lambda}$ is also determined and, as a consequence, we have obtained $\widehat{\beta}_0$. The estimated values  $\widehat{h}$ or $\widehat{k}$ and $\widehat{\beta}_0$ are then used in step 3 to fit the nonparametric component of the model. 
\end{remark}
Next, we provide the pseudocode for kernel-based SFPLM estimation; $k$NN-based estimation would be analogous:
\begin{algorithm}
	\caption{SFPLM estimation}
	\begin{algorithmic}[1]
		\For{each $h$ in the grid}
			\State Perform functional nonparametric regression to estimate $\mathbb{E}(Y_i|\mathcal{X}_i)$ and $\mathbb{E}\left(\pmb{Z}_i|\mathcal{X}_i\right)$.  \State Denote by $\widehat{m}_h^{y_i}$ and $\widehat{m}_h^{\pmb{z}_i}$ the respective estimates. Obtain $\widetilde{Y}_i=Y_i-\widehat{m}_h^{y_i}$ and $\widetilde{\pmb{Z}}_i=\pmb{Z}_i-\widehat{m}_h^{\pmb{z}_i}$. 
			 \State Construct a linear model as in (\ref{linear}).
			 \For{each $\lambda=\lambda_h$ in the grid}
			\State \hspace{\algorithmicindent} Use \texttt{grpreg} to obtain the $\pmb{\beta}_{h,\lambda}$ that minimises (\ref{expr:func_minimizar}). Denote it by $\widehat{\pmb{\beta}}_{h,\lambda}$.
			\EndFor
			\State Use  \texttt{select} to determine the optimal $\lambda_{h}$ using with a pre-specified criterion. Denote it by $\widehat{\lambda}_h$.
			\EndFor
	\State Use the same criterion to select the optimal $h$. Denote it by $\widehat{h}$. Then, $\widehat{\lambda}=\widehat{\lambda}_{\widehat{h}}$ and $\widehat{\pmb{\beta}}_0=\widehat{\pmb{\beta}}_{\widehat{h},\widehat{\lambda}}$.	
	\State Perform functional nonparametric regression using $\widehat{h}$ to fit the residuals $Y_i-\pmb{Z}_i\widehat{\pmb{\beta}}_0$ to obtain $\widehat{m}_{\widehat{h}}$.
		\end{algorithmic}
\end{algorithm}
\cite{NovoAneirosVieu2020} adapted the described procedure to fit the SFPLSIM (\ref{SFPLSIM}). In this case, conditional expectations in the expression (\ref{mod_lineal}) become $\mathbb{E}\left(Y_i|\mathcal{X}_i\right)=\mathbb{E}\left(Y_i|\left<\theta_0,\mathcal{X}_i\right>\right)$ and $\mathbb{E}\left(\pmb{Z}_i|\mathcal{X}_i\right)=\mathbb{E}\left(\pmb{Z}_i|\left<\theta_0,\mathcal{X}_i\right>\right)$, which must be estimated through functional single-index regression. For that we can use either kernel- or $k$NN-based procedures in expression (\ref{expr:StatFSIM}). In the second step, we apply the PeLS procedure, see expression (\ref{expr:func_minimizar}), but now the profile least-squares function also depends on $\theta$.

\begin{equation}
	\mathcal{Q}\left(\pmb{\beta}, \theta\right)=\frac{1}{2}\left(\widetilde{\pmb{Y}}_{\theta}-\widetilde{\pmb{Z}}_{\theta}\pmb{\beta}\right)^{\top}\left(\widetilde{\pmb{Y}}_{\theta}-\widetilde{\pmb{Z}}_{\theta}\pmb{\beta}\right)+n 
	\sum_{j=1}^{p_n}\mathcal{P}_{\lambda_{j}}\left(\beta_j\right).
	\label{expr:func_minimizar_theta}
\end{equation}
 Therefore, the minimisation must be performed over both $\pmb{\beta}$ and $\theta$. Once $\widehat{\pmb{\beta}}_0$ and $\widehat{\theta}_0$ are obtained, we estimate the real-valued link function $r$ in the same way as $m(\cdot)$ in the SFPLM, but now using one of the expressions in (\ref{expr:StatFSIM}). \cite{NovoAneirosVieu2020} demonstrated the consistency of the PeLS procedure with the SCAD penalty in the context of the SFPLSIM. They also obtained the convergence rates for the estimators and established the oracle property in this setting.

\begin{remark}
	In the case of the SFPLSIM,  step 1 involves using functional single-index regression. We follow the procedure in Section \ref{sec:ait} to obtain the set of functional directions $\Theta$.  Therefore we compute the nonparametric fit for a grid of values for the tuning parameter ($h$ or $k$) and for each $\theta\in\Theta$. As a result, each pair $(\theta,h)$ or $(\theta,k)$ leads to a different linear model in expression (\ref{linear}).
	For each linear model (\ref{linear}) (setting the values of $h$ or $k$ and $\theta$), the minimisation of expression (\ref{expr:func_minimizar}) is carried out using the function \texttt{grpreg} over a grid of values for $\lambda$. Then, we use the function \texttt{select} to determine the optimal value for $\lambda$  using a predefined objective criterion for each linear model. The same criterion is used to select the optimal value for $h$ or $k$,
	therefore we obtain the optimal $\beta$ for each $\theta\in\Theta$. After that, the $\widehat{\theta}_0$  is the $\theta\in\Theta$ that offers a lower value of the objective function $\mathcal{Q}(\cdot,\cdot)$ and $\widehat{h}$/$\widehat{k}$ and $\widehat{\beta}_0$ the associated optimal values for the remaining parameters. The estimated values  $\widehat{h}$ or $\widehat{k}$, $\widehat{\theta}_0$ and $\widehat{\beta}_0$ are then used in step 3 to fit the semiparametric component of the model. As can be observed we perform a joint selection of the tuning parameters of the model, so we must take into account the considerations in Section \ref{sec:ait}.
\end{remark}

Next, we provide the pseudocode for kernel-based SFPLSIM estimation; $k$NN-based estimation would be analogous:
\begin{algorithm}
	\caption{SFPLSIM estimation}
	\begin{algorithmic}[1]
		\For{each $\theta\in\Theta$}
		\For{each $h$ in the grid}
		\State Perform functional single-index regression to estimate $\mathbb{E}\left(Y_i|\left<\theta,\mathcal{X}_i\right>\right)$ and $\mathbb{E}\left(\pmb{Z}_i|\left<\theta,\mathcal{X}_i\right>\right)$.  \State Denote by $\hat{r}_{\theta, h}^{y_i}$ and $\hat{r}_{\theta, h}^{\pmb{z}_i}$ the respective estimates. Obtain $\widetilde{Y}_{i}=Y_i-\hat{r}_{\theta, h}^{y_i}$ and $\widetilde{\pmb{Z}}_i=\pmb{Z}_i-\hat{r}_{\theta, h}^{\pmb{z}_i}$. 
		\State Construct a linear model as in (\ref{linear}).
		\For{each $\lambda=\lambda_{h,\theta}$ in the grid}
		\State \hspace{\algorithmicindent} Use \texttt{grpreg} to obtain the $\pmb{\beta}_{h,\lambda,\theta}$ that minimises (\ref{expr:func_minimizar_theta}). Denote it by  $\widehat{\pmb{\beta}}_{h,\lambda,\theta}$.
		\EndFor
		\State Use  \texttt{select} to determine the optimal $\lambda_{h,\theta}$ using with a prespecified criterion. Denote it by  $\widehat{\lambda}_{h,\theta}$.
		\EndFor
		\State Use the same criterion to select the optimal $h=h_{\theta}$. Denote it by $\widehat{h}_{\theta}$.
		\EndFor
			\State Select the $\theta$ that minimises (\ref{expr:func_minimizar_theta}) when $\pmb{\beta}=\widehat{\pmb{\beta}}_{\widehat{h}_{\theta}, \widehat{\lambda}_{\widehat{h}_{\theta},\theta},\theta}$ is considered. Denote it by $\widehat{\theta}_0$.  Then, $\widehat{h}=\widehat{h}_{\widehat{\theta}_0}$, $\widehat{\lambda}=\widehat{\lambda}_{\widehat{h},\widehat{\theta}_0}$ and $\widehat{\pmb{\beta}}_0=\widehat{\pmb{\beta}}_{\widehat{h},\widehat{\lambda},\widehat{\theta}_0}$.	
		\State Perform functional single-index regression using $\widehat{h}$ and $\widehat{\theta}_0$ to fit the residuals $Y_i-\pmb{Z}_i\widehat{\pmb{\beta}}_0$ to obtain $\widehat{r}_{\widehat{h},\widehat{\theta}_0}$.
	\end{algorithmic}
\end{algorithm}
\subsection{Fitting SFPLM and SFPLSIM with the \texttt{fsemipar} package} \label{sec:SFPLMfsemipar}

The \texttt{fsemipar} package contains functions to estimate the SFPLM and SFPLSIM using the procedures described in Section \ref{sec:EstVS}. As in the FSIM case, it provides the flexibility to perform both kernel- and $k$NN-based fitting.
The primary related functions are summarised in the Table \ref{tab:SFPLM} along with the corresponding S3 classes designed to include essential S3 methods (\texttt{print}(), \texttt{summary}(), \texttt{plot}() and \texttt{predict}()).

\begin{table}[H]
	\centering
	\begin{tabular}{llll}
		\cmidrule[1.5pt](lrr){1-4}  
		\textbf{Model}&\textbf{Function} & \textbf{S3 class} &\textbf{Description}\\
		\cmidrule[1.5pt](lrr){1-4}
	\multirow{2}{*}{\makecell[l]{SFPLM}} &	\texttt{sfpl.kernel.fit} & `sfpl.kernel' & PeLS with kernel estimation. \\
		&\texttt{sfpl.kNN.fit} & `sfpl.kNN' & PeLS with $k$NN estimation. \\
		\cmidrule[0.5pt](lrr){1-4}
		\multirow{2}{*}{\makecell[l]{SFPLSIM}} &	\texttt{sfplsim.kernel.fit}  & `sfplsim.kernel' & PeLS with kernel estimation. \\
		&\texttt{sfplsim.kNN.fit} & `sfplsim.kNN' & PeLS with $k$NN estimation.\\
		\cmidrule[1.5pt](lrr){1-4}  
	\end{tabular}
	\caption{Summary of main functions for the SFPLM and SFPLSIM, together with the associated S3 classes.}
	\label{tab:SFPLM}
\end{table}

\subsubsection{Main fitting functions}\label{sec:MainSFPLMfsemipar}
The main fitting functions for the SFPLM and the SFPLSIM (Table \ref{tab:SFPLM}) share three required arguments: 
\begin{itemize}
	\item The functional predictor \texttt{x},  a matrix of dimension $(n,p)$ containing $n$ curves discretised in $p$ points
	\item The scalar covariates \texttt{z}, a matrix of dimension $(n,p_n)$.
	\item The scalar response variable \texttt{y}, a vector of dimension $n$.
\end{itemize} 
\begin{verbatim}
	sfpl.kernel.fit(x,z,y,...)
	sfpl.kNN.fit(x,z,y,...)
	sfplsim.kernel.fit(x,z,y,...)
	sfplsim.kNN.fit(x,z,y,...)
\end{verbatim}
\noindent However, these functions additionally accept numerous optional parameters to provide more precise customisation; some of them have been already introduced in the main fitting functions for the FSIM:
\begin{itemize}
\item \emph{Optional arguments for the PeLS procedure}. In this item, we include the arguments related to the penalisation function and the selection of the tuning parameter $\lambda$. These arguments are listed in Table \ref{tab:PeLS}. The PeLS objective function is minimised using the function \texttt{grpreg} of the package \texttt{grpreg} (\citealt{grpreg}). The four functions listed in Table \ref{tab:SFPLM} share a default option for constructing the sequence of $\lambda$ values from which $\widehat{\lambda}$ will be chosen. This sequence is generated by starting from a minimum value (expressed as a fraction of the maximum value) controlled by the user, and it can contain a customisable number of values.
More specifically, the user has the option to specify a value for the minimum $\lambda$ to face two different situations: when $n$ is smaller than \texttt{factor.pn} (default 1) times $p_n$, this value can be set using the \texttt{lambda.min.h} argument (default $0.05$); when $n$ is greater, the \texttt{lambda.min.l} argument (default $0.00001$) allows the user to set a different value. The utility of these options will be showed in Sections \ref{sec:tecator3} and \ref{sec:Sugar}. Alternatively, we can directly specify a unique value for the minimum lambda using the \texttt{lambda.min} argument. When \texttt{lambda.min} is provided, \texttt{lambda.min.h} and \texttt{lambda.min.l} are disregarded.
Furthermore, the number of $\lambda$ values in the sequence can be customized using the \texttt{nlambda} argument (default 100).
Additionally, the user has the option to provide the sequence of $\lambda$ values directly through the \texttt{lambda.seq} argument. When \texttt{lambda.seq}  is provided, all of the previously mentioned arguments are ignored.
Note that setting \texttt{lambda.seq} to 0 results in non-penalized fitting, which corresponds to the ordinary least-squares estimates.

In addition, we can choose the criterion to select $\lambda$ and the tuning parameter $h$ (kernel-based functions) or $k$ (kNN-based  functions). This can be done through the argument \texttt{criterion} (default \texttt{"GCV"}) and the available options are \texttt{"GCV", "AIC", "BIC"} and \texttt{"k-fold-CV"}. If we choose $k$-fold CV, we can specify the number or CV folds using \texttt{nfolds} (default 10) and also set the seed to obtain reproducible results using the argument \texttt{seed} (default 123).

The functions listed in Table \ref{tab:SFPLM} also allow for grouped variable selection. Using the argument \texttt{vn} (default \texttt{vn=ncol(z)}, which leads to individual penalisation), the user can directly specify the number of groups or to provide a vector of possibilities so that the routines select the optimal number of groups according to the provided \texttt{criterion}. For instance, $v_n=p_n$ means that each variable is a group (ungrouped case); whereas \texttt{vn=c(pn,2)} means that routines will choose between the ungrouped case and the case with two groups of consecutive variables (both of size $n/2$, if $n$ is even).

Finally, the user can also choose between the penalty function  SCAD/group SCAD (\texttt{grSCAD}) and  LASSO/group LASSO (\texttt{grLASSO}) using the argument $\texttt{penalty}$ (default \texttt{grSCAD}), and set the maximum number of iterations over each regularisation path though the argument \texttt{max.iter} (default $1000$).
	\begin{table}[h]
		\centering
		\begin{tabular}{p{3.6cm}lp{8.8cm}}
			\cmidrule[1.5pt](lr){1-3}  
			\textbf{Function} & \textbf{Argument} & \textbf{Description} \\
			\cmidrule[1.5pt](lr){1-3}
			\multirow{12}{*}{\makecell[l]{\small \texttt{sfpl.kernel.fit}\\ \small\texttt{sfpl.kNN.fit}\\ \texttt{sfplsim.kernel.fit}\\ \small \texttt{sfplsim.kNN.fit}}} & \texttt{lambda.min.h} & Smallest value for $\lambda$ if \texttt{n<factor.pn*pn}. \\
			& \texttt{lambda.min.l} & Smallest value for $\lambda$ if \texttt{n>factor.pn*pn}. \\
			& \texttt{lambda.min} & Smallest value for $\lambda$. \\
			& \texttt{factor.pn} & Positive integer used to set the smallest value for $\lambda$. \\
			& \texttt{nlambda} & Number of $\lambda$ values in the sequence. \\
			& \texttt{lambda.seq} & Sequence of $\lambda$ values. \\
			& \texttt{vn} & Integer/vector of integers indicating the number/numbers of groups of variables to be penalised together.\\
			& \texttt{criterion} & Criterion for selecting $\hat{\lambda}$ and $\hat{h}$/$\hat{k}$: \texttt{"GCV", "BIC", "AIC", "k-fold-CV"}.\\
		    & \texttt{nfolds} & If \texttt{criterion= "k-fold-CV"}: number of CV folds.\\ 
		     & \texttt{seed} & If \texttt{criterion= "k-fold-CV"}: Seed to obtain reproducible results.\\  
		     	& \texttt{penalty} & Penalty function: \texttt{"grLASSO", "grSCAD"}.\\ 
		     	& \texttt{max.iter} & Maximum number of iterations (across the entire path).\\    
			\cmidrule[1.5pt](lr){1-3}  
		\end{tabular}
		\caption{Optional input arguments related to the PeLS procedure.}
		\label{tab:PeLS}
	\end{table}
	\item \emph{Optional arguments for the nonparametric fit}. 
	This item refers to the arguments that are collected in Table \ref{tab:nonpar}. Functions that perform the kernel-based procedure accept the same arguments as \texttt{fsim.kernel.fit()} and \texttt{fsim.kernel.fit.optim()}. Similarly, functions implementing the \( k \)NN-based estimation accept arguments consistent with \texttt{fsim.kNN.fit()} and \texttt{fsim.kNN.fit.optim()}. The four functions in Table \ref{tab:SFPLM} accept the argument \texttt{kind.of.kernel}.
	
	Additionally, the functions for the SFPLM allow the choice of the semimetric using the argument \texttt{semimetric} (default \texttt{"deriv"}). Currently, only the semimetric based on the B-spline representation of derivatives, \texttt{semimetric="deriv"}, and the based on FPCA, \texttt{semimetric="pca"},  are implemented. The code of these functions is based on the  Frederic Ferraty routines included in the web page \url{https://www.math.univ-toulouse.fr/~ferraty/SOFTWARES/NPFDA/index.html}. If \texttt{semimetric="deriv"},  the argument \texttt{q} allows the user to control the order of the derivative (default \texttt{q=0}). If \texttt{semimetric="pca"},  the argument \texttt{q} controls the number of principal components retained (default \texttt{q=2}).
	\item  \emph{Optional arguments for B-spline expansions}. Functions for the SFPLSIM allow all the arguments listed in Table \ref{tab:bspline}; functions for the SFPLM, only those related to the B-spline expansion of the curves (\texttt{order.Bspline, nknot} and \texttt{range.grid}).
	\item \emph{Optional arguments for the parallel computation}. The functions for the SFPLSIM, which are time consuming in contrast to functions for the SFPLM, allow for parallel computation. Similar to the FSIM functions, the user can  specify the number of CPU cores to use for parallel execution by setting the argument \texttt{n.core} (see the item  \emph{Optional arguments for the parallel computation} in Section \ref{sec:mainfsim}).
\end{itemize}
	
	\subsubsection{Case study I, part II: Tecator dataset}\label{sec:tecator2}
	
	In \cite{NovoAneirosVieu2020}, the authors studied the inclusion of protein and moisture content as covariates to predict the fat content in each piece of meat. They also investigated whether the quadratic, cubic, and interaction effects of these scalar covariates should be included in the model.  This means to consider initially seven scalar covariates: denoting by $Z_1$ and $Z_2$ the protein and moisture contents, respectively, they considered as linear covariates $Z_{2j-1}=Z_1^j$ and $Z_{2j}=Z_2^j$ (for $j=1,2,3$), and $Z_{7}=Z_1Z_2$. They fitted both an SFPLM and an SFPLSIM to the data, using PeLS estimation with a SCAD penalty combined with a kernel-based technique. In addition, they used the BIC criterion for selecting the tuning parameters. In this context, we can illustrate and compare functions in Table \ref{tab:SFPLM} by extending that application, thereby demonstrating the wide range of capabilities of the \texttt{fsemipar} package.
	
As an example, the following code shows the implementation of the SFPLSIM using the function \texttt{sfplsim.kernel.fit()}. In addition to the data, we have provided some specific input arguments to improve the fit: we have reduced the interval in which the bandwidth is selected (setting \texttt{max.q.h=0.35}) discarding the largest values (given that by default \texttt{num.h}=10, reducing these interval will lead to more precision in the selection of $\widehat{h}$ without increasing the computational cost), we have also increased the minimum value of the sequence for selecting $\lambda$ (\texttt{lambda.min=0.01}) and the number of iterations (\texttt{max.iter=5000}) trying to avoid convergence problems in this context of high dependence between predictors; we also chose the BIC as the criterion for selecting the tuning parameters, which leads to slightly faster computation than GCV.
As in Section \ref{sec:tecator}, we set \texttt{nknot.theta=4}, \texttt{nknot=20} and \texttt{range.grid=c(850,1050)}.
		\begin{verbatim}
			>z1<-Tecator$protein       
			>z2<-Tecator$moisture
			>z.com<-cbind(z1,z2,z1^2,z2^2,z1^3,z2^3,z1*z2)
			>fit2<-sfplsim.kernel.fit(x=x[train,], z=z.com[train,], y=y[train], 
			max.q.h=0.35,lambda.min.l=0.01,max.iter=5000, nknot.theta=4,criterion="BIC",
			nknot=20,range.grid=c(850,1050)) 
		\end{verbatim}	
The fitted object resulting from the functions listed in Table \ref{tab:SFPLM} contains valuable information about the performed estimation. This includes the fitted values for the response (\texttt{fitted.values}), the residuals (\texttt{residuals}), and the estimates for the linear coefficients (\texttt{beta.est}). It also provides the selected values for the tuning parameters, such as $\hat{h}$ (\texttt{h.opt}) or $\hat{k}$ (\texttt{k.opt}), and $\hat{\lambda}$ (\texttt{lambda.opt}). Additionally, the object includes the optimal value of the criterion used to select $\hat{h}/\hat{k}$ and $\hat{\lambda}$ (\texttt{IC}), as well as the optimal value of the objective function of the PeLS procedure (see \ref{expr:func_minimizar}) (\texttt{Q}). In the case of SFPLSIM, it also contains the coefficients of $\widehat{\theta}$ in the B-spline basis (\texttt{theta.est}), among other details.

\begin{figure}[h] 
		\centering
		\begin{subfigure}{0.7\textwidth}
			\includegraphics[width=1\linewidth]{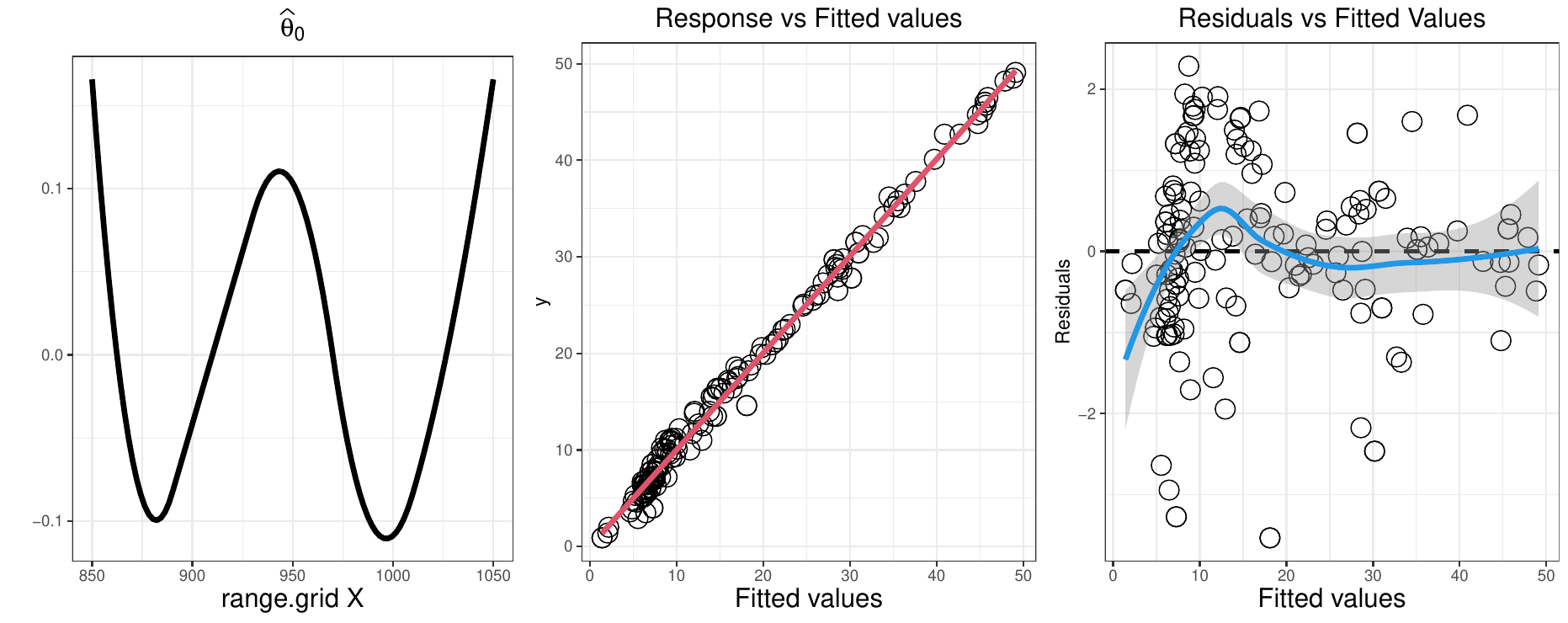}
			\caption{\texttt{sfplsim.kernel.fit()} ($\widehat{n}_r=4$)}
		\end{subfigure}
	
	\vspace{1cm}
	\centering
		\begin{subfigure}{0.7\textwidth}
			\includegraphics[width=1\linewidth]{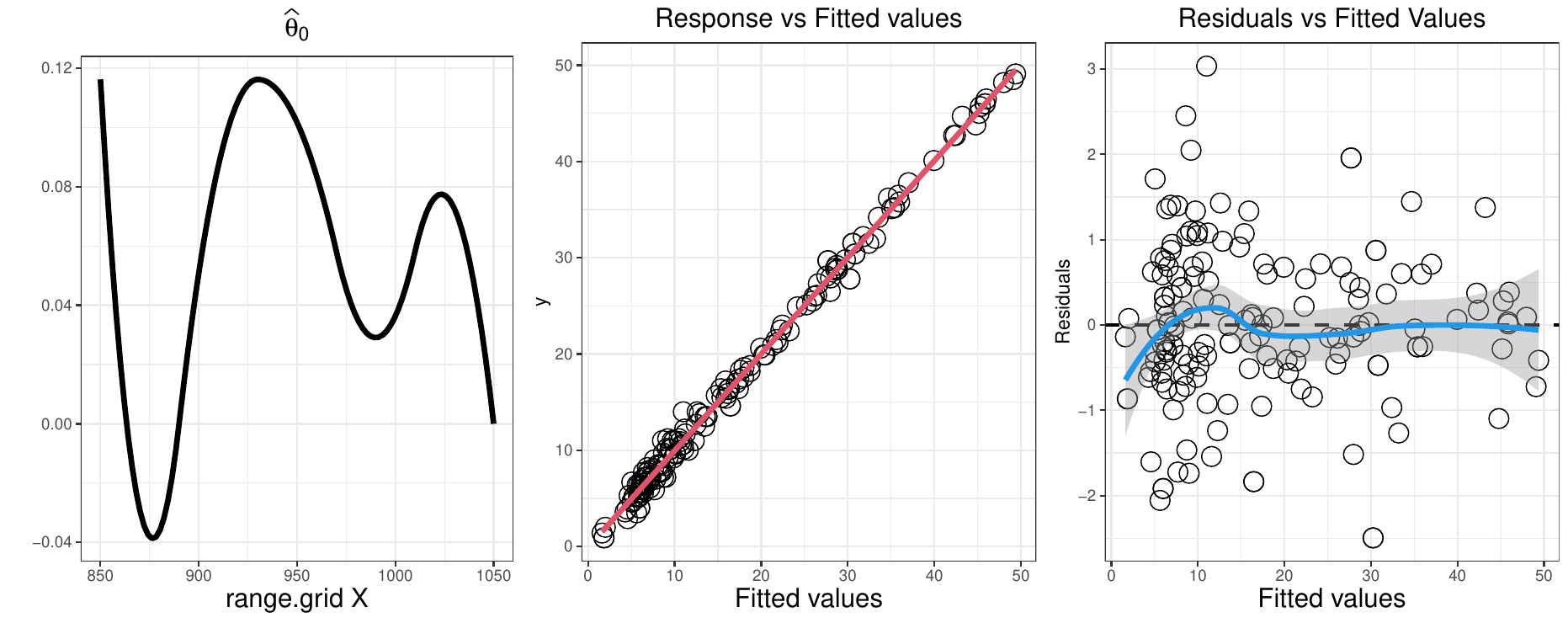}
			\caption{\texttt{sfplsim.kNN.fit()} ($\widehat{n}_r^*=4$)}
		\end{subfigure}

		\caption{Outputs of the S3 method \texttt{plot()} applied to the fitted objects of functions for the SFPLSIM, using the corresponding BIC selectors for the tuning parameters and $n_r$.}
		\label{fig:sfplsim}
	\end{figure}

\begin{figure}[h] 
	\centering
	\begin{subfigure}{0.48\textwidth}
		\includegraphics[width=\linewidth]{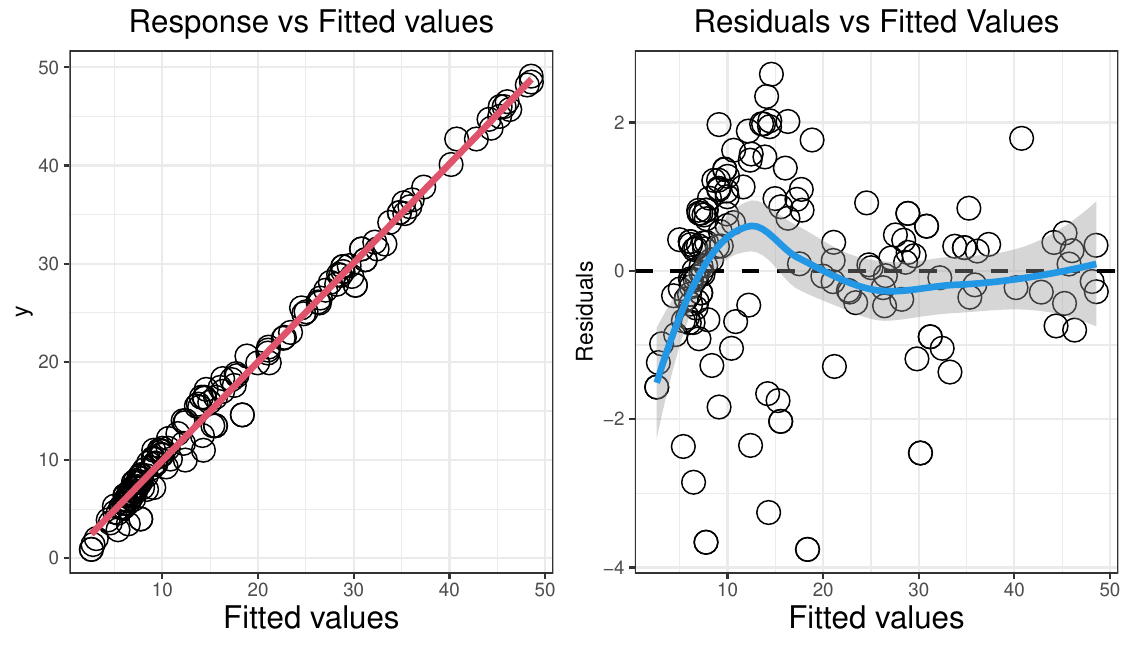}
		\caption{\texttt{sfplm.kernel.fit()}}
	\end{subfigure}
	\begin{subfigure}{0.48\textwidth}
		\includegraphics[width=\linewidth]{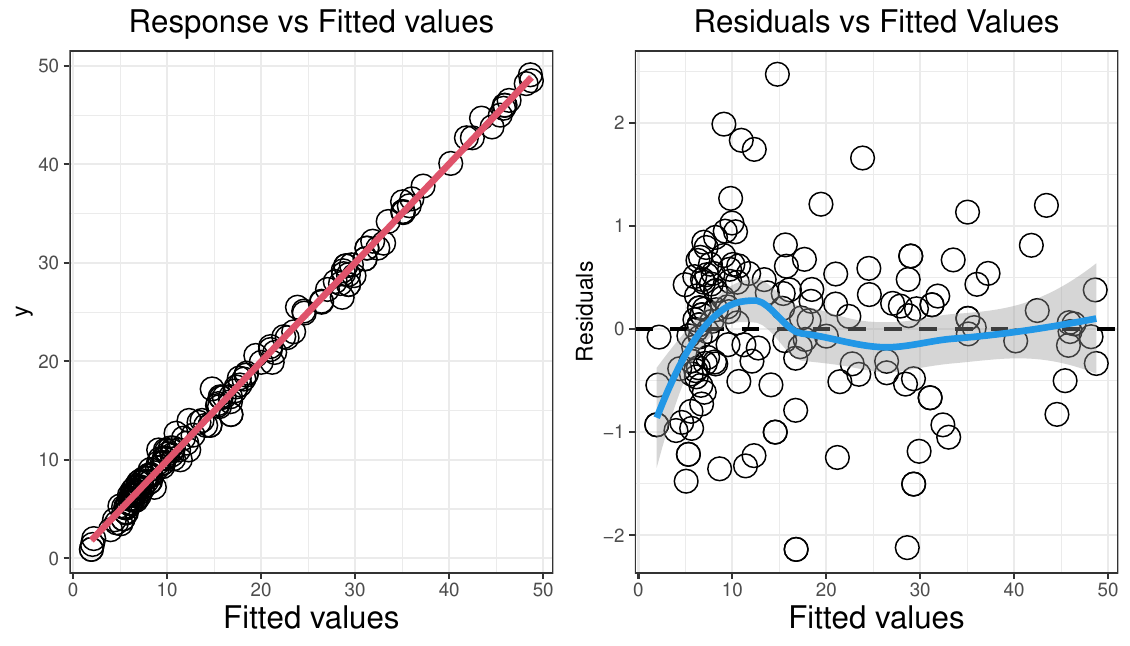}
		\caption{\texttt{sfplm.kNN.fit()}}
	\end{subfigure}
	
	\caption{Outputs of the S3 method \texttt{plot()} applied to the fitted objects of functions for the SFPLM, using the corresponding BIC selectors for the tuning parameters.}
	\label{fig:sfplm}
\end{figure}

In the case of \texttt{sfplsim.kernel.fit()} (class \texttt{sfplsim.kernel}), the fitted object can be used in the functions \texttt{print.sfplsim.kernel()} and \texttt{summary.sfplsim.kernel()} to display summaries of the fitted model. It can also  be used in \texttt{predict.sfplsim.kernel()} to obtain predictions for new values of the covariates provided by the arguments \texttt{newdata.x} (curves) and \texttt{newdata.z} (scalar predictors). This function implements two possibilities for prediction:
\begin{itemize}
	\item If \texttt{option=1}, we maintain all the estimations (\texttt{h.opt},  \texttt{beta.est} and \texttt{theta.est}) to predict the functional single-index component of the model.
	
	\item If \texttt{option=2}, we maintain \texttt{beta.est} and  \texttt{theta.est}, while the tuning parameter (\texttt{h.opt}) is selected again to predict the functional single-index component of the model. This selection is performed using the LOOCV criterion in the associated FSIM.
\end{itemize}
 For each option, the MSEP is also obtained when the user provides the actual responses via the \texttt{y.test} argument. The remaining functions of Table \ref{tab:SFPLM} offer analogous possibilities for prediction. 
		\begin{verbatim}
		> summary(fit2)
		*** SFPLSIM fitted using penalized least-squares combined with kernel estimation with
		 Nadaraya-Watson weights ***
		
		-Call: sfplsim.kernel.fit(x = x[train, ], z = z.com[train, ], y = y[train], 
		nknot.theta = 4, max.q.h = 0.35, range.grid = c(850, 1050), nknot = 20,  
		lambda.min= 0.01, criterion = "BIC", max.iter = 5000)
		
		-Bandwidth (h): 27.11844
		-Theta coefficients in the B-spline basis: 0.1656316 -0.1656316 0 0.1656316 -0.1656316 
		0 0.1656316
		-Linear coefficients (beta): 0.6852112 -1.98299 0 0.01086126 -0.001282229 0 0
		-Number of non-zero linear coefficients: 4
		-Indexes non-zero beta-coefficients: 1 2 4 5
		-Lambda: 0.01244154
		-IC: 523.0534
		-Q: 102.0231
		-Penalty: grSCAD
		-Criterion: BIC
		-vn: 7
		
		> predict(fit2,newdata.x=x[161:215,],newdata.z=z.com[161:215,],y.test=y[161:215],
		option=2)$MSEP
		[1] 1.262723
		\end{verbatim}
		Regarding the prediction, Table \ref{tab:SFPLSIMMSEP} displays the MSEP and selected scalar covariates corresponding to each primary fitting function when using the respective BIC selectors for $n_r$ (only in the SFPSLIM), $\lambda$, and $h$, or $k$.  In the functions for the SFPLM, the results were obtained using the semimetric based on the derivatives (\texttt{semimetric="deriv"}, the option by default) and as \texttt{x}  contains already the second derivative of the curves, we used the option \texttt{q=0} (also, the option by default). 
		\begin{verbatim}			
	sfpl.kNN<-sfpl.kNN.fit(x=x[train,],y=y[train],z=z.com[train,], step=1, min.knn=10,
	max.knn=15,criterion="BIC",range.grid=c(850,1050),lambda.min=0.01,nknot=20,max.iter=
	5000)	
		\end{verbatim}
		 As can be observed from Table \ref{tab:SFPLSIMMSEP}, under the conditions described, the best result in MSEP is obtained with the SFPLM estimated using PeLS with $k$NN estimation, which retains only 3 scalar covariates in the model. 
		\begin{table}[h]
			\centering
			\begin{tabular}{p{4cm}p{3.5cm}p{1.5cm}p{1.5cm}p{1.5cm}}  
				\cmidrule[1.5pt](lr){1-5} 
				& & & \multicolumn{2}{c}{\textbf{MSEP}}\\
				\cmidrule[1pt](lr){4-5} 
				\textbf{Function} & \textbf{Selected variables} & \textbf{$\widehat{n}_r$} & \textbf{1} & \textbf{2} \\  
				\cmidrule[1.5pt](lr){1-5} 
				\texttt{sfpl.kernel.fit()} & $Z_1,Z_2$ & - & 1.317 & 1.317 \\  
				\texttt{sfpl.kNN.fit()} &  $Z_2,Z_3, Z_4$ & - & 0.797 & 0.797\\ 
				\texttt{sfplsim.kernel.fit()} & $Z_1,Z_2,Z_4,Z_5$ & 4 & 1.260 & 1.263 \\ 
				\texttt{sfplsim.kNN.fit()} & $Z_1,Z_2,Z_3,Z_4$ & 4 & 1.349 & 1.349 \\
				\cmidrule[1.5pt](lr){1-5}  
			\end{tabular}
			\caption{MSEP obtained for each function using the respective BIC selectors for tuning parameters ($h$/$k$, $\lambda$; also $n_r$ in the SFPLSIM fitting functions). }
			\label{tab:SFPLSIMMSEP}
		\end{table}
	Finally, the output of the S3 method \texttt{plot()} for the functions in Table \ref{tab:SFPLM} represents two diagnostic plots in addition to the estimated functional index for the SFPLSIM fitting functions. In Figures \ref{fig:sfplsim} and \ref{fig:sfplm}, we provide the outputs of the 4 main functions, obtained under the same conditions as Table \ref{tab:SFPLSIMMSEP}.
	
\section{Multiple scalar predictors with functional origin and a functional covariate}
In Section 4, we focus on SoF regression including as predictors multiple scalar variables with linear effect coming from the discretisation of a curve. Section 4.1 introduces the concept of \emph{points of impact} and some modelling alternatives based on linear or  partial linear approaches when more than one functional variable is available. In Section 4.2, we discuss the techniques implemented in \texttt{fsemipar} for estimation and  selection of points of impact in such models. In Section 4.3 we present the associated functions of the package and illustrate their use through two real chemometric datasets.

\subsection{Linear and partial linear modelling approaches}
In several studies, the interpretability of the outcomes prompted researchers to revisit the concept of discretising functional objects. They discovered that the discretised values of a random curve \(\zeta(t)\), $\zeta(t_1), ..., \zeta(t_{p_n})$, might hold insights that are not accessible via the continuous curve \(\zeta(t)\), and vice versa (see, for instance, \citealt{McKeagueSen2010}). 
So, in certain applications we want to sift through the discretised observations of a curve \(\zeta(t)\) and identify the points of the domain at which the curve most influence a scalar variable of interest $Y$. In some papers (for example, \citealt{Kneipetal2016} or \citealt{NovoVieuAneiros2021}, among others), these points of the domain are called \emph{points of impact}, so we will follow such terminology along the paper.
In the following lines, we summarise the modelling alternatives involving the discretised values of a curve (scalar variables with functional origin) implemented in the package \texttt{fsemipar}:
\begin{itemize}
\item  The multiple linear model (MLM) studied in \citet{AneirosVieu2014}, which is given by the expression: 
\begin{equation}
	Y_i=\beta_{00}+\sum_{j=1}^{p_n}\beta_{0j}\zeta_i(t_j)+\varepsilon_i,  \quad i=1,\dots, n, \label{expr:MLM}
\end{equation}
where $\zeta_i$ is a random curve defined on some interval $[a,b]$ and is observed in the points $a\leq t_1<\dots<t_{p_n}\leq b$. In addition, $\left(\beta_{00},\beta_{01},\dots,\beta_{0p_n}\right)^{\top}$ is a vector of unknown real coefficients. 

\item The multi-functional partial linear model (MFPLM) examined in \citet{AneirosVieu2015}. This model follows the expression 
\begin{equation}
	Y_i=\sum_{j=1}^{p_n}\beta_{0j}\zeta_i(t_j)+m(\mathcal{X}_i)+\varepsilon_i, \quad i=1,\dots, n,\label{expr:MFPLM}
\end{equation}
where $\mathcal{X}_i$ denotes another functional variable valued in some semi-metric space, which influences the response nonparametrically.

\item The multi-functional partial linear single-index model (MFPLSIM) studied in \citet{NovoVieuAneiros2021}, defined by the following relationship:
\begin{equation}
	\label{expr:MFPLSIM}
	Y_i=\sum_{j=1}^{p_n}\beta_{0j}\zeta_i(t_j)+r\left(\left<\theta_0,\mathcal{X}_i\right>\right)+\varepsilon_i \quad i=1,\dots, n. 
\end{equation}
Here, $\mathcal{X}_i$ is valued in a separable Hilbert space and its effect on the response is given through a functional single-index component.
\end{itemize}
In models (\ref{expr:MLM})-(\ref{expr:MFPLSIM}), it is assumed that $p_n\rightarrow\infty$ as $n\rightarrow\infty$, and only a
few of the discretised points $\zeta(t_j)$, $j=1,\dots,p_n$, influence the response $Y$. We could consider applying to these models the same techniques described in Section \ref{sec:EstVS} to estimate and select relevant variables in the linear component.
However, 
these methods are often inadequate for two main reasons. First, there exists a strong dependence between variables, a consequence of their functional origin. Second, the sheer number of observations ($n<<p_n$) makes it difficult to obtain results in a reasonable time frame. Then, it is necessary to develop specialised methods tailored to these particular scenarios. Therefore, in the following section, we provide a short review of the techniques implemented in \texttt{fsemipar} for estimating these models.
\subsection{Selection of points of impact and model estimation}\label{sec:SIP}
\cite{AneirosVieu2014} proposed an algorithm called Partitioning Variable Selection (PVS) for the MLM (\ref{expr:MLM}). This two-step procedure is based on the idea that observations $\mathcal{\zeta}_i(t_j)$ and $\mathcal{\zeta}_i(t_k)$, where $t_j$ and $t_k$ are closely spaced points in the domain of the curve, contain similar information about the response. The method proceeds as follows: 
\begin{enumerate}
	\item In the first step, it considers a reduced linear model incorporating only a few covariates, $w_n$, 
 covering the entire discretisation interval of $\zeta_i(t)$. For this purpose: 
 \begin{itemize}\item We assume, without loss of generality, that $p_n=q_nw_n$.
 	\item The $w_n$ variables taken into account are $\zeta_i(t_{(2k-1)q_n/2})$ with  $k\in\{1,\dots,w_n\}$. 
 	\item The remaining $p_n-w_n$ covariates from the original set are directly discarded. \end{itemize}
 A PeLS procedure (see Section \ref{sec:EstVS}) is applied to the reduced model. In this manner, the dependence among covariates is mitigated before the variable selection is performed.
	\item In the second step,  a linear model is considered that includes the  variables selected in the first step and those in their neighbourhood. Specifically:
	\begin{itemize}
		\item Let's denote $\widehat{S}_{1n}=\{k = 1,\dots,w_n, \ \widehat{\beta}_{(2k-1)q_n/2}\not=0\}$.
		\item The set of covariates considered in the second step is $$\mathcal{R}=\cup_{k\in \widehat{S}_1}\{\zeta_i(t_{(k-1)q_n+1}),\zeta_i(t_{(k-1)q_n+2}) \dots, \zeta_i(t_{kq_n})\}.$$
	\end{itemize} 
A PeLS procedure is applied again to the resultant model. By doing so, relevant information that may have been overlooked in the first step is reconsidered. 
\end{enumerate}
The outputs of the second step of the PVS algorithm are the estimates for the MLM (\ref{expr:MLM}). The selected points of impact are those belonging to the set
$\widehat{S}_{2n}=\{j= 1,\dots,p_n, \ \textrm{such that} \ \zeta_i(t_j)\in\mathcal{R} \ \textrm{and}  \ \widehat{\beta}_{0j}\not=0 \ \textrm{in the second step}\}$, and $\widehat{\beta}_{0j}$ is the estimate of the second step if $\zeta_i(t_j)\in\mathcal{R}$ and $0$ otherwise. 
The PVS algorithm requires dividing the sample to execute the two steps. The natural choice involves using half of the sample in the first step and the other half in the second step, that is $n_1=n_2=n/2$, although  in some applications this may not be the optimal option. The consistency of this procedure was showed in \cite{AneirosVieu2014}, who also provided the rates of convergence of the estimators.

 Note that the PVS algorithm has an extra tuning parameter: the number of covariates in the reduced model, $w_n$, in the first step of the method. This parameter can be selected using the same objective criterion to select the parameter $\lambda$ in the PeLS procedure involved.

 In addition, the first step of the PVS algorithm relies on the assumption  $p_n=w_n q_n$. When this assumption fails, we can use the solution proposed by \citet{AneirosVieu2015}, which is based on consider non-fixed  $q_n=q_{n,k}$ values (for $k=1,\dots,w_n$)  when $p_n/w_n$ is not an integer number. Specifically:
\begin{equation*}
	q_{n,k}= \left\{\begin{aligned}
		&[p_n/w_n]+1&  \quad k\in\{1,\dots,p_n-w_n[p_n/w_n]\},\\
		&[p_n/w_n]&\quad k\in\{p_n-w_n[p_n/w_n]+1,\dots,w_n\}, \\
	\end{aligned}
	\right.
\end{equation*}
where $[z]$ denotes the integer part of $z\in\mathbb{R}$.

\cite{AneirosVieu2015} extended the PVS algorithm to estimate the MFPLM (\ref{expr:MFPLM}). The idea was to transform the MFPLM (\ref{expr:MFPLM}) into a linear model using the same strategy described in Section \ref{sec:EstVS}. Once an approximate linear model was obtained, they applied the PVS algorithm to that model. They proved the consistency of the method and obtained the rates of convergence for the estimators. 

\cite{NovoVieuAneiros2021} addressed the MFPLSIM (\ref{expr:MFPLSIM}).  In this case, there were two additional challenges when extending the PVS: first, estimating the functional index $\theta_0$ was computationally  intensive, and second, it required a relatively large sample size (see \citealt{AneirosNovoVieu2022}). They then explored two variants of the PVS algorithm: the improved algorithm for sparse multi-functional regression (IASSMR), which essentially adapts  the PVS to the MFPLSIM, and the fast algorithm for sparse multi-functional regression (FASSMR), which employs only the first step of the PVS. The motivation for the latter variant  was to reduce the computational cost for large discretisation sizes ($p_n$) and improve performance in situations with small sample sizes, as the sample no longer needs to be divided. Both the IASSMR and the FASSMR require the transformation of the expression (\ref{expr:MFPLSIM}) into a linear model, using the strategy described in Section \ref{sec:EstVS}. In \cite{NovoVieuAneiros2021}, the authors demonstrated the consistency of both methods and derived the convergence rates for the estimators of the former.

\subsection{Fitting the MLM, MFPLM and MFPLSIM with the \texttt{fsemipar} package} \label{sec:ImpactPointsemipar}

The \texttt{fsemipar} package implements the algorithms described in Section \ref{sec:SIP} to estimate models (\ref{expr:MLM})-(\ref{expr:MFPLSIM}). In the case of models with partial linear structure, it provides the flexibility to perform both kernel- and $k$NN-based fitting.
The primary related functions are summarised in the Table \ref{tab:MFPLSIM} along with the corresponding S3 classes.

\begin{table}[h]
	\centering
	\begin{tabular}{llll}
		\cmidrule[1.5pt](lrr){1-4}  
		\textbf{Model}&\textbf{Function} & \textbf{S3 class} &\textbf{Description}\\
		\cmidrule[1.5pt](lrr){1-4}
		\multirow{1}{*}{\makecell[l]{MLM}} &	\texttt{PVS.fit} & `PVS' & PVS algorithm. \\
		\cmidrule[1pt](lrr){1-4}
		\multirow{2}{*}{\makecell[l]{MFPLM}} &	\texttt{PVS.kernel.fit} & `PVS.kernel' & PVS algorithm with kernel estimation. \\
		&\texttt{PVS.kNN.fit} & `PVS.kNN' & PVS algorithm with $k$NN estimation. \\
		\cmidrule[0.5pt](lrr){1-4}
		\multirow{4}{*}{\makecell[l]{MFPLSIM}} &	\texttt{IASSMR.kernel.fit}  & `IASSMR.kernel' & IASSMR with kernel estimation. \\
		&\texttt{IASSMR.kNN.fit} & `IASSMR.kNN' & IASSMR with $k$NN estimation.\\
		&\texttt{FASSMR.kernel.fit} & `FASSMR.kernel' & FASSMR with kernel estimation.\\
		&\texttt{FASSMR.kNN.fit} & `FASSMR.kNN' & FASSMR with $k$NN estimation.\\
		\cmidrule[1.5pt](lrr){1-4}  
	\end{tabular}
	\caption{Summary of main functions for MLM, MFPLM and MFPLSIM together with the associated S3 classes.}
	\label{tab:MFPLSIM}
\end{table}

\subsubsection{Main fitting functions}
The main fitting functions for MLM, MFPLM and MFPLSIM (Table \ref{tab:MFPLSIM}) share the following required arguments: 
\begin{itemize}
	\item The functional predictor \texttt{z} (pointwise effect), a matrix of dimension $(n,p_n)$ containing $n$ curves discretised in $p_n$ points.
	\item The scalar response variable \texttt{y}, a vector of dimension $n$.
\end{itemize} 
\begin{verbatim}
	PVS.fit(z,y,...)
	\end{verbatim}
In the case of the MFPLM and MFPLSIM, the functions also require:
\begin{itemize}
	\item The functional predictor \texttt{x} (continuous effect), a matrix of dimension $(n,p)$ containing $n$ curves discretised in $p$ points.
\end{itemize}
\begin{verbatim}
	PVS.kernel.fit(x,z,y,...)
	PVS.kNN.fit(x,z,y,...)
	IASSMR.kernel.fit(x,z,y,...)
	IASSMR.kNN.fit(x,z,y,...)
	FASSMR.kernel.fit(x,z,y,...)
	FASSMR.kNN.fit(x,z,y,...)
\end{verbatim}
\noindent These functions additionally accept numerous optional parameters. Many of them have been already introduced in Section \ref{sec:MainSFPLMfsemipar}:
\begin{itemize}
	\item \emph{Optional arguments for the two steps of the algorithm}. In functions tailored to PVS and IASSMR (which are two-step procedures), users can control the indices of the data used for the first and second steps with the arguments \texttt{train.1} (default \texttt{1:ceiling(n/2)}) and \texttt{train.2} (default \texttt{(ceiling(n/2)+1):n}), respectively. Additionally, for the seven functions listed in Table \ref{tab:MFPLSIM}, the argument \texttt{wn} (default \texttt{c(10,15,20)}) allows users to provide a sequence of possible values for $w_n$ in the reduced model of step 1.
	\item \emph{Optional arguments for the PeLS procedure}. These arguments are listed in Table \ref{tab:PeLS} and commented in Section \ref{sec:MainSFPLMfsemipar}. In this case, the \texttt{criterion} chosen by the user to select the tuning parameters of the estimation/variable selection procedure is also used to select $\widehat{w}_n$ from the values provided in \texttt{wn}.
\end{itemize}
In addition, functions related to the MFPLM and MFPLSIM admit the following optional arguments:
\begin{itemize}	
	\item \emph{Optional arguments for the nonparametric fit}. 
	This item refers to the arguments collected in Table \ref{tab:nonpar}. Kernel-based functions accept the same arguments as \texttt{fsim.kernel.fit()} and \texttt{fsim.kernel.fit.optim()}, and \( k \)NN-based ones accept arguments consistent with \texttt{fsim.kNN.fit()} and \texttt{fsim.kNN.fit.optim()}.
	Additionally, the functions for the MFPLM admit the arguments \texttt{semimetric} and \texttt{q} (see \emph{Optional arguments for the nonparametric fit} in Section \ref{sec:MainSFPLMfsemipar}).
	\item  \emph{Optional arguments for B-spline expansions}. Functions for the MFPLSIM allow all the arguments listed in Table \ref{tab:bspline}; functions for the MFPLM, only those related to the B-spline expansion of the curves (\texttt{order.Bspline, nknot} and \texttt{range.grid}).
	\item \emph{Optional arguments for the parallel computation}. The functions for the MFPLSIM allow for parallel computation, so they admit the argument \texttt{n.core} (see the item  \emph{Optional arguments for the parallel computation} in Section \ref{sec:mainfsim}).	
\end{itemize}

\subsubsection{Case study I, part III: Tecator dataset}\label{sec:tecator3}
The purpose of this section is twofold: to illustrate the implementation of the models (\ref{expr:MLM})--(\ref{expr:MFPLSIM}) using the functions in Table \ref{tab:MFPLSIM} and to show the predictive advantage of including impact points in the modelling of the Tecator dataset. 

Although MFPLM (\ref{expr:MFPLM}) (so as to the MFPLSIM (\ref{expr:MFPLSIM})) is tailored to address situations with more than one functional covariate, in \cite{AneirosVieu2015}, the authors applied it in the case $\zeta(t)=\mathcal{X}(t)$. They model the Tecator dataset (one functional covariate) using the model: 
	\begin{equation*}
		Y_i=\sum_{j=1}^{100}\beta_{0j}\mathcal{X}_i(t_j)+m(\mathcal{X}_i)+\varepsilon_i, \quad i=1,\dots, 160,\label{expr:MFPLM2}
	\end{equation*}
where $\mathcal{X}_i$ is the second derivative of the absorbance spectrum of each piece of meat, which was recorded in 100 equally spaced wavelengths in the interval 850-1050. 
To illustrate and compare the functions in Table \ref{tab:MFPLSIM}, we extended that application, fitting also the models:
\begin{align*}
	Y_i = \sum_{j=1}^{100} \beta_{0j} \mathcal{X}_i(t_j) + r(\langle \mathcal{X}_i, \theta_0 \rangle) + \varepsilon_i, \quad i = 1, \dots, 160. \\
	Y_i = \beta_{00}+\sum_{j=1}^{100} \beta_{0j} \mathcal{X}_i(t_j) + \varepsilon_i, \quad i = 1, \dots, 160.
\end{align*}

As an example, the following code shows the implementation of the MFPLSIM using the functions \texttt{FASSMR.kernel.fit()} and \texttt{IASSMR.kernel.fit()}. In addition to the data, we have provided some specific input arguments.  Following the comments made in Section \ref{sec:tecator2}, we set \texttt{max.q.h=0.35}, \texttt{criterion="BIC"}, \texttt{nknot.theta=4},  \texttt{nknot=20},  and \texttt{range.grid=c(850,1050)}. We increased the smallest value in the sequence to select $\widehat{\lambda}$ (\texttt{lambda.min=0.03}), trying to obtain a simpler model. Note that in this case, if we use half of the sample in each stage of the two-step procedures (that is, $n_1 = n_2 = 80$ by setting \texttt{train.1=1:80} and \texttt{train.2=81:160}), then $n_1>w_n$ and $n_2$ is likely to exceed the cardinality of $\mathcal{R}$. Consequently, it may not be worthwhile to specify \texttt{lambda.min.h}, \texttt{lambda.min.l}, and \texttt{factor.pn} (refer to \emph{Optional Arguments for the PeLS Procedure} in Section \ref{sec:MainSFPLMfsemipar}) to adjust the level of penalization in each step of the IASSMR. Typically, we assign a value to \texttt{factor.pn} that is $\geq 1$, aiming to impose greater penalization in situations where the sample size is smaller than the number of covariates in the model. 
\begin{verbatim}
fit3 <- FASSMR.kernel.fit(x=x[train,],z=x[train,], y=y[train], nknot.theta=4, 
max.q.h=0.35, range.grid=c(850,1050), nknot=20,criterion="BIC", lambda.min=0.03)

fit4<- IASSMR.kernel.fit(x=x[train,],z=x[train,], y=y[train],train.1=1:80,train.2=81:160,
nknot.theta=4,max.q.h=0.35, range.grid=c(850,1050), nknot=20,criterion="BIC",
lambda.min=0.03)
\end{verbatim}
The objects \texttt{fit3} and \texttt{fit4} belong to an S3 class (see Table \ref{tab:MFPLSIM}), which implements S3 methods. In addition to the information listed in Section \ref{sec:tecator2},  these objects include the selected value for $\widehat{w}_n$ (\texttt{w.opt}).  
As in the Sections \ref{sec:tecator} and \ref{sec:tecator2}, we can display a summary of the fitted models using \texttt{summary.FASSMR.kernel()} (or \texttt{print.FASSMR.kernel()}) and  \texttt{summary.IASSMR.kernel()} (or \texttt{print.IASSMR.kernel()}), respectively. To make  predictions for the fitted object \texttt{fit3}, we can use \texttt{predict.FASSMR.kernel()}. This function implements the two possibilities for prediction outlined in Section \ref{sec:tecator2}. To make predictions for \texttt{fit4}, we can use \texttt{predict.IASSMR.kernel()}, which offers the following possibilities:
\begin{itemize}
	\item If \texttt{option=1}, we retain all the estimates (\texttt{h.opt}, \texttt{theta.est}, and \texttt{beta.est}) to predict the functional single-index component of the model. Since we use the estimates from the second step of the algorithm, only \texttt{train.2} was used as the training sample for prediction. Therefore, \texttt{h.opt} might not be suitable for predicting the functional single-index component of the model.
	\item If \texttt{option=2}, we keep \texttt{theta.est} and \texttt{beta.est}, while the tuning parameter $\widehat{h}$ is selected anew for predicting the functional single-index component of the model. This selection is performed using the LOOCV criterion in the associated FSIM and the complete training sample (i.e., \texttt{train=c(train.1,train.2)}).
	\item If \texttt{option=3}, we maintain only the indices of the relevant variables selected by the IASSMR. We re-estimate the linear coefficients and the functional index using \texttt{sfplsim.kernel.fit}, without penalization (by setting \texttt{lambda.seq=0}) and using the entire training sample (\texttt{train=c(train.1,\newline train.2)}). This method provides two predictions (and MSEPs):
	\begin{enumerate}[a)]
		\item The prediction associated with \texttt{option=1} for the \texttt{sfplsim.kernel} class.
		\item The prediction associated with \texttt{option=2} for the \texttt{sfplsim.kernel} class.
	\end{enumerate}
\end{itemize}
Note that the objects resulting from the functions \texttt{IASSMR.kNN.fit}, \texttt{PVS.kernel.fit} and \texttt{PVS.kNN.fit} offer analogous possibilities for prediction. For more details, see Section \ref{sec:Sugar} or refer to the manual of the \texttt{fsemipar} package (\citealt{AneirosNovo2023}).

Table \ref{tab:MFPLSIMMSEP} displays the MSEP and the selected points of impact for each primary fitting function, using the respective BIC selectors for $\lambda$, $w_n$, and $h$/$k$ in the MFPLM and MFPLSIM models. The points of impact, selected from the original set of 100, are represented in red on a timeline for enhanced visualization and comparison. The column $\widehat{s}_n$ indicates the number of points of impact selected by each function. For the MFPLM functions, the results were obtained using \texttt{semimetric="deriv"}. For comparative purposes, the same values for common input arguments were used across the seven functions in Table \ref{tab:MFPLSIM}. Additionally, $h$ and $k$ were selected using the same grid for each pair of functions employing kernel/$k$NN estimation, setting \texttt{min.knn=2}, \texttt{max.knn=10}, and \texttt{step=1} in the $k$NN case.

It is remarkable that the MLM does not offer good predictive power in this case. This could be expected since the relationship between the response and the predictor is non-linear, as pointed out in \cite{NovoAneirosVieu2019}, among others. Another important observation is that the MFPLSIM appears to offer the lowest MSEP under our conditions, although the MFPLM, estimated using the PVS with $k$NN smoothing, is also competitive in terms of MSEP. Among the algorithms for estimating the MFPLSIM, it is important to note that even the 1-step procedure offers a very competitive result, especially when combined with kernel estimation. In the two-step procedures, estimating again the tuning parameters (\texttt{option=3}) offers the higher predictive power.
\begin{table}[H]
	\centering
	\scalebox{0.95}{ \begin{tabular}{p{1.5cm}p{3cm}p{5.7cm}p{0.5cm}p{1cm}p{1cm}p{1cm}p{1cm}}
			\cmidrule[1.5pt](lr){1-8}  
			& & & \multicolumn{4}{r}{\textbf{MSEP}}\\
			\cmidrule[1pt](lr){5-8} 
			\textbf{Model} & \textbf{Function} & \textbf{Points of impact} & \textbf{$\widehat{s}_n$} & \textbf{1}&\textbf{2} & \textbf{3a} & \textbf{3b} \\
			\cmidrule[1.5pt](lr){1-8} 
			{\footnotesize MLM} & {\footnotesize \texttt{PVS.fit()}} & \impactline{37/37,43/43,44/44} 
			& 3 &19.46 &- &- &- \\ 
			\cmidrule[0.5pt](lrr){1-8}
			\multirow{2}{*}{{\footnotesize MFPLM}} & {\footnotesize \texttt{PVS.kernel.fit()}} & \impactline{29,32,39,79,80,87,99}
			& 7 &6.94 &5.14 &4.66 & 4.66\\ 
			& \texttt{{\footnotesize PVS.kNN.fit()}} & \impactline{31,43,71,72,74} 
			& 5 &2.65&1.84 & 1.29 &1.29 \\ 
			\cmidrule[0.5pt](lrr){1-8}
			\multirow{4}{*}{{\footnotesize MFPLSIM}} & {\footnotesize \texttt{FASSMR.kernel.fit()}} &  \impactline{3,28,33,38,43,78,83,88,93}
			& 9 &1.60& 1.79 &- &- \\
			& {\footnotesize \texttt{FASSMR.kNN.fit()}}&  \impactline{3,33,38,43,68,73,93}
			&7&2.25& 2.25 &- &- \\
			& {\footnotesize \texttt{IASSMR.kernel.fit()}} &  \impactline{2,3,31,35,39,42}
			& 6&1.66 & 1.69 & 1.10&1.15 \\
			& {\footnotesize \texttt{IASSMR.kNN.fit()}} &  \impactline{31,36,41,73,76,89,93} 
			& 7 &1.50& 1.66 & 1.04& 1.04 \\
			\cmidrule[1.5pt](lr){1-8}  
	\end{tabular}}
	\caption{MSEP obtained for each function using the respective BIC selectors for the tuning parameters ($\lambda$ and $w_n$; also $h$ or $k$ in models with functional nonparametric/single-index component).}
	\label{tab:MFPLSIMMSEP}
\end{table}
Focusing on the selection of impact points, we observe four distinct clusters of impact points. It seems that all algorithms consistently select observations in the second quartile of the timeline (wavelengths between $900$ and $950\ nm$). Additionally, some algorithms select impact points at the beginning of the first quartile, while others choose them at the end of the fourth quartile. Another subset of algorithms selects points towards the end of the third quartile. If we combine this information with the MSEP, the best choice appears to be made by the functions \texttt{IASSMR.kNN.fit()} and \texttt{IASSMR.kernel.fit()}.

A comparison of Tables \ref{tab:fsimMSEP}, \ref{tab:SFPLSIMMSEP}, and \ref{tab:MFPLSIMMSEP} reveals that the lowest MSEP value is obtained when the protein and moisture contents are included in the model (using the SFPLM estimated using \texttt{sfpl.kNN.fit()}). However, acquiring these measurements necessitates chemical analyses of the meat samples, which we aim to circumvent by utilizing spectrometric information. A comparison between Table \ref{tab:fsimMSEP} and Table \ref{tab:MFPLSIMMSEP} clearly demonstrates that incorporating both continuous and pointwise effects of the functional predictor not only enhances predictive power but also preserves interpretability: each variable is associated with an interpretable parameter in the model.

\begin{figure}[H] 
	\centering
	\begin{subfigure}{0.25\textwidth}
		\includegraphics[width=1\linewidth]{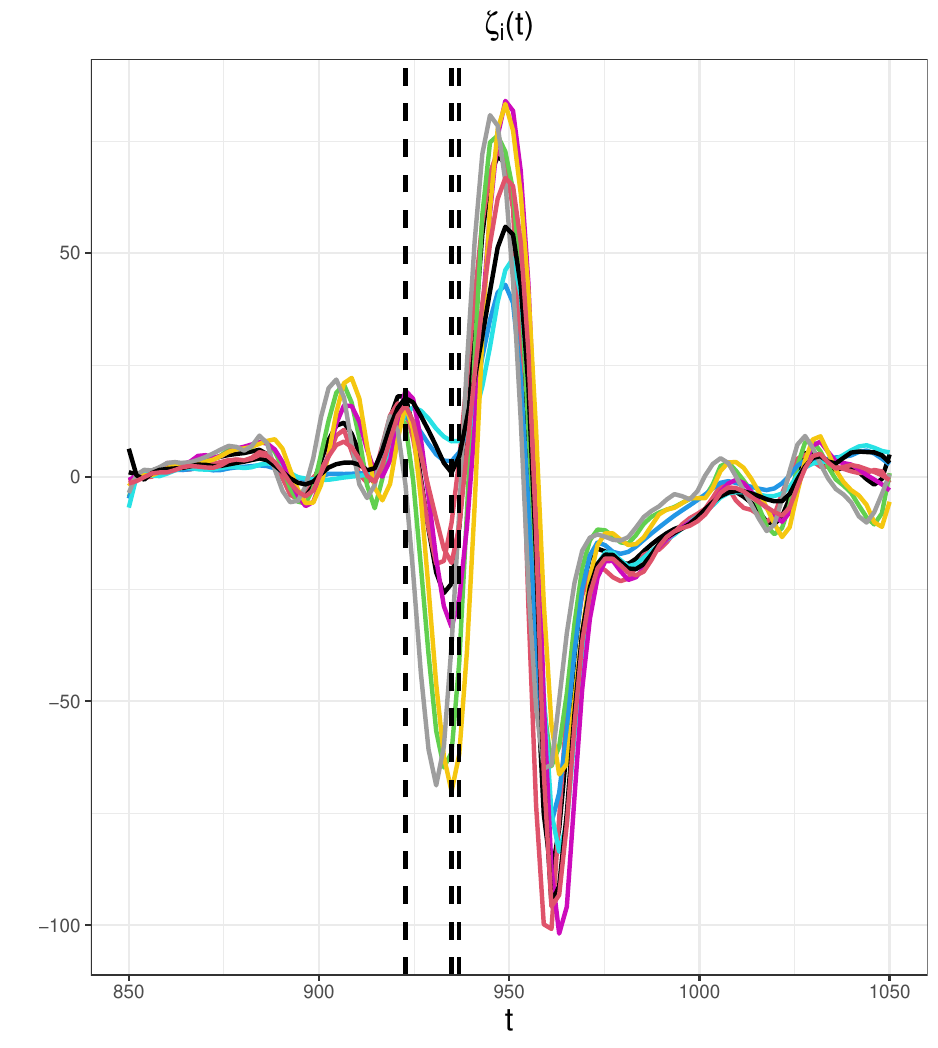}
		\caption{\texttt{PVS.fit()}}
	\end{subfigure}
	\begin{subfigure}{0.25\textwidth}
		\includegraphics[width=1\linewidth]{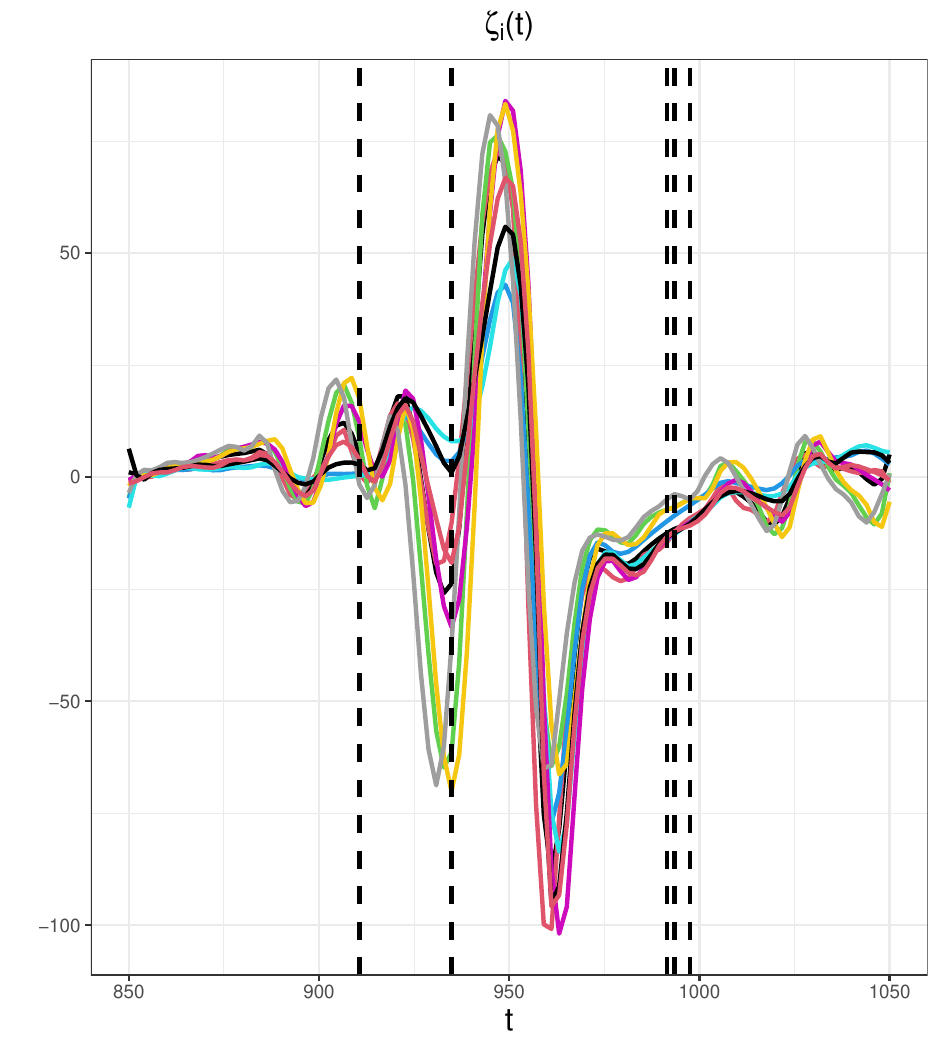}
		\caption{\texttt{PVS.kernel.fit()}}
	\end{subfigure}
		\begin{subfigure}{0.25\textwidth}
		\includegraphics[width=1\linewidth]{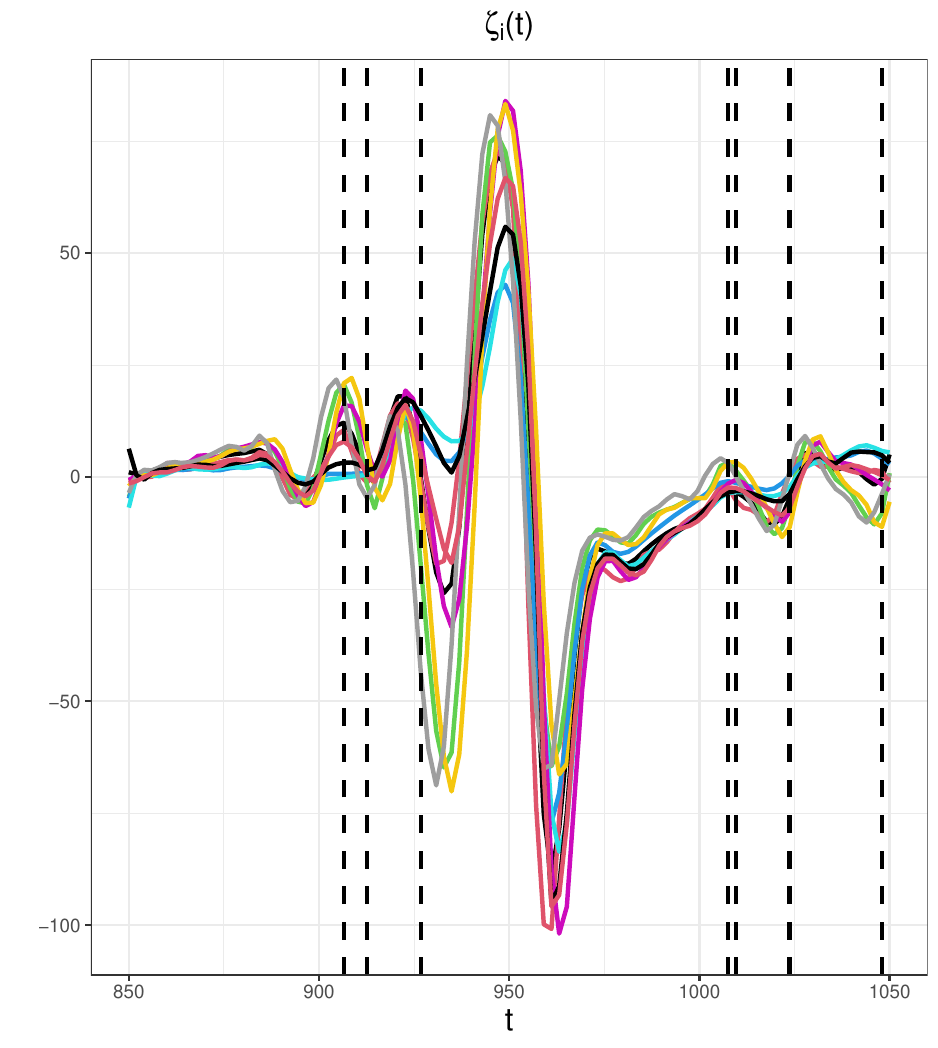}
		\caption{\texttt{PVS.kNN.fit()}}
	\end{subfigure}
	\begin{subfigure}{0.25\textwidth}
	\includegraphics[width=1\linewidth]{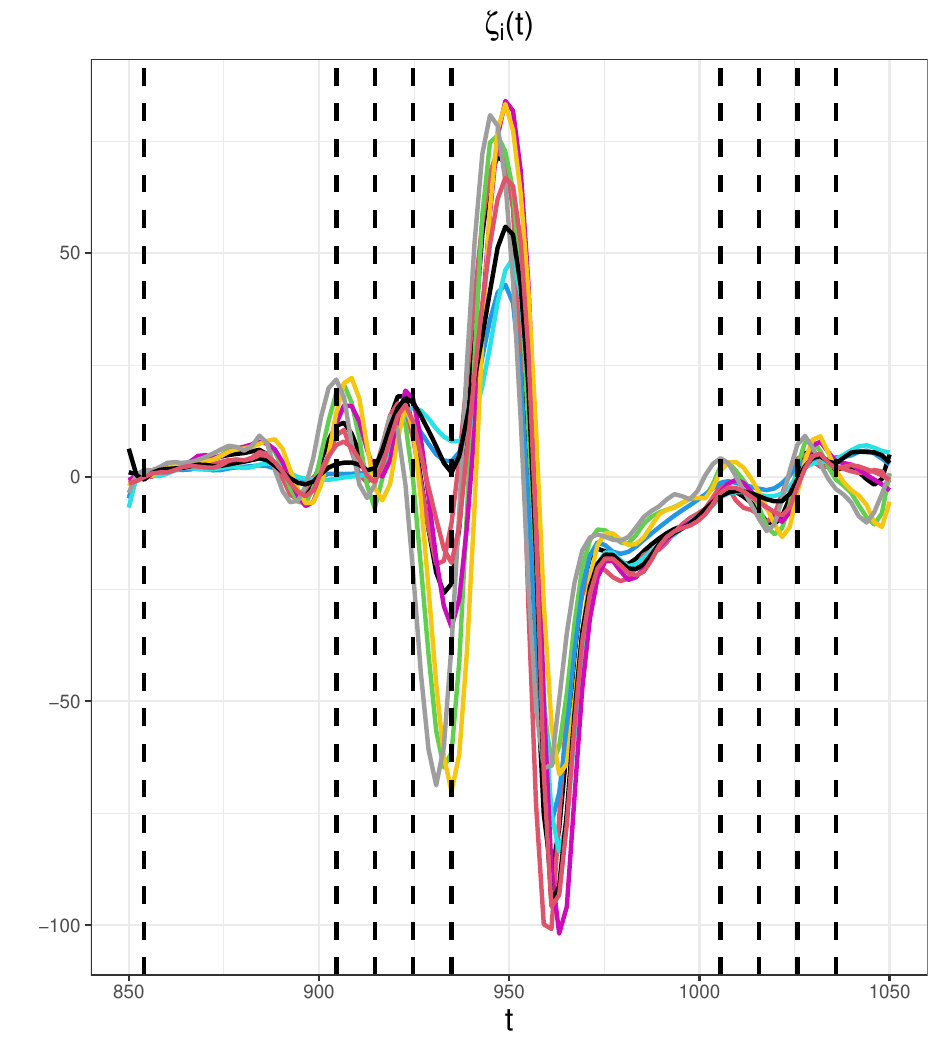}
	\caption{\texttt{FASSMR.kernel.fit()}}
\end{subfigure}
	\begin{subfigure}{0.25\textwidth}
	\includegraphics[width=1\linewidth]{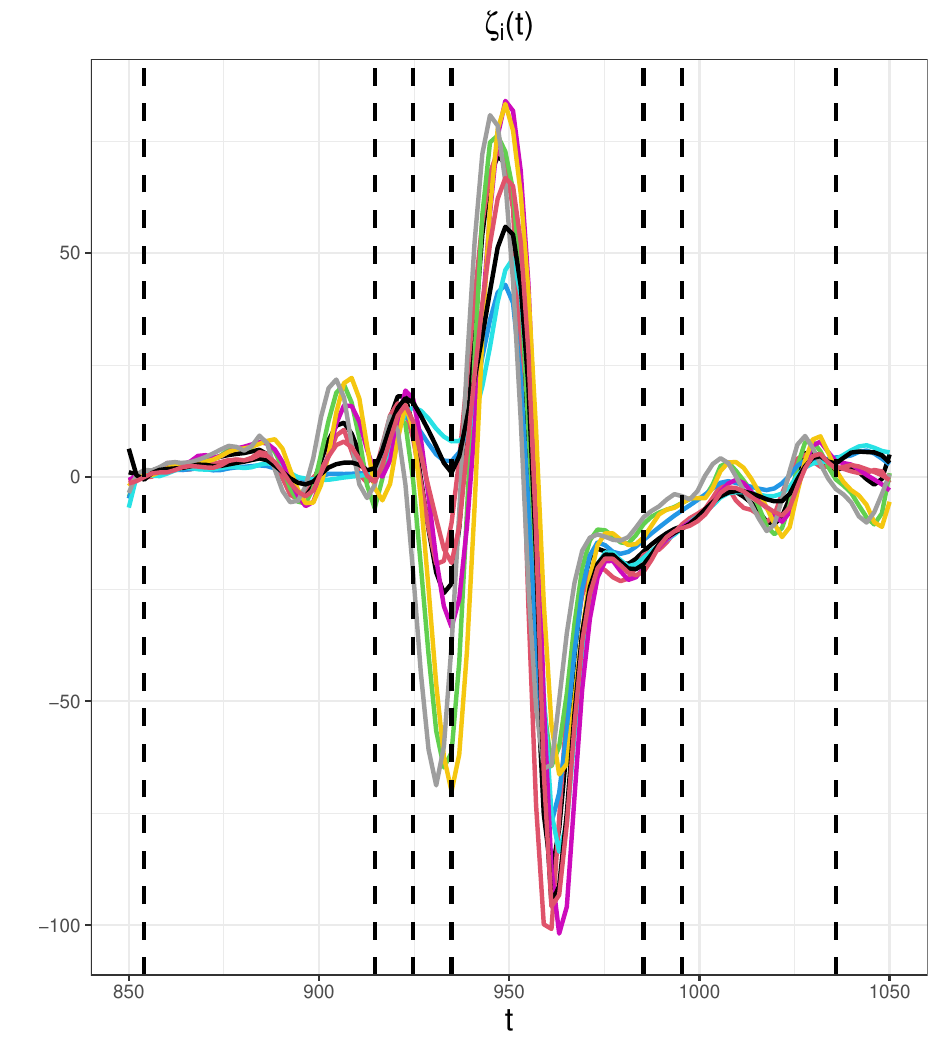}
	\caption{\texttt{FASSMR.kNN.fit()} }
\end{subfigure}
	\begin{subfigure}{0.25\textwidth}
	\includegraphics[width=1\linewidth]{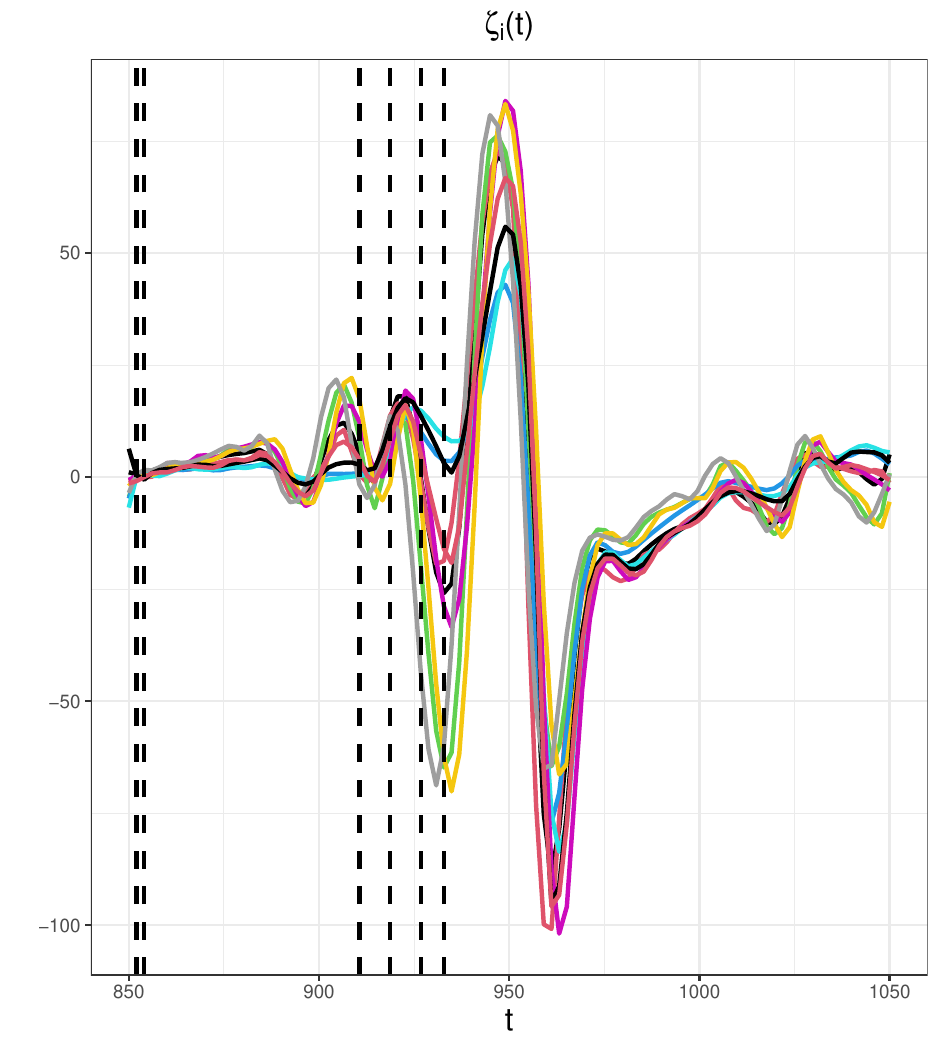}
	\caption{\texttt{IASSMR.kernel.fit()}}
\end{subfigure}
	\begin{subfigure}{0.25\textwidth}
	\includegraphics[width=1\linewidth]{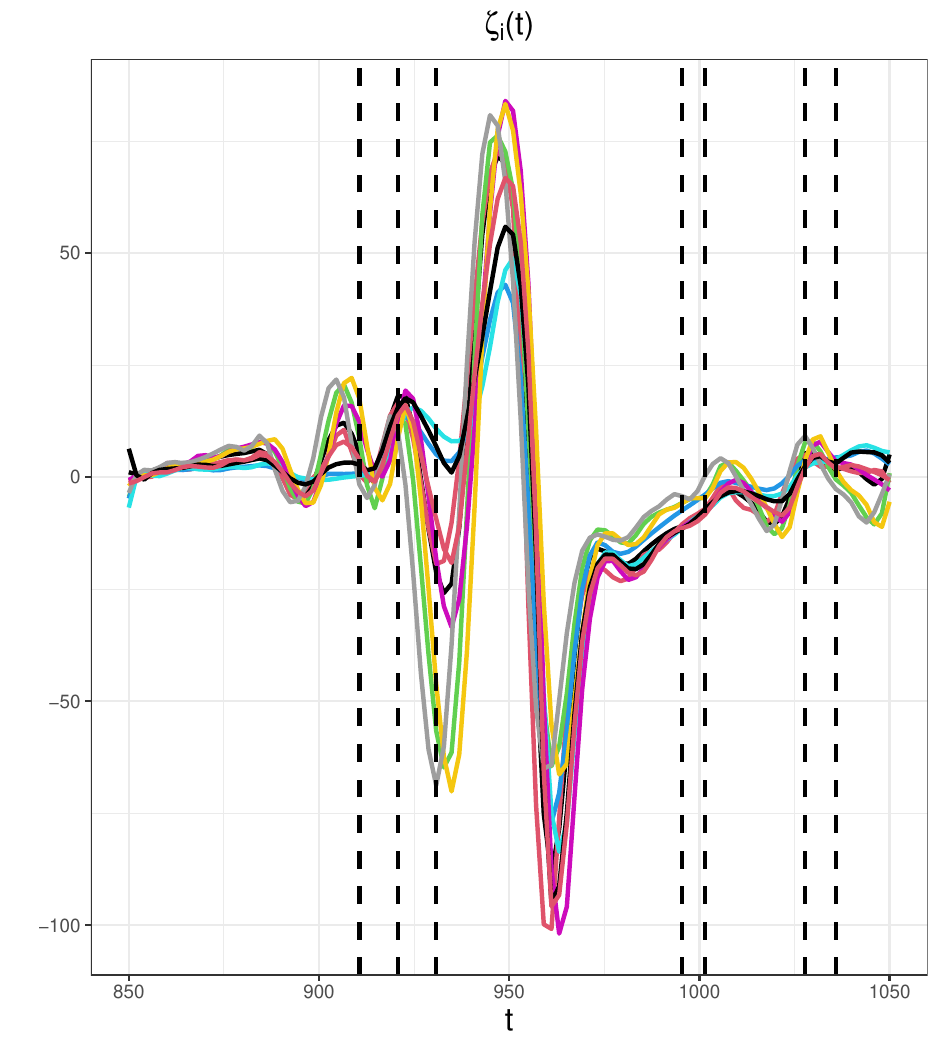}
	\caption{\texttt{IASSMR.kNN.fit()}}
\end{subfigure}
	\caption{Outputs of the S3 method \texttt{plot()} applied to the fitted objects of functions for the MLM, MFPLM and MFPLSIM, using the corresponding BIC selectors for the tuning parameters.}
	\label{fig:MFPLSIM}
\end{figure}

Finally, the output of the S3 method \texttt{plot()} for the functions in Table \ref{tab:MFPLSIM} represents the curves with their impact points (in dashed vertical lines), the same two diagnostic plots as the functions of Section \ref{sec:SFPLMfsemipar}, and the estimated functional index for the MFPLSIM fitting functions. In Figure \ref{fig:MFPLSIM}, we provide the outputs of the 7 main functions (indicating \texttt{option=1} to get just the first plot), obtained under the same conditions as Table \ref{tab:MFPLSIMMSEP}.

\subsubsection{Case study II: Sugar dataset}\label{sec:Sugar}

The purpose of this section is to demonstrate the implementation of the models (\ref{expr:MFPLM}) and (\ref{expr:MFPLSIM}) when dealing with two functional covariates. Additionally, it addresses scenarios where the discretisation size may exceed the sample size in the second step of the algorithms. 

\begin{figure}[h] 
	\centering
	\includegraphics[width=0.5\textwidth]{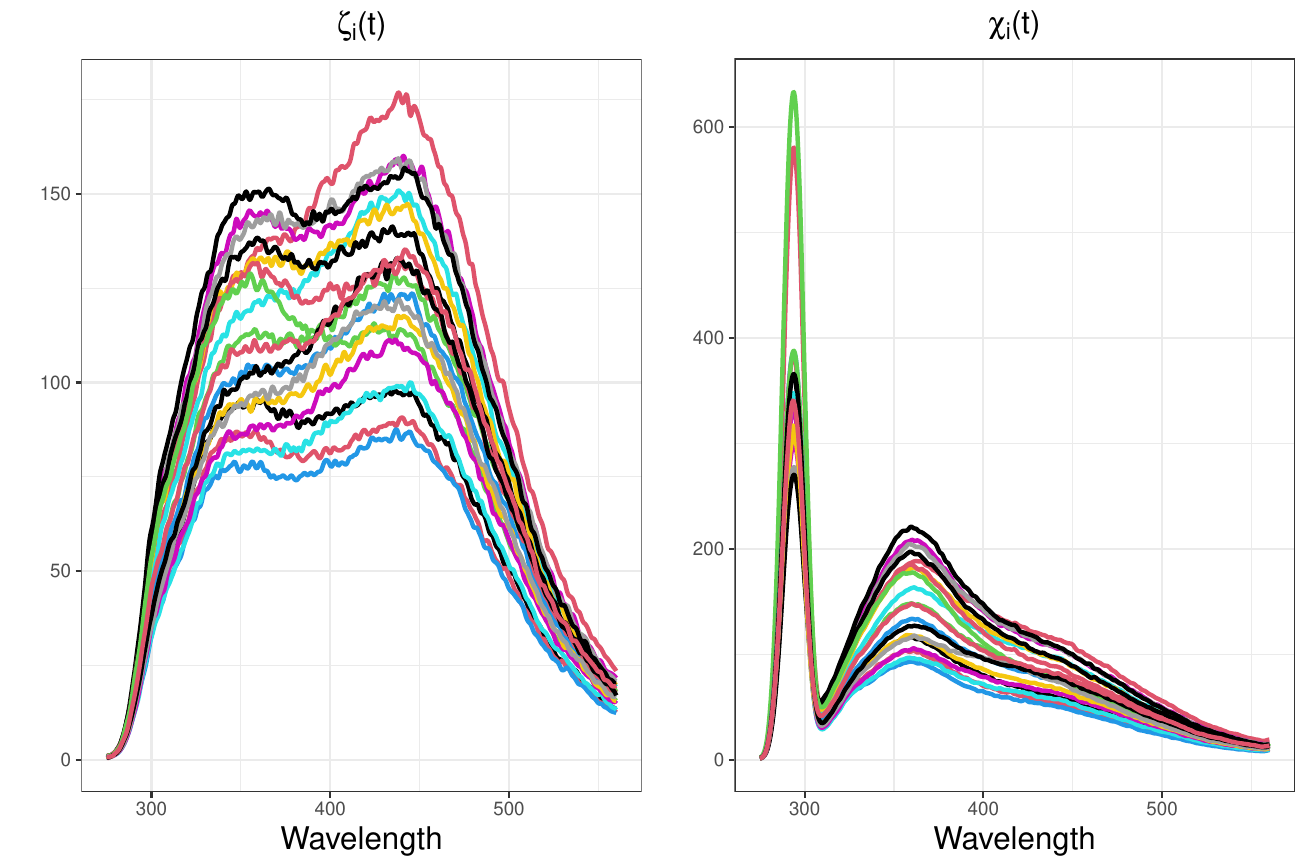}
	\caption{Sample of 20 spectrometric curves measured at excitation wavelengths $240\,\text{nm}$ (left panel) and $290\,\text{nm}$ (right panel).}
	\label{fig:sugar}
\end{figure}

The Sugar dataset contains information about $268$ samples obtained at a sugar plant in Scandinavia (see \url{https://ucphchemometrics.com/sugar-process-data/} for details).
For each sample, the absorbance spectra from $275\,\text{nm}$ to $560\,\text{nm}$ were measured in $0.5\,\text{nm}$ intervals ($p_n=571$). But in this case, unlike Tecator dataset,  measurements at several different excitation wavelengths were recorded, generating various functional variables. In this application, we are going to consider two, those taken into account in \cite{NovoVieuAneiros2021}: the absorbance spectra at excitation wavelengths $240\,\text{nm}$ (the first functional variable, which will be denoted as $\zeta$) and at excitation wavelengths $290\,\text{nm}$ (the second functional variable, which will be named $\mathcal{X}$). Figure \ref{fig:sugar} shows a sample of 20 curves from each functional variable. 

 In addition, Sugar dataset also contains the ash content of sugar.
The target of the statistical analysis is to predict the ash content of sugar having the measurements of the absorbance spectra at those two excitation wavelengths. For that, we used models (\ref{expr:MFPLM}) and (\ref{expr:MFPLSIM}). In the bi-functional models, to select the most accurate effect of each functional predictor, continuous or pointwise-linear, we followed the study performed in \cite{NovoVieuAneiros2021}. They concluded that the absorbance spectra at excitation wavelengths $240\,\text{nm}$ ($\zeta(t)$) should be included with the pointwise-linear effect. Then, we fitted models
\begin{align*}
		Y_i&=\sum_{j=1}^{571}\beta_{0j}\zeta_i(t_j)+m(\mathcal{X}_i)+\varepsilon_i,  \quad i = 1, \dots, 216, \\
\end{align*}
and
\begin{align*}		
	Y_i &= \sum_{j=1}^{571} \beta_{0j} \zeta_i(t_j) + r(\langle \mathcal{X}_i, \theta_0 \rangle) + \varepsilon_i, \quad i = 1, \dots, 216. 
\end{align*}
From the original set of $268$ samples, $2$ were discarded as extreme outliers. As a consequence, the available sample size is $266$. To compare procedures we split the sample into two subsamples: a training sample composed of the first $n=216$ observations and a testing sample containing the $n_{test}=50$ last. 
\begin{verbatim}
	>data(Sugar)
	>y<-Sugar$ash
	>x<-Sugar$wave.290
	>z<-Sugar$wave.240
	>index.y.25 <- y > 25
	>index.atip <- index.y.25
	>(1:268)[index.atip]
	[1]  71 129
	>x.sug <- x[!index.atip,]
	>z.sug<- z[!index.atip,]
	>y.sug <- y[!index.atip]
	>train<-1:216
\end{verbatim}
The following code demonstrates the implementation of the MFPLM using the functions \texttt{PVS.kernel.\newline fit()} and \texttt{PVS.kNN.fit()}. We have chosen a semimetric based on derivatives, specified as \texttt{semimetric=\newline"deriv"} with \texttt{q=0} (default options). In each stage of the two-step procedures, we use half of the sample, setting $n_1=n_2=108$ by defining \texttt{train.1=1:108} and \texttt{train.2=109:216}. Consequently, $n_1 > w_n$, but $n_2$ may be smaller than the cardinality of $\mathcal{R}$ (denoted as $\sharp\mathcal{R}$). In such cases, it is advisable to specify \texttt{lambda.min.h}, \texttt{lambda.min.l}, and \texttt{factor.pn} to adjust the level of penalisation in each step of the PVS. Therefore, we set \texttt{lambda.min.l=0.05}, \texttt{lambda.min.h=0.08}, and \texttt{factor.pn=1}. This ensures that if $n_2<\sharp\mathcal{R}$, then \texttt{lambda.min=0.08}; otherwise, \texttt{lambda.min=0.05}.
\begin{verbatim}
>fit5 <- PVS.kNN.fit(x=x.sug[train,],z=z.sug[train,], y=y.sug[train],train.1=1:108,
train.2=109:216, lambda.min.l=0.05,lambda.min.h=0.08, max.knn=10,nknot=20,
criterion="BIC")

>fit6 <- PVS.kernel.fit(x=x.sug[train,],z=z.sug[train,], y=y.sug[train], train.1=1:108,
train.2=109:216, lambda.min.l=0.05,lambda.min.h=0.08, num.h = 10, max.q.h=0.35, 
nknot=20,criterion="BIC")
\end{verbatim} 
The objects \texttt{fit5} and \texttt{fit6} belong to an S3 class (see Table \ref{tab:MFPLSIM}), which implements S3 methods. The functionalities of these methods were detailed throughout the case studies in Sections \ref{sec:tecator}, \ref{sec:tecator2}, and \ref{sec:tecator3}. They include displaying a summary of the fitted models using \texttt{summary.PVS.kNN()} and \texttt{summary.PVS.kernel()}, respectively (alternatively, \texttt{print.PVS.kNN()} and \texttt{print.PVS.kernel()} can be used); obtaining graphical representations of the outputs through \texttt{plot.PVS.kNN()} and \texttt{plot.PVS.\newline kernel()}; and making predictions for the fitted objects using \texttt{predict.PVS.kNN()} and \texttt{predict.PVS.\newline kernel()}, respectively. These latter functions implement the three prediction possibilities outlined in Section \ref{sec:tecator3}. In addition, \texttt{predict.PVS.kNN()} implements \texttt{option=4}, which involves reselecting $k$ for predicting the functional nonparametric component of the model, but this time in a local manner. This selection is performed using the LOOCV criterion in the functional nonparametric model, applied to the complete training sample (i.e., \texttt{train=c(train.1,train.2)}).

Table \ref{tab:MFPLSIMMSEP2} shows the MSEP, setting \texttt{option=2}, and $\widehat{s}_n$ for each primary fitting function listed in Table \ref{tab:MFPLSIM} allowing for 2 curves in the model (i.e., the MFPLM and MFPLSIM fitting functions). The results were obtained using the respective BIC selectors for $\lambda$, $w_n$, and $h$/$k$ in two different scenarios: (i) \texttt{lambda.min=0.05} (i.e., \texttt{lambda.min.h=lambda.min.l=0.05}) and (ii) \texttt{lambda.min=0.05} and \texttt{lambda.min=0.08}, with \texttt{factor.pn=1}. It should be noted that in the 1-step procedures (\texttt{FASSMR.kernel\newline.fit()} and \texttt{FASSMR.kNN.fit()}), the same results will be obtained in both scenarios (i) and (ii) as long as $n>$\texttt{factor.pn}*$w_n$. For comparative purposes, the same values for common input arguments were used across all six functions. Additionally, $h$ and $k$ were selected using the same input options: setting \texttt{max.knn=10} for $k$NN-based estimation functions and \texttt{max.q.h=0.35} for kernel-based ones. The remaining arguments were set to their default options.

It is noteworthy that both the MFPLM and MFPLSIM fitting functions improve their predictive ability by setting \texttt{lambda.min.h=0.08}. Additionally, under this setting,  they tend to select an equal or lower number of impact points, leading to a simpler model. The functions fitting the MFPLSIM show a more significant improvement in MSEP and a greater reduction in the number of impact points selected. However, the best result in terms of MSEP is provided by the MFPLM fitted using \texttt{PVS.kNN.fit()}. 
\begin{table}[H]
	\centering
	\begin{tabular}{p{2cm}p{3.9cm}p{1.7cm}p{1.7cm}p{1.7cm}p{1.7cm}}
		\cmidrule[1.5pt](lr){1-6}  
		\textbf{Model} & \textbf{Function} & \multicolumn{2}{l}{\texttt{lambda.min=0.05}} & \multicolumn{2}{l}{\begin{tabular}[l]{p{1.7cm}p{1.7cm}}\texttt{lambda.min.l=0.05} \\
		  \texttt{lambda.min.h=0.08}\end{tabular}} \\
		\cmidrule[0.5pt](lr){3-6} 
			& & \textbf{MSEP}&
			 \textbf{$\hat{s}_n$}&\begin{tabular}{l}\textbf{MSEP}\end{tabular}&
			 \begin{tabular}{l}\textbf{$\hat{s}_n$}\end{tabular} \\
	\cmidrule[1.5pt](lr){1-6}
		\multirow{2}{*}{MFPLM} & \texttt{PVS.kernel.fit()} & 2.00 &10& 1.92&9\\ 
		& \texttt{PVS.kNN.fit()} & 1.20 &7&1.19&7\\ 
		\cmidrule[0.5pt](lr){1-6}
		\multirow{4}{*}{MFPLSIM} & \texttt{FASSMR.kernel.fit()} & 4.70  &6&4.70&6\\
		& \texttt{FASSMR.kNN.fit()} & 3.60  &8&3.60&8\\
		& \texttt{IASSMR.kernel.fit()} &3.28  & 11& 2.44&5\\
		& \texttt{IASSMR.kNN.fit()} &3.29 &8&1.85&4\\
		\cmidrule[1.5pt](lr){1-6}  
	\end{tabular}
	\caption{MSEP obtained for each function using the respective BIC selectors for the tuning parameters ($\lambda$, $w_n$ and $h$ or $k$).}
	\label{tab:MFPLSIMMSEP2}
\end{table}

\begin{table}[H]
	\centering
	\begin{tabular}{p{1.6cm}p{3.4cm}p{5.1cm}p{5.3cm}}
		\cmidrule[1.5pt](lr){1-4}  
		\textbf{Model} & \textbf{Function} & \multicolumn{1}{l}{\begin{tabular}{l}\texttt{lambda.min=0.05}\end{tabular}} & \multicolumn{1}{l}{\begin{tabular}{l}\texttt{lambda.min.l=0.05} \\
				\texttt{lambda.min.h=0.08}\end{tabular}} \\
		\cmidrule[0.5pt](lr){3-4} 
		& & \begin{tabular}{r}\textbf{Points of impact}\end{tabular} &
		\begin{tabular}{r}\textbf{Points of impact} \end{tabular}\\
		\cmidrule[1.5pt](lr){1-4}
		\multirow{2}{*}{MFPLM} & \texttt{PVS.kernel.fit()} &\impactlines{1, 4, 6, 10, 17, 37, 154, 232, 251, 401}
			 &\impactlines{1,4,6,10,17,37,154,232,401}\\ 
		& \texttt{PVS.kNN.fit()} & \impactlines{1, 7, 11, 13, 17, 232, 425} & \impactlines{1, 7, 11, 13, 17, 232, 425}\\
		\cmidrule[0.5pt](lr){1-4}
		\multirow{4}{*}{MFPLSIM} & \texttt{FASSMR.kernel.fit()} &  \impactlines{20, 96, 134, 248, 324, 476} &\\
		& \texttt{FASSMR.kNN.fit()} & \impactlines{20, 58, 96, 134, 172, 324, 400, 476}  & 
			\\
		& \texttt{IASSMR.kernel.fit()} & \impactlines{1, 5, 13, 18, 204, 215, 232, 359, 495, 511, 565} &\impactlines{1, 13, 17, 18, 348}
			 \\
		& \texttt{IASSMR.kNN.fit()} &\impactlines{40, 139, 157, 164, 200, 420, 566, 568
		} &\impactlines{1, 5, 17, 348}\\
		\cmidrule[1.5pt](lr){1-4}  
	\end{tabular}
	\caption{Impact points obtained for each function using the respective BIC selectors for the tuning parameters ($\lambda$, $w_n$ and $h$ or $k$).}
	\label{tab:MFPLSIMIP}
\end{table}

To gain a deeper understanding, Table \ref{tab:MFPLSIMIP} displays (as explained in Section \ref{sec:tecator3}) the impact points selected by the six main functions in scenarios (i) and (ii). In the case of 1-step functions, the selection remains the same under both scenarios, as previously detailed. Comparing the results of each method in the two settings, it is noteworthy that the PVS functions obtain almost the same impact point selection: exactly the same using the \texttt{PVS.kNN.fit()} function, and with just one additional impact point using \texttt{PVS.kernel.fit()}. However, for the IASSMR functions, there is a significant change in the impact point selection, both in the number and position of the impact points. This fact explains the substantial change in MSEP in the two scenarios.

Analysing the results in scenario (ii) for the two-step functions, we observe four distinct clusters of impact points. It appears that all algorithms consistently select a set of observations at the beginning of the first quartile of the timeline. Additionally, both IASSMR algorithms select an impact point in the middle of the third quartile, while the PVS algorithms choose it at the end of the third quartile. Futhermore, PVS procedures select another point at the end of the second quartile and \texttt{PVS.kNN.fit()} also selects an impact point at the beginning of the second quartile. 

Combining the information from Tables \ref{tab:MFPLSIMMSEP2} and \ref{tab:MFPLSIMIP}, the best choice of points appears to be made by the function \texttt{PVS.kNN.fit()}. Furthermore, in this real data analysis, it seems that the second step plays a crucial role in the IASSMR algorithms, as the results provided in MSEP and impact point selection by the FASSMR are substantially worse.

\section{Discussion}
This article introduces a new R package, \texttt{fsemipar} (see \citealt{AneirosNovo2023}), tailored for SoF semiparametric regression. The package implements various models with at least one functional predictor. The effects included for functional covariates are functional single-index, functional nonparametric, and pointwise linear. The link function of functional single-index components and the functional operator of the functional nonparametric ones are estimated using kernel smoothing techniques. This includes not only kernel-based estimation but also $k$NN-based estimation, allowing for both global and local smoothing. In the case of the pointwise linear effect of curves, the package includes algorithms to handle the high dependence between the discretized observations, allowing the selection of points of impact. This feature has proven to be highly beneficial from a predictive perspective in the real data applications included in the article.

The package \texttt{fsemipar} also allows the inclusion of scalar variables as predictors with a linear effect, enabling simultaneous estimation and variable selection using penalised least-squares procedures. Through the real data applications, the article demonstrates the capabilities of the package, providing a comprehensive review of its features. Despite the large number of optional arguments for each function, all functions require only the data as a mandatory argument, facilitating easy use.

Despite the wide range of possibilities it currently offers, the package will continue to be updated, striving to enhance its functionalities. This includes optimizing the computational cost of functions that estimate functional single-index components, exploring implementations in $\rm{C}_{++}$ using the \texttt{Rcpp} or \texttt{RcppArmadillo} packages. It also involves incorporating goodness-of-fit tests for functional semiparametric regression (see, e.g., \citealt{Chan2023}), and increasing the flexibility of the included models. Enhancements may involve increasing the flexibility in the effect of scalar variables. For instance, this could include accounting for measurement errors (see, e.g., \citealt{Zhuetal2020}) and varying coefficients (see, e.g., \citealt{Wuetal2010}), or handling responses that are missed at random (see, e.g., \citealt{Lingetal2019}).

	\bibliographystyle{apalike}
	
	\bibliography{Rpackage}

\begin{thebibliography}{}

\bibitem[Ait-Sa{\"i}di et~al., 2008]{AitSaidietal2008}
Ait-Sa{\"i}di, A., Ferraty, F., Kassa, R., and Vieu, P. (2008).
\newblock Cross-validated estimations in the single-functional index model.
\newblock {\em Statistics}, 42(6):475--494.

\bibitem[Amato et~al., 2006]{Amatoetal2006}
Amato, U., Antoniadis, A., and De~Feis, I. (2006).
\newblock Dimension reduction in functional regression with applications.
\newblock {\em Computational Statistics \& Data Analysis}, 50(9):2422--2446.

\bibitem[Aneiros et~al., 2015]{AneirosFerratyVieu2015}
Aneiros, G., Ferraty, F., and Vieu, P. (2015).
\newblock Variable selection in partial linear regression with functional
  covariate.
\newblock {\em Statistics}, 49(6):1322--1347.

\bibitem[Aneiros and Novo, 2024]{AneirosNovo2023}
Aneiros, G. and Novo, S. (2024).
\newblock {\em fsemipar: Estimation, Variable Selection and Prediction for
  Functional Semiparametric Models}.
\newblock R package version 1.1.1.

\bibitem[Aneiros et~al., 2022]{AneirosNovoVieu2022}
Aneiros, G., Novo, S., and Vieu, P. (2022).
\newblock Variable selection in functional regression models: A review.
\newblock {\em Journal of Multivariate Analysis}, 188:104871.

\bibitem[Aneiros and Vieu, 2014]{AneirosVieu2014}
Aneiros, G. and Vieu, P. (2014).
\newblock Variable selection in infinite-dimensional problems.
\newblock {\em Statistics {$\&$} Probability Letters}, 94:12--20.

\bibitem[Aneiros and Vieu, 2015]{AneirosVieu2015}
Aneiros, G. and Vieu, P. (2015).
\newblock Partial linear modelling with multi-functional covariates.
\newblock {\em Computational Statistics}, 30(3):647--671.

\bibitem[Aneiros-P{\'e}rez and Vieu, 2006]{AneirosVieu2006}
Aneiros-P{\'e}rez, G. and Vieu, P. (2006).
\newblock Semi-functional partial linear regression.
\newblock {\em Statistics \& Probability Letters}, 76(11):1102--1110.

\bibitem[Boente and Vahnovan, 2017]{Boente2017}
Boente, G. and Vahnovan, A. (2017).
\newblock Robust estimators in semi-functional partial linear regression
  models.
\newblock {\em Journal of Multivariate Analysis}, 154:59--84.

\bibitem[Breheny and Huang, 2011]{Breheny2011}
Breheny, P. and Huang, J. (2011).
\newblock {Coordinate descent algorithms for nonconvex penalized regression,
  with applications to biological feature selection}.
\newblock {\em The Annals of Applied Statistics}, 5(1):232 -- 253.

\bibitem[Breheny and Huang, 2015]{grpreg}
Breheny, P. and Huang, J. (2015).
\newblock Group descent algorithms for nonconvex penalized linear and logistic
  regression models with grouped predictors.
\newblock {\em Statistics and Computing}, 25:173--187.

\bibitem[Brockhaus et~al., 2020]{FDBoost2020}
Brockhaus, S., R\"ugamer, D., and Greven, S. (2020).
\newblock Boosting functional regression models with {FDboost}.
\newblock {\em Journal of Statistical Software}, 94(10):1--50.

\bibitem[Burba et~al., 2009]{Burbaetal2009}
Burba, F., Ferraty, F., and Vieu, P. (2009).
\newblock k-{Nearest Neighbour} method in functional nonparametric regression.
\newblock {\em Journal of Nonparametric Statistics}, 21(4):453--469.

\bibitem[Cardot et~al., 1999]{Cardotetal1999}
Cardot, H., Ferraty, F., and Sarda, P. (1999).
\newblock Functional linear model.
\newblock {\em Statistics and Probability Letters}, 45(1):11--22.

\bibitem[Chan et~al., 2023]{Chan2023}
Chan, L., Delsol, L., and Goia, A. (2023).
\newblock A link function specification test in the single functional index
  model.
\newblock {\em Advances in Data Analysis and Classification}.

\bibitem[Collomb, 1979]{Collomb1979}
Collomb, G. (1979).
\newblock Estimation de la r\'egression par la m\'ethode des k points les plus
  proches: propri\'et\'es de convergence ponctuelle, ({F}rench).
\newblock {\em Comptes Rendus de l'Acad\'emie des Sciences}, pages 245--247.

\bibitem[Dai and Genton, 2018]{DaiGenton2018}
Dai, W. and Genton, M.~G. (2018).
\newblock Multivariate functional data visualization and outlier detection.
\newblock {\em Journal of Computational and Graphical Statistics},
  27(4):923--934.

\bibitem[De~Boor, 1978]{deBoor2001}
De~Boor, C. (1978).
\newblock {\em A practical guide to splines}, volume~27.
\newblock springer-verlag New York.

\bibitem[Devroye et~al., 1994]{Devroyeetal1994}
Devroye, L., Gyorfi, L., Krzyzak, A., and Lugosi, G. (1994).
\newblock {On the Strong Universal Consistency of Nearest Neighbor Regression
  Function Estimates}.
\newblock {\em The Annals of Statistics}, 22(3):1371 -- 1385.

\bibitem[Fan, 1997]{Fan97}
Fan, J. (1997).
\newblock Comments on \textquotedblleft {Wavelets} in {Statistics}: A
  review\textquotedblright\ by {A. Antoniadis}.
\newblock {\em Journal of the Italian Statistical Society}, 6:131.

\bibitem[Fan and Li, 2001]{FanLi2001}
Fan, J. and Li, R. (2001).
\newblock Variable selection via nonconcave penalized likelihood and its oracle
  properties.
\newblock {\em Journal of the American Statistical Association},
  96(456):1348--1360.

\bibitem[Febrero-Bande et~al., 2017]{Febreroetal2017}
Febrero-Bande, M., Galeano, P., and González-Manteiga, W. (2017).
\newblock Functional principal component regression and functional partial
  least-squares regression: An overview and a comparative study.
\newblock {\em International Statistical Review}, 85(1):61--83.

\bibitem[Febrero-Bande and Gonz{\'a}lez-Manteiga, 2013]{Febrero2013}
Febrero-Bande, M. and Gonz{\'a}lez-Manteiga, W. (2013).
\newblock Generalized additive models for functional data.
\newblock {\em Test}, 22:278--292.

\bibitem[Febrero-Bande and {Oviedo de la Fuente}, 2012]{FebreroOviedo2012}
Febrero-Bande, M. and {Oviedo de la Fuente}, M. (2012).
\newblock Statistical computing in functional data analysis: The {R} package
  {fda.usc}.
\newblock {\em Journal of Statistical Software}, 51(4):1--28.

\bibitem[Ferraty et~al., 2013]{Ferratyetal2013}
Ferraty, F., Goia, A., Salinelli, E., and Vieu, P. (2013).
\newblock Functional projection pursuit regression.
\newblock {\em Test}, 22:293--320.

\bibitem[Ferraty et~al., 2003]{Ferratyetal2003}
Ferraty, F., Peuch, A., and Vieu, P. (2003).
\newblock Mod{\'e}le {\'a} indice fonctionnel simple.
\newblock {\em Comptes Rendus Math{\'e}matique de l'Acad{\'e}mie des Sciences
  Paris}, 336(12):1025--1028.

\bibitem[Ferraty and Vieu, 2006]{FerratyVieu2006}
Ferraty, F. and Vieu, P. (2006).
\newblock {\em Nonparametric Functional Data Analysis, Theory and Practice}.
\newblock Springer Series in Statistics. Springer-Verlag, New York.

\bibitem[Galeano et~al., 2015]{Galeanoetal2015}
Galeano, P., Joseph, E., and Lillo, R.~E. (2015).
\newblock The mahalanobis distance for functional data with applications to
  classification.
\newblock {\em Technometrics}, 57(2):281--291.

\bibitem[Goldsmith et~al., 2023]{refund2023}
Goldsmith, J., Scheipl, F., Huang, L., Wrobel, J., Di, C., Gellar, J.,
  Harezlak, J., McLean, M.~W., Swihart, B., Xiao, L., Crainiceanu, C., and
  Reiss, P.~T. (2023).
\newblock {\em refund: Regression with Functional Data}.
\newblock R package version 0.1-32.

\bibitem[Greven and Scheipl, 2017]{Greven2017}
Greven, S. and Scheipl, F. (2017).
\newblock A general framework for functional regression modelling.
\newblock {\em Statistical Modelling}, 17(1-2):1--35.

\bibitem[Kara-Zaitri et~al., 2017a]{karaetal2017b}
Kara-Zaitri, L., Laksaci, A., Rachdi, M., and Vieu, P. (2017a).
\newblock Data-driven k{NN} estimation in nonparametric functional data
  analysis.
\newblock {\em Journal of Multivariate Analysis}, 153:176--188.

\bibitem[Kara-Zaitri et~al., 2017b]{karaetal2017a}
Kara-Zaitri, L., Laksaci, A., Rachdi, M., and Vieu, P. (2017b).
\newblock Uniform in bandwidth consistency for various kernel estimators
  involving functional data.
\newblock {\em Journal of Nonparametric Statistics}, 29(1):85--107.

\bibitem[Kneip et~al., 2016]{Kneipetal2016}
Kneip, A., Po{\ss}, D., and Sarda, P. (2016).
\newblock {Functional linear regression with points of impact}.
\newblock {\em The Annals of Statistics}, 44(1):1 -- 30.

\bibitem[Li et~al., 2019]{lietal2019}
Li, Y., Huang, C., and Härdle, W.~K. (2019).
\newblock Spatial functional principal component analysis with applications to
  brain image data.
\newblock {\em Journal of Multivariate Analysis}, 170:263--274.
\newblock Special Issue on Functional Data Analysis and Related Topics.

\bibitem[Ling et~al., 2019]{Lingetal2019}
Ling, N., Kan, R., Vieu, P., and Meng, S. (2019).
\newblock Semi-functional partially linear regression model with responses
  missing at random.
\newblock {\em Metrika}, 82(1):39--70.

\bibitem[Ling and Vieu, 2018]{LingVieu2018}
Ling, N. and Vieu, P. (2018).
\newblock Nonparametric modelling for functional data: selected survey and
  tracks for future.
\newblock {\em Statistics}, 52(4):934--949.

\bibitem[Ling and Vieu, 2021]{LingVieu2020}
Ling, N. and Vieu, P. (2021).
\newblock On semiparametric regression in functional data analysis.
\newblock {\em WIREs Computational Statistics}, 13(6):e1538.

\bibitem[McKeague and Sen, 2010]{McKeagueSen2010}
McKeague, I.~W. and Sen, B. (2010).
\newblock {Fractals with point impact in functional linear regression}.
\newblock {\em The Annals of Statistics}, 38(4):2559 -- 2586.

\bibitem[McLean et~al., 2014]{McLeanetal2014}
McLean, M.~W., Hooker, G., Staicu, A.-M., Scheipl, F., and Ruppert, D. (2014).
\newblock Functional generalized additive models.
\newblock {\em Journal of Computational and Graphical Statistics},
  23(1):249--269.
\newblock PMID: 24729671.

\bibitem[Nadaraya, 1964]{Nadaraya1964}
Nadaraya, E.~A. (1964).
\newblock On estimating regression.
\newblock {\em Theory of Probability and Application}, 9:141--142.

\bibitem[Novo et~al., 2019]{NovoAneirosVieu2019}
Novo, S., Aneiros, G., and Vieu, P. (2019).
\newblock Automatic and location-adaptive estimation in functional single-index
  regression.
\newblock {\em Journal of Nonparametric Statistics}, 31(2):364--392.

\bibitem[Novo et~al., 2021a]{NovoAneirosVieu2020}
Novo, S., Aneiros, G., and Vieu, P. (2021a).
\newblock Sparse semiparametric regression when predictors are mixture of
  functional and high-dimensional variables.
\newblock {\em TEST}, 30:481--504.

\bibitem[Novo et~al., 2021b]{NovoVieuAneiros2021}
Novo, S., Vieu, P., and Aneiros, G. (2021b).
\newblock Fast and efficient algorithms for sparse semiparametric bi-functional
  regression.
\newblock {\em Australian and New Zealand Journal of Statistics}, 63:606--638.

\bibitem[Preda and Saporta, 2005]{PredaSaporta2005}
Preda, C. and Saporta, G. (2005).
\newblock {PLS} regression on a stochastic process.
\newblock {\em Computational Statistics {$\&$} Data Analysis}, 48(1):149--158.

\bibitem[{R Core Team}, 2022]{RCoreTeam}
{R Core Team} (2022).
\newblock {\em R: A Language and Environment for Statistical Computing}.
\newblock R Foundation for Statistical Computing, Vienna, Austria.

\bibitem[Ramsay et~al., 2022]{Ramsayetal2022}
Ramsay, J.~O., Graves, S., and Hooker, G. (2022).
\newblock {\em fda: Functional Data Analysis}.
\newblock R package version 6.0.5.

\bibitem[Ramsay and Silverman, 2005]{RamsaySilverman2005}
Ramsay, J.~O. and Silverman, B. (2005).
\newblock {\em Functional Data Analysis}.
\newblock Springer Series in Statistics. Springer-Verlag, New York, 2nd
  edition.

\bibitem[Reiss et~al., 2017]{Reissetal2017}
Reiss, P.~T., Goldsmith, J., Shang, H.~L., and Ogden, R.~T. (2017).
\newblock Methods for scalar-on-function regression.
\newblock {\em International Statistical Review}, 85(2):228--249.

\bibitem[Rice et~al., 2020]{Riceetal2020}
Rice, G., Wirjanto, T., and Zhao, Y. (2020).
\newblock Tests for conditional heteroscedasticity of functional data.
\newblock {\em Journal of Time Series Analysis}, 41(6):733--758.

\bibitem[Scheipl et~al., 2022]{CranTask2022}
Scheipl, F., Arnone, E., Hooker, G., Tucker, J.~D., and Wrobel, J. (2022).
\newblock {\em CRAN Task View: Functional Data Analysis}.
\newblock Version 2022-03-21.

\bibitem[Tibshirani, 1996]{Tibshirani1996}
Tibshirani, R. (1996).
\newblock Regression shrinkage and selection via the {LASSO}.
\newblock {\em Journal of the Royal Statistical Society: Series B},
  58:267--288.

\bibitem[Wang et~al., 2016]{Wang2016}
Wang, G., Feng, X.-N., and Chen, M. (2016).
\newblock Functional partial linear single-index model.
\newblock {\em Scandinavian Journal of Statistics}, 43(1):261--274.

\bibitem[Wang et~al., 2007]{Wangetal2007}
Wang, L., Chen, G., and Li, H. (2007).
\newblock {Group SCAD regression analysis for microarray time course gene
  expression data}.
\newblock {\em Bioinformatics}, 23(12):1486--1494.

\bibitem[Watson, 1964]{Watson1964}
Watson, G. (1964).
\newblock Smooth regression analysis.
\newblock {\em Sankhy\~a: The Indian Journal of Statistics Series A},
  26:359--372.

\bibitem[Wu et~al., 2010]{Wuetal2010}
Wu, Y., Fan, J., and M{\"u}ller, H.-G. (2010).
\newblock {Varying-coefficient functional linear regression}.
\newblock {\em Bernoulli}, 16(3):730 -- 758.

\bibitem[Yuan and Lin, 2005]{YuanLin2005}
Yuan, M. and Lin, Y. (2005).
\newblock {Model Selection and Estimation in Regression with Grouped
  Variables}.
\newblock {\em Journal of the Royal Statistical Society Series B: Statistical
  Methodology}, 68(1):49--67.

\bibitem[Zhu et~al., 2020]{Zhuetal2020}
Zhu, H., Zhang, R., and Zhu, G. (2020).
\newblock Estimation and inference in semi-functional partially linear
  measurement error models.
\newblock {\em Journal of Systems Science and Complexity}, 33:1179--1199.

\end{thebibliography}
	\end{document}